\documentclass[11pt]{article}
\usepackage{graphicx} 

\usepackage[dvipsnames,table]{xcolor}
\usepackage{pgfplots}
\pgfplotsset{compat=1.18}
\usepackage{graphicx,color}
\usepackage[T1]{fontenc}
\usepackage{times}
\usepackage[utf8]{inputenc}
\usepackage[longnamesfirst]{natbib}
\usepackage{float,graphicx,ucs}
\graphicspath{{figures/}}
\usepackage[outdir=figures/]{epstopdf}
\usepackage[brazil,portuguese,english]{babel}
\usepackage{setspace} % must come before hyperref
\usepackage[hidelinks]{hyperref}
\usepackage{authblk,amsfonts,amsmath,amsthm,amssymb,latexsym,bbm}
\usepackage{upgreek,palatino,geometry,booktabs,listings} 
\usepackage{mathtools}
\usepackage{longtable,caption,subcaption,threeparttable,verbatim}
\usepackage[inline]{enumitem}
\usepackage[title]{appendix}
\usepackage[nottoc]{tocbibind}
\usetikzlibrary{calc,patterns,angles,quotes}
\usepackage{cleveref}
\def\b{\beta}
\def\g{\gamma}
\def\d{\delta}
\def\e{\varepsilon}

\def\th{\theta}

\def\k{\kappa}

\def\m{\mu}

\def\G{\Gamma}
\def\D{\Delta}
\def\Th{\Theta}
\def\L{\Lambda}

\def\HH{\mathcal{H}}

\def\WW{\mathcal{W}}

\DeclareMathOperator{\real}{\mathbb{R}}
\DeclareMathOperator{\one}{\mathbbm{1}}

\def\ul{\underline}
\newcommand{\Abs}[1]{\left\lvert #1 \right\rvert}

\newcommand{\Paren}[1]{\left( #1 \right)}

\newcommand{\Brac}[1]{\left[ #1 \right]}

\newcommand{\Set}[1]{\left\{ #1 \right\}}

\newcommand{\de}{\mathop{}\!\mathrm{d}}

\makeatletter
\newcommand{\leqnos}{\tagsleft@true\let\veqno\@@leqno}
\newcommand{\reqnos}{\tagsleft@false\let\veqno\@@eqno}
\reqnos
\makeatother

\newtheoremstyle{plainew}% name of the style to be used
  {5pt}% measure of space to leave above the theorem. E.g.: 3pt
  {5pt}% measure of space to leave below the theorem. E.g.: 3pt
  {\itshape}% name of font to use in the body of the theorem
  {0pt}% measure of space to indent
  {\bfseries}% name of head font
  {.}% punctuation between head and body
  { }% space after theorem head; " " = normal interword space
  {\thmname{#1}\thmnumber{ #2}\thmnote{ (#3)}}

\makeatletter
\g@addto@macro\normalsize{%
  \setlength\abovedisplayskip{6pt plus 2pt minus 2pt}%
  \setlength\belowdisplayskip{6pt plus 2pt minus 2pt}%
  \setlength\abovedisplayshortskip{0pt plus 2pt}%
  \setlength\belowdisplayshortskip{3pt plus 2pt minus 1pt}%
}
\makeatother

\theoremstyle{plainew}
\newtheorem{proposition}{Proposition}

\theoremstyle{plainew}
\newtheorem{theorem}{Theorem}

\theoremstyle{plainew}
\newtheorem{assumption}{Assumption}

\theoremstyle{plainew}
\newtheorem{corollary}{Corollary}

\theoremstyle{remark}

\theoremstyle{definition}

\theoremstyle{remark}

\theoremstyle{plainew}
\newtheorem{lemma}{Lemma}

\theoremstyle{plainew}
\newtheorem*{lemma*}{Lemma}

\theoremstyle{plainew}

\theoremstyle{plainew}
\newtheorem{claim}{Claim}

\renewenvironment{proof}[1][\proofname]{{\bfseries #1. }}{\qed}
\DeclareMathOperator\supp{supp}
\DeclareMathOperator*{\argmax}{arg\,max}

\usepackage{datetime}
\newdateformat{specialdate}{\THEDAY~\monthname[\THEMONTH] \THEYEAR}

\title{Who and How? \\ Adverse Selection and Flexible Moral Hazard\thanks{We are deeply indebted to Hector Chade for valuable conversations and insightful suggestions. We are grateful to Roberto Corrao, Daniel Gottlieb, Daniel Kr\"ahmer, Paula Onuchic,  Alessandro Pavan, Doron Ravid and various seminar audiences for thoughtful feedback and discussions.}}

\date{\specialdate\today}

\author{Henrique Castro-Pires\thanks{Harvard Business School, \texttt{hcastropires@hbs.edu}.},\qquad  Deniz Kattwinkel\thanks{University College London, \texttt{d.kattwinkel@ucl.ac.uk}.},\qquad  Jan Knoepfle\thanks{Queen Mary University of London, \texttt{jan.knoepfle@gmail.com}.}}

\begin{document}

\maketitle 

\begin{abstract}
\noindent
 We characterize incentive-compatible mechanisms in environments with hidden types and flexible hidden actions. Our approach introduces extended recommendation schedules that also prescribe actions off path, after misreports. This yields a tractable and complete characterization of incentive compatibility, which includes a generalized integral monotonicity condition capturing the interaction between adverse selection and moral hazard. We illustrate the usefulness of the characterization in a range of applications, including franchise agreements, employment contracts, and optimal taxation, covering both risk neutrality and risk aversion, with and without limited liability. Among other results, we identify a broad class of environments in which menus of affine contracts are optimal for a profit maximizing principal.
 
\noindent  \textbf{JEL}: D82, D86
\end{abstract}

 % \newpage
% \tableofcontents

\defcitealias{georgiadis2024flexible}{GRS}
\section{Introduction}

Many contracting environments feature two informational frictions: agents have private information about their characteristics and take imperfectly observed or non-contractible  actions. An incentive-compatible menu of contracts must therefore accomplish two tasks at once---\emph{screen types} and \emph{incentivize behavior}. This joint structure arises in a wide range of contexts, including  compensation design, investment financing, and tax policy. In these cases, the chosen menu determines both \emph{who} selects into each contract and \emph{how} any agent behaves once he has chosen a contract.

Despite the prevalence of hidden types and hidden actions, a complete characterization of incentive compatibility has remained elusive. The core difficulty is the possibility of \emph{double deviations}: when types are private and actions are hidden, an agent may select a contract intended for a different type and then optimally adjust his action. As a result, screening incentives affect continuation effort choices, while effort incentives in turn shape which types prefer to mimic others. This interaction alters the structure of the incentive constraints and has rendered the joint problem analytically intractable outside special cases.\footnote{See \cite{castro2024disentangling} for a recent discussion of the relatively limited literature that jointly analyzes hidden types and hidden actions.}

This paper's main contribution is to provide a necessary-and-sufficient characterization of incentive-compatible mechanisms in environments with both hidden information and hidden actions. We study a framework in which the agent's action consists of flexibly choosing an output distribution at a type-dependent cost. The type and the distribution choice are privately observed by the agent; the principal only observes the selected contract and the realized outcome. Modeling actions as flexible distributions rather than a choice from a fixed parametric family allows for a tractable and general representation of the joint screening-and-incentive problem. We illustrate the scope of the characterization through applications to joint ventures, employment contracts, and optimal taxation, including settings with risk aversion and limited liability.

In our model, a principal offers a menu of contracts, that is, a collection of output-contingent wage schedules $w : X \to \mathbb{R}$. The agent privately observes his type $\theta$, selects a wage schedule from the menu, and then privately chooses a distribution $\mu$ over outputs $x$. The agent's preferences are additively separable in wages and costs: $\int_X u \circ w(x)\, \mu(\de x) - C(\theta,\mu)$, allowing for risk neutrality (linear $u$) or risk aversion (concave $u$). 
We impose two regularity conditions on the cost function $C(\th, \mu)$: differentiability in $\theta$, which yields an envelope representation of information rents, and G\^{a}teaux differentiability in $\mu$, which yields a first-order characterization of obedience \cite[in the sense of][]{georgiadis2024flexible}.

Applying the generalized revelation principle \citep{myerson1982optimal}, posting any  menu is outcome equivalent to a direct, truthful revelation mechanism where, for each type $\th$, the principal designs a contract $w_\th$ and recommends a distribution of outputs $\mu_\th$, such that agents are better off reporting truthfully and choosing the recommended distribution. Incentive compatibility (IC) then requires preventing three types of deviations: (i) misreporting, where the agent reports a false type but follows the recommended distribution for that report; (ii) disobedience, where the agent reports truthfully but deviates from the recommended distribution; and (iii) double deviations, where the agent both misreports and chooses a different distribution than the one recommended after his misreport. Pure adverse selection involves only constraints preventing (i) and pure moral hazard only (ii). In our setting, the interaction between the two gives rise to nontrivial cross-effects that must be addressed simultaneously to deter double deviations (iii). Our main result, Theorem~\ref{thm:Characterization}, provides four conditions that are jointly necessary and sufficient for incentive compatibility, with each condition having a transparent economic interpretation. 

The first condition is an \emph{envelope} representation of the agent's indirect utility function. As in pure screening problems, truthtelling implies that the indirect utility of the agent must be absolutely continuous with respect to his true type. Moreover, its derivative depends solely on the allocation delivered to that type. In our setting, this translates to dependence on the distribution recommended to each type, and allows us to write the information rents of each type solely as a function of the implemented profile of distributions.

The second condition refers to \emph{on-path obedience}. As in pure moral-hazard problems, an incentive-compatible mechanism must guarantee that agents are willing to follow the recommendations once they have truthfully reported their types. By \cite{georgiadis2024flexible}, obedience in a flexible moral-hazard setting is characterized by a first-order condition that determines the wage schedule up to a free type-dependent constant that shifts utility levels uniformly across output realizations. In the joint problem, the envelope condition in the previous step determines this utility-constant, so that each profile of recommended distributions completely characterizes the wages that ensure on-path obedience and deter local misreports after which the agent follows the resulting recommendation. We must, however, still consider the possibility of global misreports and double deviations. 

To address double deviations systematically, we extend the recommended distribution schedule to off-path histories, where $\mu_{\theta,\hat{\theta}}$ specifies the distribution recommended to a true type $\theta$ who selects the contract corresponding to report $\hat{\theta}$. This leads to the third condition in our characterization: \emph{off-path obedience}. For any profile of on-path distributions, the previous conditions determine the associated menu of wage schedules. Off-path obedience requires that, for each $\hat{\theta}\neq \theta$, the distribution $\mu_{\theta,\hat{\theta}}$ is optimal for type $\theta$ under the wage schedule designed for $\hat{\theta}$. Importantly, the off-path recommendations do not introduce an additional instrument: for each pair $(\theta,\hat\theta)$ we specify with $\mu_{\theta,\hat\theta}$ the agent's best responses to the contract $w_{\hat\theta}$. We impose a mild existence condition so that these best responses are well-defined and the extension is without loss.

The fourth condition deters non-local misreports and double deviations. Using the envelope representation, the constructed wage contracts, and the extended recommendations, we compare the utility from reporting truthfully and obeying the recommendation with the utility from misreporting and optimally adjusting the distribution. A mechanism is incentive compatible if and only if, for every type-misreport combination, the former is at least as large as the latter. 
By an envelope argument, we can represent this as a \emph{generalized integral monotonicity (GIM)} constraint: an integral inequality comparing the partial cost-derivative terms $\partial C(\th,\mu_\th)/\partial \th$ on-path to the corresponding terms $\partial C(\th,\mu_{\th,\hat{\theta}})/ \partial \th$ after misreports. When actions are contractible (so that after reporting $\hat\theta$ the agent is forced to take $\mu_{\hat\theta}$),
we have $\mu_{\theta,\hat\theta}=\mu_{\hat\theta}$ and (GIM) collapses to the standard integral monotonicity constraint from pure adverse selection.\footnote{If, in addition to effort being contractible, we further assume a single-crossing condition, for instance, $C(\th,\mu) = h(\th)\kappa(\mu)$ with a decreasing $h$, then (GIM) becomes the usual monotonicity constraint requiring $\kappa(\mu_{\th})$ to be increasing in $\th$. }

The first two conditions, envelope and on-path obedience, are necessary to prevent each type of deviation---misreporting or disobeying---separately. Given that these first two conditions must hold, the last two conditions, off-path obedience and (GIM), are then necessary and sufficient to prevent any other type of double deviation. As a result, (IC) is equivalent to these four conditions holding jointly.

Our characterization can be applied to analyze a broad class of incentive problems. On the agent's side, it accommodates both risk neutrality and risk aversion, as well as the presence or absence of limited liability or other bounds on transfers. Because the theorem characterizes incentive compatibility itself, it is independent of the principal's objective and can therefore be embedded in arbitrary optimization problems. Importantly, the result is not merely descriptive but constructive. For any profile of output distributions that satisfies the characterization, and after fixing the utility of a reference type (for instance, by setting the lowest type's utility equal to the outside option), there exists an associated menu of contracts with uniquely determined on-path payments that implements it.

%%%%%%%%% description of Examples
We illustrate the usefulness of the characterization in a series of applications. When the agent is risk neutral, transfers are unconstrained, and effort costs are multiplicatively separable in the type and the chosen distribution, we solve a relaxed problem that ignores the GIM constraint and identify a regularity condition under which relaxed optimum is globally incentive compatible. The optimum can then be implemented by a menu of affine contracts (Proposition~\ref{prop:multsep-linear-contracts}). Affine compensation is not imposed ex ante; it follows from the optimality conditions of the principal and the obedience constraints of the agent.

Two technologies illustrate the implications of Proposition~\ref{prop:multsep-linear-contracts}. In Example~1, the agent pays a relative-entropy cost to move output away from a baseline distribution, and the implementing contracts take the form of franchise agreements: an upfront fee in exchange for a fixed share of revenue. Private information lowers that share below its first-best level, and more productive types buy larger shares at higher fees. In Example~2, costs depend only on the expectation of a single index of output, and the curvature of that index determines the shape of the optimal distributions. When the index is convex, output dispersion is costly, the optimal distributions are degenerate, and they can also be implemented by target contracts that pay a fixed wage when output meets a performance standard and impose a penalty otherwise, much like efficiency wages. When the index is concave, dispersion is cheap, the optimal distributions are binary, and they can be implemented by single-bonus contracts. With an S-shaped index, low types take target contracts and high types take bonus contracts.

Example~3 turns to a risk-averse agent with logarithmic utility and relative-entropy costs. The principal internalizes that stronger incentives for one type raise the rents of every more productive type. Under risk neutrality, this force merely flattens the slope of an affine contract. Under log utility, information rents depend on the spread in log wages while the wage bill depends on the spread in wages, so the higher on the wage scale a dollar of extra pay is added, the less rent it generates for higher types. The principal therefore concentrates incentives at high outputs, and the optimal contract is strictly convex for every type except the highest, for whom there are no more productive types to leave rents to.

In Examples~1--3, the GIM constraint is slack at the relaxed optimum. Examples~4 and~5 turn to cases in which it binds. Example~4 presents an environment where regularity is violated and output is binary. With hidden types alone, implementability requires only that the probability of high output be weakly increasing across types. With moral hazard, on-path obedience links each success probability to the associated wage spread. GIM also requires the implied bonus for high output to be weakly increasing in type. The pure adverse-selection solution violates this requirement, and the optimal mechanism with both frictions pools types on the same bonus contract, which increases the  downward distortion for low types and decreases it for high types, compared to pure adverse selection. Example~5 adds agent-side limited liability to Example~1 and shows that GIM can bind even under regularity. The optimal contracts then provide inefficiently high-powered incentives over the range of strictly positive wages.

Finally, Example~6 applies the characterization to a Mirrlees-style optimal taxation problem with flexible moral hazard. When mean-preserving income spreads are costly, the optimum can be implemented with a single nonlinear tax schedule despite the joint presence of hidden information and hidden actions, establishing a version of the taxation principle in this environment.\footnote{The taxation principle does not follow immediately from the degeneracy of on-path distributions; the discussion following Proposition~\ref{prop:MirrleesExample} explains why. See \cite{poggi2025} for a discussion of when the taxation principle does (or does not) hold.}

\paragraph{Related literature.} The problem of optimal contracting with joint hidden information and hidden actions has a long history in the mechanism-design and regulation literature \citep[e.g.,][]{arrow1984economics}. A central difficulty is that, due to the possibility of double deviations, local incentive conditions are generally not sufficient. Accordingly, much of the classic literature characterizes only local implications of incentive compatibility \citep[e.g.,][]{baronBesanko1987, faynzilbergKumar1997}, or studies special cases in which the two information frictions can be handled separately \citep[e.g.,][]{laffontTirole1986, picard1987, melumadReichelstein1989, CGR1992Restud}.

Several recent contributions obtain implications for optimal mechanisms by identifying environments in which the problem simplifies. \citet{OT2013JET, gottliebMoreira2022,castroPiresMoreira2021} derive conditions under which the optimal mechanism collapses to a single contract for all types. Most closely related, \cite*{castro2024disentangling} analyze a joint adverse-selection/moral-hazard model in a traditional (parametric) moral-hazard environment. They  solve a relaxed program that imposes truthtelling and on-path obedience while ignoring the double-deviation constraints, and provide sufficient conditions under which the relaxed solution is fully incentive compatible. Our contribution is complementary: by working with flexible moral hazard, we obtain a necessary-and-sufficient characterization of incentive compatibility, explicitly capturing double deviations by a condition that can be enforced directly rather than verified ex post through sufficient conditions on the solution of a relaxed problem.

The characterization of incentive constraints preventing double deviations is closely connected to the role played by integral monotonicity in multidimensional \citep[]{R1987JMathE,CE2023JET, yang2022selling} and dynamic \citep[]{PST2014ECMA} screening. Our contribution is to show that in our setting with flexible actions, the double-deviation constraints can be summarized by a generalized integral monotonicity (GIM) condition, which extends integral monotonicity to environments with off-path action adjustments. This perspective connects our characterization to dynamic applications combining adverse selection and moral hazard, such as managerial compensation and turnover \citep[]{GP2012JPE, GARRETT2015775} and Mirrleesian taxation in settings with evolving private information \citep[]{MP2020WP,MP2025WP}.

On the flexible-moral-hazard side, \citet{holmstromMilgrom1987} introduce environments in which the agent can influence the distribution of outcomes. Subsequent work studies particular distributional technologies, including technologies in which mean-preserving spreads are costless \citep{barron2020} or based on $f$-divergences \citep{hebert2018, mattssonWeibull2023, bonhamRiggsCragun2024}. We build directly on \citet*[henceforth GRS]{georgiadis2024flexible} who provide a first-order characterization of implementable distributions in the pure flexible moral-hazard problem. We combine this implementability analysis with screening by extending classical recommendation schedules in the sense of \citet{myerson1982optimal} to incorporate recommendations following misreports, which yields a tractable representation of incentive compatibility in the joint problem.

Finally, a growing literature applies flexible distributional choice to problems in corporate finance, regulation, and bargaining \citep{ krahmer2025optimal, krahmer2025security, KRAHMER2026106205}. These papers feature different informational structures---in particular, the distributional-choice stage typically precedes (or is otherwise separated from) the adverse-selection component.\footnote{For related problems with flexible \textit{information acquisition} preceding screening, see \cite{mensch2022screening, mensch2026monopoly}.} \cite{K2026Restud} studies optimal taxation of very rich households in a flexible moral-hazard setting, focusing mostly on how marginal tax rates at the top of the income distribution vary with risk aversion. In a related environment with moment-based flexible moral-hazard costs and a binary productivity type, \citet{Liu2025} studies optimal contract design and derives sufficient conditions under which a profit-maximizing principal optimally offers a single full-range contract. Our focus is different: we provide a general characterization of incentive compatibility in environments with both hidden information and hidden actions, which can be combined with arbitrary principal objectives and constraints.

\paragraph{Roadmap.} Section~2 describes the environment and assumptions. Section~3 considers the benchmarks with each friction separately. Section~4 states and interprets our characterization theorem and shows how it can be used to verify incentive compatibility and construct implementing contracts. Section~5 presents applications. Section~6 discusses extensions and the scope of the approach.

%%%%%%%%%%%%%%%%%%%%%%%%%%%%%%%

\section{Model}\label{sec:model}

\paragraph{Environment.}
A principal contracts with an agent who generates costly output.  
Output is denoted by $x \in X $. The set $X \subset \real$ is either finite or a compact interval $[\ul  x, \bar x]$.\footnote{The restriction to a finite or connected set simplifies the exposition. The characterization result holds for any compact $X$.}   
The agent's action is to choose a probability measure over outputs
$\mu\in \D(X)$, where $\D(X)$ is the set of Borel probability measures, endowed with the weak${}^{*}$ topology. 
The agent's cost for output distributions, $C\colon \Th \times \D(X)\to \real_+$, depends on his private type $\theta\in \Th = [\ul \th, \bar \th]$, distributed with cdf $F \in \D(\Th)$, which admits a density $f>0$. His von Neumann-Morgenstern utility function over payments $u\colon\real\to\real$ is strictly increasing, continuous, and weakly concave.
 When the agent of type $\th$ chooses distribution $\mu$ and receives payment $w$, his payoff is
\begin{align}\label{eq:agent-utility}
    \int u\circ w(x)\mu(dx) - C(\th,\mu).
\end{align}

The cost function 
$C(\th,\mu)$
is weak${}^{*}$–continuous and convex in~$\mu$ for every~$\theta$
and bounded on its domain.
Throughout we impose the following conditions on the cost function $C$.

We say that $C(\theta,\cdot)$ is \emph{Gâteaux differentiable} at
$\mu\in\Delta(X)$ if there exists a continuous function
$c_\mu(\theta,\cdot):X\to\mathbb R$ such that
\[
\lim_{\varepsilon\searrow 0}
\frac{C\!\bigl(\theta,(1-\varepsilon)\mu+\varepsilon\mu'\bigr)
      -C(\theta,\mu)}{\varepsilon}
=
\int_X c_\mu(\theta,x)\,[\mu'-\mu](dx)
\qquad\text{for every }\mu'\in\Delta(X).
\]

\begin{assumption}\label{Ass:Gat}
For every $\theta\in\Theta$ and every bounded measurable $g:X\to\mathbb{R}$, $C(\theta,\cdot)$ is Gâteaux differentiable at every distribution $\mu\in\Delta(X)$ that maximizes $\int_X g(x)\,\mu(dx)-C(\theta,\mu)$.
\end{assumption}

Assumption~\ref{Ass:Gat} slightly relaxes the corresponding assumption in \citetalias{georgiadis2024flexible}, who require Gâteaux differentiability at every $\mu\in\D(X)$; their characterization, however, only uses differentiability at the distribution being characterized, so their results carry over.\footnote{The relaxation accommodates costs such as the Kullback--Leibler divergence, whose Gâteaux derivative is not defined at some distributions that cannot be implemented, like distributions assigning zero probability to some outcome in the support of the baseline distribution.} The most important difference, however, is that our cost function depends on the type $\th$. 
Assumption \ref{Ass:Diff} concerns the dependence of the cost on $\th$:
\begin{assumption}\label{Ass:Diff}
    $C(\cdot,\mu)$ is continuously differentiable for all $\mu\in \D(X)$. Further, there exists an integrable function $b:\Theta\rightarrow \mathbb{R}_{+}$ such that  $\Abs{ \frac{\partial C(\th,\mu)}{\partial \th} } \leq b(\th)$ for all $\mu\in\Delta(X)$ and almost all $\th\in \Th$.
\end{assumption}
In most examples we consider, higher types $\th$ refer to more productive agents with lower effort costs. However, assuming that $C(\cdot, \mu)$ be decreasing for all distributions $\mu \in \D(X)$ is too restrictive for some cases of interest. For example, consider the Kullback-Leibler divergence $C(\th, \mu) = D_{KL}(\mu \| \nu_\th)$ with respect to type-dependent baseline distributions   $\nu_\th$ that are increasing in $\th$ in the sense of first order stochastic dominance. In \Cref{sec:applications}, we discuss different specifications of how costs and types interact when presenting applications illustrating our results.   

\paragraph{Contracts.}
The game between the principal and the agent unfolds as follows: First, the principal commits to a menu of wage contracts \( W \subseteq \mathcal{W}\), where $\mathcal{W}$ is the set of all feasible contracts. Each feasible contract \( w \colon X \to \real \) is measurable, bounded, and potentially constrained by minimum and maximum allowed payments. That is,
\[\mathcal{W}=\{w \ | \  w:X \to \real, \; sup \ |w(X)| <+\infty, \; \underline{w} \leq w(x) \leq x+\overline{w}\},\]
where \( \underline{w} \) and \( \overline{w} \) belong to the extended reals and capture potential limited liability restrictions.\footnote{For instance, $\underline{w} = \overline{w}=0$ represents the usual bilateral limited liability case where wages cannot be negative nor exceed the generated output. Conversely,  $\underline{w}=-\infty$ and $\overline{w}=+\infty$ is the case without any limited liability constraints.} Then, the agent selects one contract from the menu or rejects all of them, in which case he gets his outside option with payoff $0$. 
If the agent selects a contract, he privately chooses a distribution $\mu \in \D(X)$. 
Finally, $x\sim \m$ realizes publicly and the principal pays $w(x)$ to the agent. 
Denote by $
    U(\theta,w,\mu)$, type~$\theta$'s expected utility from choosing distribution $\mu$ under contract $w$, given in  equation \eqref{eq:agent-utility}. 
    
\section{Benchmarks: the separate problems}

\label{sec:SepProbs}
We begin by analyzing the set of mechanisms that are implementable when only one friction---either moral hazard or adverse selection---is present at a time. Each friction introduces a distinct implementability constraint: obedience in the case of moral hazard and truthtelling in the case of adverse selection. These constraints not only appear in the general characterization with both frictions present; considering them in isolation also helps illustrate the effects of their interaction.

\subsection{Pure Adverse Selection}
Consider now the pure adverse-selection problem, where the principal does not observe the agent's type $\th$, but the output distribution $\mu$ is contractible. By the revelation principle, it is without loss to restrict attention to direct truthful revelation mechanisms. That is, schedules of the form $\{w_{\th}, \mu_{\th}\}_{\th\in\Th}$, in which each $w_{\theta}\in \mathcal{W}$ and the principal elicits a type report $ \th$ from the agent and imposes action $\mu_{\th}\in \D(X)$. We ask what collections  $\{ w_\th,  \mu_\th\}_{\th \in \Th}$ of wages and imposed output distributions satisfy truthtelling (TT):
\begin{align}
      &U(\th, w_\th, \mu_\th) \; \ge  \;   U(\th, w_{\hat \th}, \mu_{\hat \th}),  \hspace{15pt} \text{ for all } \th, \hat \th \in \Th. \tag{TT} \label{eq:pureAS_TT}
\end{align}
We can describe a mechanism based on the action $\mu_{\th}$ assigned and the indirect utility delivered $V(\th):= U(\th,w_{\th},\mu_{\th})$ to each type. Usual arguments in the adverse-selection literature\footnote{See \Cref{pureAS} in Supplementary Appendix \ref{sec:proofsAS} for a formal proof.} imply that truthtelling is equivalent to the following two conditions holding for all $\th \in \Th$:
\begin{enumerate}
    \item \textbf{Envelope:} 
         \vspace*{-.2em}
    $$V(\theta) = V(\underline{\theta})-\int_{\underline{\theta}}^{\theta} \frac{\partial}{\partial \th} C(t,\mu_{t})\de t,$$
    
    \item \textbf{Integral Monotonicity:} 
         \vspace*{-.2em}
    $$\int_{\hat{\theta}}^{\theta}\Big[- \frac{\partial C(t,\mu_{t})}{\partial \th }+ \frac{\partial C(t,\mu_{\hat{\theta}})}{\partial \th }\Big]\de t \geq 0 \quad \text{for all } \ \hat{\theta} \in \Th.$$
\end{enumerate}
The Envelope condition has the usual interpretation of the information rents that must be left to more productive types depending on the allocation---here output distributions---of less productive types. Integral Monotonicity has appeared in the literature on implementability in static environments where usual differentiability or single-crossing conditions are violated or allocations are high-dimensional (see \citealp{R1987JMathE,CE2023JET} and \citealp{yang2022selling}) as well as in dynamic environments (see \citealp{PST2014ECMA}).
A more familiar monotonicity constraint appears, for example, when $C(\th,\mu)= h(\th) \kappa(\mu)$ for some strictly decreasing $h:\Th\to (0,\infty)$. Then, the allocation problem becomes essentially one-dimensional and implementability is characterized by the pointwise condition that $\kappa(\mu_{ \th})$ be nondecreasing in $\th$, i.e. higher types must get costlier distributions.

The participation (IR) constraint, $ U(\th, w_\th, \mu_\th)  \ge  0$, is equivalent to $V(\th)\geq 0$. Wages must satisfy 
\begin{align}\label{wagesAS}
\int u\circ w_{\th}(x)\mu_{\th}(\de x) = V(\th)+C(\th,\mu_{\th}).
\end{align}
In general, many wage functions can achieve the desired utility level when there is no moral hazard. If the principal wants to minimize expected payments, then a constant wage is (weakly) optimal under pure adverse selection because the agent is (weakly) risk averse.

Finally, a profile of distributions is not implementable by any profile of wage functions if the distributions do not satisfy the integral monotonicity constraint, or if there exists no profile of wage functions such that equation \eqref{wagesAS} and potential limited liability constraints are satisfied.

\subsection{Pure Flexible Moral Hazard}
Consider now the pure moral-hazard problem, where the principal observes $\th$ but not the agent's chosen distribution. The principal can then directly condition the contract offered on the agent's type. We ask what collections $\{ w_\th,  \mu_\th\}_{\th \in \Th}$ of wages $w_\th \in \WW$ and recommended output distributions $\mu_\th \in \D(X)$ satisfy obedience (OB):
\begin{align}
    & U(\th, w_\th, \mu_\th) \; \ge \; U(\th, w_{\th}, \mu') 
    \hspace{20pt} \text{for all } \mu' \in \D(X)\ \text{and } \th\in \Th. \tag{OB} \label{eq:pureMH_OB} 
\end{align}
By Proposition 2 in \citetalias{georgiadis2024flexible}, for any $\th \in \Th$, obedience is satisfied if and only if
\begin{align}\label{wageMH}
u\circ w_{\theta}(x) \begin{cases}
       \le  c_{\mu_{\theta}}(\th, x)+m_{\th}
    \quad \text{for all } x \in X,
    \\
 =  c_{\mu_{\theta}}(\th, x)+m_{\th}  \quad \mu_{\theta}\text{-almost all } x\in X;
\end{cases}
\end{align}
for some output-independent constant $m_{\th}$. Thus, obedience pins down the on-path utilities from wages up to an additive constant. Below, when an implementing wage is said to be determined by a recommended distribution, this means almost-surely determined under that distribution. 
The values the additive constant $m_\th$  may take are determined by the participation (IR) constraint  $ U(\th, w_\th, \mu_\th)  \ge  0$, or, equivalently,   
\begin{align}
 m_{\th} \geq C(\th,\mu_{\th})-\int_X c_{\mu_{\th}}(\th,x)\mu_{\th}(\de x) \hspace{20pt} \text{ for all } \th \in \Th. \label{IRMH}  
\end{align}
Hence, a mechanism $\{ w_\th,  \mu_\th\}_{\th \in \Th}$ satisfies (OB) and (IR) if and only if it satisfies \eqref{wageMH} and \eqref{IRMH}.

Recall that, if any, limited liability constraints are directly captured by the set of available contracts $\mathcal{W}$. Hence, they need not be explicitly stated when checking if a mechanism satisfies (OB) and (IR). 
Alternatively, equations \eqref{wageMH} and \eqref{IRMH} can be used to determine what distributions $\mu_{\th}$ are implementable for type $\th$ by some wage function $w_{\th}$ that satisfies, for instance, the limited liability constraints $\ul w \leq w(x) \leq \overline{w} +x$ for all $x\in X$. This is the case if and only if there exists a constant $m_{\th}$ satisfying
\small
\begin{align}\label{LL}
\max \Big\{C(\th,\mu_{\th})-\int c_{\mu_{\th}}(\th,x)\mu_{\th}(\de x)\, ,\sup_{x \in X}  \big\{ u(\ul w) - c_{\mu_{\th}}(\th, x)\big\} \Big\} \; \leq \; \ m_{\th} \; \leq \; \inf_{x\in \supp{\mu_\th}}\big\{u(\overline{w} +x)-c_{\mu_{\th}}(\th,x)\big\}.
\end{align}
\normalsize
The first inequality guarantees the constant $m_{\th}$ is large enough to satisfy \eqref{IRMH} and the agent's side of the limited liability constraint, $w_\th \ge \ul w$. The second inequality guarantees the principal's side of the limited liability constraint, $w_\th(x) \le \overline{w} + x$.\footnote{See Supplementary Appendix \ref{sec:proofsMH} for the formal statement and proof of such a claim.} Note that, as in \citetalias{georgiadis2024flexible}, if $u$ is unbounded above and the principal has no limited liability constraint---equivalently, $\overline{w} =+ \infty$---any profile of distributions at which the required cost Gâteaux derivatives exist can be implemented, as there always exists a large enough $m_{\th}$ satisfying all inequalities. In contrast, with a smaller $\overline{w}$ not all profiles are implementable, as the set of $m_\th$ satisfying \eqref{LL} may be empty for some $\mu_\th \in \D(X)$.

\section{Characterization in the joint problem}

Consider the joint case where the principal does not observe the type and output distributions are not contractible. Invoking the generalized revelation principle \citep{myerson1982optimal}, it is without loss to focus on direct revelation mechanisms $\{w_{{\theta}}, \mu_{\th}\}_{{\theta}\in\Th}$ with $(w_{\theta}, \mu_{\th }) \in \WW \times \D(X)$ for all $\th$, such that it is optimal for the agent to truthfully reveal his type $\th$ and choose the recommended distribution $\mu_{ \th}$. We refer to mechanisms in this class as \textit{incentive-compatible} (IC) mechanisms:  
\begin{equation}\label{eq:IC}
    U(\th, w_\th, \mu_\th) \; \ge  \;   U(\th, w_{\hat \th}, \mu') \quad \text{ for all } \hat \th \in \Th, \text{ and all } \mu'\in \D(X). \tag{IC}
\end{equation}

Incentive compatibility \eqref{eq:IC} rules out any profitable deviation, jointly with respect to reports and actions. In particular, \eqref{eq:IC} implies the on-path obedience \eqref{eq:pureMH_OB} and truthtelling \eqref{eq:pureAS_TT} conditions from the pure moral-hazard and adverse-selection problem, respectively. 
While \eqref{eq:pureAS_TT} and \eqref{eq:pureMH_OB} are necessary conditions for \eqref{eq:IC}, they are not generally sufficient. 
One of the main challenges in determining optimal contracts with hidden types and hidden actions is the potential for \textit{double deviations}, where the agent first reports another type $\hat \th \ne \th $ and then chooses an action different from the recommendation $\mu_{\hat\th}$ associated with report $\hat{\th}$. 

To analyze all deviations in a systematic way, we extend the typical direct revelation mechanisms: in addition to the schedule $\{\mu_{\hat \th}\}$ that recommends  a distribution for each (truthful) report, we specify a recommended distribution for each combination of the agent's (unverifiable) true type $\th$ and  his report $\hat \th$. Thus, an \emph{extended recommendation schedule} is given by $\{\mu_{\th, \hat{\th}}\}_{(\th, \hat{\theta})\in\Th^2}$, where $\mu_{\th, \hat{\th}}$ is the distribution the principal recommends to a true type~$\theta$ if he reports $\hat{\theta}$ (i.e., selects contract $w_{\hat{\theta}}$). Hence, extended recommendation schedules impose obedience  both on path, $ U(\th, w_\th, \mu_{\th,\th}) \ge U(\th,w_\th, \mu')$, and off path, $ U(\th, w_{\hat \th}, \mu_{\th,\hat \th}) \ge U(\th,w_{\hat \th}, \mu')$. 

It is not a priori clear that imposing this additional structure on a mechanism is without loss. If type $\th$ misreports $\hat \th$, there might not exist an optimizer among the set of output distributions. We impose \Cref{Ass:Existence} below to ensure that an optimizer $\mu_{\th,\hat \th}$ always exists so that extending the recommendation schedule does not restrict the set of feasible outcomes. 
 Since  this is an assumption on an endogenous object, we then provide sufficient conditions on model primitives that imply \Cref{Ass:Existence}.
\begin{assumption}\label{Ass:Existence}
For all feasible contracts $w \in \WW\colon$
$$\arg\max_{\mu \in \D(X)} U(\th, w, \mu) \ne \varnothing, \qquad \text{ for all } \th \in \Th.$$    
\end{assumption}
If the principal wants to implement distribution $\mu_\th$ for type $\th$, the existence of a maximizer of $U(\th, w_\th, \cdot)$ follows from incentive compatibility. The substantive part of \Cref{Ass:Existence} is to ensure that even after a misreport $\hat \th \ne \th$, the supremum of $\{U(\th, w_{\hat \th}, \mu)\}_{\mu \in \D(X)}$ is attained at some $\mu_{\th,\hat \th} \in \D(X)$. This will allow for a convenient characterization of incentive compatibility in \Cref{thm:Characterization} below by explicitly controlling the most attractive continuation play after a misreport.

\paragraph{Sufficient conditions.} The following results provide simple sufficient conditions for \Cref{Ass:Existence} to be without loss. Readers may skip them on a first reading and proceed directly to \Cref{thm:Characterization}.
\begin{lemma}\label{lm:SemiContWages}
    If $w\colon X \to \real$ is upper-semicontinuous, then $\arg\max_{\mu \in \D(X)} U(\th, w, \mu) \ne \varnothing$ for all $\th \in \Th$.
\end{lemma}
All proofs not contained in the main text are relegated to Appendix \ref{sec:proofs}. Given \Cref{lm:SemiContWages}, each of the following conditions on the environment either implies \Cref{Ass:Existence} directly or that the restriction is without loss for implementability. 
\begin{corollary}\label{cor:finiteX-existence}
    If the space of outputs $X$ is finite, then for any $w \in \WW$ and all  $\th \in \Th\colon$ $\arg\max_{\mu \in \D(X)} U(\th, w, \mu) \ne \varnothing$.
\end{corollary}
If the output space $X= \{x_1, \cdots, x_n\}$ is finite, then $\int_X u \circ w(x) \mu(\de x) = \sum_i u \circ w(x_i) \mu(x_i) $ is trivially continuous in $\mu$ for all $w\colon X \to \real$, implying that $\mu \mapsto U(\th, w, \mu)$ is continuous.  

We say that $C(\th,\cdot)$ is FOSD (first-order stochastic dominance) increasing if whenever $\mu'$ dominates $\mu$ in FOSD, then $C(\theta,\mu')\ge C(\theta,\mu)$.
\begin{lemma}\label{lm:MonotoneWages} 
Assume $C(\th,\cdot)$ is FOSD-increasing for all $\th \in \Th$. 
    Take any mechanism $\{ w_\th,  \mu_\th\}_{\th \in \Th}$ that satisfies \eqref{eq:IC}. 
    The same outcome can be implemented with a menu of wage contracts $ \{ \overline{w}_\th\}_{\th \in \Th} $ such that for all $\th$, the wage $\overline{w}_\th\colon X\to \real$ is nondecreasing and right-continuous. 
    That is, $\mu_\th \Paren{ \{x \colon  \overline{w}_\th(x) \ne w_\th(x) \}} = 0$ for all $\th\in \Th$; and  $\{ \overline{w}_\th,  \mu_\th\}_{\th \in \Th}$ satisfies \eqref{eq:IC}.
\end{lemma}
Thus, if $C(\th,\cdot)$ is FOSD-increasing, it is without loss to consider nondecreasing and right-continuous (hence upper-semicontinuous) wage contracts.
 Given monotonicity of the cost in output, it is intuitive that the agent facing wages $w$ puts zero mass on any outcome $x$ for which there exists an outcome $y <x$ with $w(y)\ge w(x)$. Similarly, the left-limit at an upward jump of $w$ is dominated. The proof of \Cref{lm:MonotoneWages} also confirms that using monotone and right-continuous wages does not create additional profitable deviations. 

We turn to our main characterization result. 
\begin{theorem}\label{thm:Characterization}
A mechanism $\{w_\th,\mu_\th\}_{\th\in\Th}$ is incentive compatible if and only if there exists an extended recommendation schedule $\{\mu_{\th,\hat\th}\}_{(\th,\hat\th)\in\Th^2}$ with $\mu_{\th,\th}=\mu_\th$ and an absolutely continuous function $V$ such that $V(\th) = U(\th, w_\th, \mu_\th)$ and the following conditions are jointly satisfied for all $\th\in\Th$:
\begin{enumerate}
    \item \textbf{Envelope:} \label{it:Envelope}
         \vspace*{-.2em}
    $$V(\theta) = V(\underline{\theta})-\int_{\underline{\theta}}^{\theta} \frac{\partial}{\partial \th} C(t,\mu_{t})\de t;$$
    
    \item \textbf{On-path Obedience:} \label{it:On-path-OB}
         \vspace*{-.2em}
$$ u\circ w_{\theta}(x) \begin{cases}
       \le  c_{\mu_{\theta}}(\th, x)+V(\theta)+C(\th,\mu_{\theta})-\int_X c_{\mu_{\theta}}(\th,y)\mu_{\theta}(\de y)
    \quad \text{for all } x \in X,
    \\
 = c_{\mu_{\theta}}(\th, x)+V(\theta)+C(\th,\mu_{\theta})-\int_X c_{\mu_{\theta}}(\th,y)\mu_{\theta}(\de y) \quad \mu_{\theta}\text{-almost all } x\in X;
\end{cases}
$$
     \item \textbf{Off-path Obedience:} \label{it:Off-path-OB}
     \vspace*{-.2em}
\begin{equation*}
    \mu_{\th, \hat{\th}} \in \underset{\nu\in \D(X)}{\argmax} \int_X \Big[u\circ w_{\hat{\theta}}(x)-c_{\mu_{\th, \hat{\th}}}(\th,x)\Big]\nu(\de x) \qquad \text{ for all } \hat\th \ne \th;
\end{equation*}

    \item \textbf{Generalized Integral Monotonicity:} \label{it:GIM}
         \vspace*{-.2em}
    $$\int_{\hat{\theta}}^{\theta}\Big[- \frac{\partial C(t,\mu_{t})}{\partial \th }+ \frac{\partial C(t,\mu_{t,\hat{\theta}})}{\partial \th }\Big]\de t \geq 0 \quad \text{for all } \ \hat{\theta}.$$
\end{enumerate}
\end{theorem}

Introducing extended recommendation schedules enables us to explicitly keep track of the relevant continuation strategy for each type after any misreport. Combining this approach with flexible moral hazard, which gives necessary and sufficient first-order conditions for wages to implement any action, is particularly useful: the wages that implement any on-path recommendation  
 schedule $\mu_{\hat \th} = \mu_{\hat \th, \hat \th}$ directly pin down the off-path recommendations $\mu_{\th,\hat\th}$ for all true types $\th$. 

 The envelope and first-order conditions in items \ref{it:Envelope} and \ref{it:On-path-OB} are  the standard conditions in the isolated adverse-selection \citep[item \ref{it:Envelope}, see e.g.][]{milgrom2002envelope} and flexible moral-hazard problems \citepalias[item \ref{it:On-path-OB}, see][]{georgiadis2024flexible}.

Items \ref{it:Off-path-OB} and \ref{it:GIM} jointly deal with the interplay between adverse selection and moral hazard. \Cref{it:Off-path-OB} ensures that $\mu_{\th,\hat\th}$ is indeed an optimal continuation strategy after type $\th$ reports $\hat\th$. 
Note that imposing the obedience condition in \cref{it:Off-path-OB} for $\hat \theta = \th$ would imply the condition on wages in \cref{it:On-path-OB}, with one additional degree of freedom: to satisfy obedience alone, the term $V(\theta)+C(\th,\mu_{\theta})-\int_X c_{\mu_{\theta}}(\th,y)\mu_{\theta}(\de y)$ could be replaced by any constant (independent of $x$). The truthtelling constraint from the hidden-type component determines this constant in \cref{it:On-path-OB}. 

Finally, \cref{it:GIM} generalizes the monotonicity constraint in pure adverse-selection problems. 
To see the parallels, suppose the agent's action $\mu$ was contractible so that deviations from the recommended choice can be deterred. Then $\mu_{\th, \hat \th} = \mu_{\hat \th}$ for any true type $\th$ and Generalized Integral Monotonicity in \Cref{thm:Characterization} reduces to the Integral Monotonicity condition under pure adverse selection in \Cref{sec:SepProbs}.
In the pure case, we argued that the monotonicity constraint simplifies to pointwise monotonicity of $\k(\mu_{\hat \th})$ in $\hat \th$ when the cost function is $C(\th, \mu) = h(\th) \k(\mu)$ for some strictly decreasing $h\colon \Th \to (0,\infty)$.
In the joint problem, however, the monotonicity condition in \cref{it:GIM} must take into account the potential for double deviations in which $\mu_{\th, \hat \th}$ may differ from $\mu_{\hat \th}$. The following observation illustrates how this leads to a different pointwise monotonicity condition.
\begin{corollary}\label{cor:GenMon-MultSeparableCost}
Suppose \( C(\theta, \mu) = h(\theta)\kappa(\mu) \), where \( h\colon \Th \to (0,\infty) \) is strictly decreasing. If 
\begin{align*}
    \k (\mu_{\th, \hat \th}) \quad \text{ is nondecreasing in the report } \hat \th \quad \text{ for all } \th \in \Th,
\end{align*}
then the generalized integral monotonicity condition in \cref{it:GIM} of Theorem~\ref{thm:Characterization} holds.
\end{corollary}
The monotonicity condition above states that if a type $\th$ (mis)reports a more productive type, then type $\th$ must find it optimal to choose a costlier distribution. This condition means that contracts associated with higher types must not only implement ``higher'' allocations but also provide higher-powered incentives: pretending to be more (less) productive implies that the agent faces a contract that generates incentives for a more (less) costly continuation. 

While this property is sufficient to satisfy generalized integral monotonicity, it is not necessary, even under the multiplicative cost specification. For example, consider a menu consisting of two distinct single-bonus contracts, $w_{\hat \th_1}(x) = b_1 \one_{\{x \ge x_1\}}$ and $w_{\hat \th_2}(x) = b_2 \one_{\{x \ge x_2\}}$,  with $b_1 < b_2$ and $x_1 < x_2$. It may well be that an agent of a fixed type exerts costly effort to choose distribution $\d_{x_1}$ in order to collect $b_1$ after reporting $\hat \th_1$, but will not exert effort (choose $\d_{0}$) after reporting $\hat\th_2$ if $x_2$ is sufficiently high. This would violate the monotonicity described above if $\hat \th_1 < \hat \th_2$. Nonetheless, such a binary menu can still be incentive-compatible and, hence, satisfy generalized integral monotonicity. The detailed argument is provided in Example 7 in Online Appendix~\ref{app:nonnencessity}.

%%%%%%%%%%%%%%%%%%%%%%%%%%%%%%%%%%%%%%%%%%%%%%%%%%%%%%%%%

\subsection{How to use Theorem \ref{thm:Characterization}}
Theorem \ref{thm:Characterization} provides a practical framework for solving problems with combined moral hazard and adverse selection. At first glance, one might expect the flexible moral-hazard framework to complicate matters by enlarging the space of potential deviations. However, the characterization in GRS shows that this added flexibility can sharpen implementability rather than obscure it. In contrast to the parametric effort setting, where multiple contracts might induce the same effort level, the flexible approach tightly links distributions to contracts: each feasible distribution for a given type determines a payment function on its support that is unique up to an additive constant. Once adverse selection is introduced, this constant is pinned down by the envelope condition that ensures truthtelling, thereby fully identifying on-path payments, while providing upper bounds to payments off the support of the implemented distributions. These off-support payments can then be chosen at the minimal feasible level.\footnote{If instead payments are not bounded below, one can set them at a sufficiently low level so that these outputs remain off the support for true type that chooses that given contract.}

This link simplifies the incentive compatibility analysis. Once a profile of on-path recommendations is fixed, Theorem \ref{thm:Characterization} can be used both as an algorithm to verify whether the profile is implementable and, if so, to construct a menu that implements it; and as a method to solve analytically for the optimal contracts. We describe the two approaches below.

\paragraph{Algorithm for IC verification.} Fix a candidate profile of on-path distributions $\{\mu_\theta\}_{\theta\in\Theta}$ and lowest type's utility $V(\ul \th)$. Theorem \ref{thm:Characterization} yields the following procedure to verify incentive compatibility and construct the required contracts:
\begin{enumerate}
    \item \textbf{Indirect Utility Calculation (Envelope Condition):}  
    Apply \cref{it:Envelope} to compute the indirect utility $V(\theta)$ that must be delivered to each type based on the proposed distribution profile $\{\mu_\theta\}_{\th\in \Th}$.
    
    \item \textbf{Wage Function Determination (On-path Obedience):}  
    Use \cref{it:On-path-OB} to solve for the wage function $w_\theta$ that supports each distribution $\mu_\theta$, ensuring obedience on the equilibrium path, and choose sufficiently low off-support payments.
    
    \item \textbf{Off-path Recommendations (Off-path Obedience):}  
    Given the computed wage functions, apply \cref{it:Off-path-OB} to determine the optimal deviation distributions $\mu_{\theta,\hat{\theta}}$ that would arise if a type $\theta$ were to misreport as type $\hat{\theta}$.\footnote{Optimal continuation distributions need not be unique. However, for each report $\hat{\th}$, the partial derivative $\hat\theta$, $C(\theta,\mu_{\theta,\hat\theta})\partial \th$ is the same across all optimal continuation distributions for almost every $\theta$. Thus, the GIM condition, and hence implementability, is independent of the off-path best response selection.}
    
    \item \textbf{Incentive Compatibility Check (Generalized Integral Monotonicity):}  
    Finally, verify whether the generalized integral monotonicity condition (\cref{it:GIM}) holds. If it does, the proposed on-path action profile satisfies global incentive compatibility. If it does not, such a profile is not implementable.
\end{enumerate}

\noindent
Taken together, these steps not only check the feasibility of a candidate distribution profile $\{\mu_\theta\}$, but also construct the contract menu that implements it---including wage functions on each $\mu_\th$'s support---while explicitly computing its cost to the principal. Each step involves only standard operations, such as integration and optimization over known functions, making the overall procedure well-suited for numerical implementation.

\paragraph{Solving optimal contracting problems analytically.} The same mapping can be used to obtain analytical solutions to optimization problems subject to incentive constraints. Since items~\ref{it:Envelope}--\ref{it:On-path-OB} express transfers as functions of the on-path distributions, the principal's problem can be rewritten as a problem of choosing $\{\mu_\theta\}$ subject to (i) feasibility/transfer bounds and (ii) generalized integral monotonicity (\cref{it:GIM}), with off-path behavior summarized by $\{\mu_{\theta,\hat\theta}\}$ (\cref{it:Off-path-OB}). In applications, one can proceed in two ways: (a) solve a relaxed problem that drops \cref{it:GIM} and then verify when the relaxed solution satisfies generalized integral monotonicity, thereby yielding a fully incentive-compatible optimum; or (b) impose \cref{it:GIM} directly, exploiting cases in which generalized integral monotonicity simplifies (e.g., to a standard monotonicity condition on a one-dimensional statistic). Each approach will be illustrated with the examples that follow.

\section{Applications}\label{sec:applications}

We illustrate how to apply \Cref{thm:Characterization} across several economically relevant settings, with particular emphasis on joint ventures
and employment contracts. Throughout the section, assume the agent's cost is multiplicatively separable:
\[
C(\theta,\mu)=h(\theta)\kappa(\mu),
\]
with $h\colon\Theta\to\mathbb{R}_{++}$ continuously differentiable and $h'(\theta)<0$. Recall that the agent's outside option is normalized to 0, so we impose the IR constraint $V(\th) \ge 0$ for all $\th$ throughout.\footnote{We assume the principal’s payoff upon rejection is sufficiently low that an optimal mechanism induces participation by all types.}
The first five examples study a risk-neutral principal who maximizes expected profits. We first use the relaxed-problem approach, applying it to risk-neutral
agents in Proposition~\ref{prop:multsep-linear-contracts} and Examples~1--2, and to a risk-averse agent in Example~3. We then consider settings in which
generalized integral monotonicity binds and must be imposed directly (Examples~4--5). The final application turns to optimal taxation (Example~6). Solutions to the optimization problems are identified up to Lebesgue-almost-every type. % Naturally, Incentive compatibility remains imposed for \textit{every} type and   \textit{every}  report.

\subsection{Profit Maximization}

Throughout this subsection, the principal is risk neutral and maximizes expected profit:
\[
\int_{\underline\theta}^{\overline\theta}
\int_X [x-w_\theta(x)]\,\mu_\theta(\de x)\,\de F(\theta).
\]
The examples differ in the agent's risk preferences, the technology $\kappa$, and the limited liability constraints imposed on the transfers.

\subsubsection{Relaxed Problem Approach}

\subsubsection*{Risk-neutral agent}

We first consider a risk-neutral agent with unconstrained wages: $u(w)=w$ and $\underline{w}=-\infty$, $\overline{w}=+\infty$.

Write $m(\mu):=\int_{X}x\,\mu(dx)$ for the mean of $\mu$, and define the \emph{virtual cost} and \emph{incentive power} of type $\theta$ as
\[
\Phi(\theta):=h(\theta)-h'(\theta)\frac{1-F(\theta)}{f(\theta)},
\qquad \text{ and } \qquad 
\beta(\theta):=\frac{h(\theta)}{\Phi(\theta)}.
\]

Using the envelope condition in \Cref{thm:Characterization} and integration by parts, the principal's profit from any incentive-compatible mechanism satisfies
\begin{equation}\label{eq:multsep-virtual-profit}
\int_{\underline\theta}^{\overline\theta}\int_X [x-w_\theta(x)]\,\mu_\theta(\de x)\,\de F(\theta)
=
\int_{\underline\theta}^{\overline\theta}
\bigl[m(\mu_\theta)-\Phi(\theta)\kappa(\mu_\theta)\bigr]\,\de F(\theta) - 
V(\underline\theta).
\end{equation}
If we ignore generalized integral monotonicity, the principal's relaxed problem decomposes pointwise across types:
\begin{equation}\label{eq:multsep-relaxed}
\mu_\theta^*\in
\arg\max_{\mu\in\Delta(X)}
\Bigl\{m(\mu)-\Phi(\theta)\kappa(\mu)\Bigr\}
\qquad\text{for each }\theta\in\Theta.
\end{equation}
At the optimum, we have $V^*(\ul \th) = 0 $. Thus, the allocation induces the rent schedule
\[
V^*(\theta)=\int_{\underline\theta}^{\theta}\big[-h'(t)\big]\,\kappa(\mu_t^*)\,\de t.
\]

Under a natural regularity condition, the solution to this relaxed problem satisfies the relaxed constraint and will be globally optimal:
\begin{assumption}\label{ass:applications-regularity}
The function
\[
\theta\mapsto \beta(\theta)=\frac{h(\theta)}{h(\theta)-h'(\theta)\frac{1-F(\theta)}{f(\theta)}}
\]
is weakly increasing on $\Theta$.
\end{assumption}

\begin{proposition}\label{prop:multsep-linear-contracts}
Let $\{\mu_\theta^*\}_{\theta\in\Theta}$ solve the relaxed problem \eqref{eq:multsep-relaxed}, and let $V^*$ be defined as above. If \Cref{ass:applications-regularity} holds, then the affine wage schedule
\[
w_\theta^*(x)=a(\theta)+\beta(\theta)x,
\]
where
\[a(\theta):=V^*(\theta)+h(\theta)\kappa(\mu_\theta^*)-\beta(\theta)m(\mu_\theta^*),\]
implements $\mu_\theta^*$ on path. Moreover, for each $(t,\hat\theta)\in\Theta^2$, let
\[
\mu_{t,\hat\theta}^*
\in
\arg\max_{\nu\in\Delta(X)}
\Bigl\{\beta(\hat\theta)m(\nu)-h(t)\kappa(\nu)\Bigr\}.
\]
Then the extended mechanism
\[
\bigl\{w_\theta^*, V^{*}(\theta), \mu_{\th ,\hat\theta}^*\bigr\}_{(\th,\hat\theta)\in\Theta^2}
\]
satisfies all conditions in \Cref{thm:Characterization}. In particular, the solution to the relaxed problem is optimal in the original problem.
\end{proposition}

Importantly, the linearity of wages is not imposed ex ante. It is a consequence of optimality in the relaxed problem. At any relaxed maximizer $\mu_\theta^*$, the variational inequality for \eqref{eq:multsep-relaxed} implies that $x-\Phi(\theta)\kappa_{\mu_\theta^*}(x)$ is constant on the support of $\mu_\theta^*$, which yields an affine wage after rescaling by $h(\theta)/\Phi(\theta)$ and choosing the intercept to satisfy the envelope condition. The monotonicity condition in \Cref{ass:applications-regularity} plays the role of a strengthened  regularity assumption: it guarantees that higher reports induce steeper incentives and therefore costlier off-path continuation distributions. By \Cref{cor:GenMon-MultSeparableCost}, generalized integral monotonicity then follows automatically. We illustrate \Cref{prop:multsep-linear-contracts} with two classes of examples.

\paragraph{Example 1 (KL divergence).}\label{sec:ExampleKL}
Let $X=\{x_1,\dots,x_n\}$ be finite and suppose
\[
\kappa(\mu)=D_{KL}(\mu\|\nu):=\sum_{i=1}^n \log\!\left(\frac{\mu(x_i)}{\nu(x_i)}\right)\mu(x_i),
\]
where $\nu\in\Delta(X)$ has full support. The relaxed problem \eqref{eq:multsep-relaxed} becomes
\[
\max_{\mu\in\Delta(X)}
\left\{
\sum_{i=1}^n x_i\mu(x_i)-\Phi(\theta)D_{KL}(\mu\|\nu)
\right\},
\]
and has the unique solution
\begin{equation}\label{eq:KL-on-path}
\mu_\theta^*(x_i)
=
\frac{\nu(x_i)\exp\!\bigl(x_i/\Phi(\theta)\bigr)}{\sum_{j=1}^n \nu(x_j)\exp\!\bigl(x_j/\Phi(\theta)\bigr)},
\qquad i=1,\dots,n.
\end{equation}

By \Cref{prop:multsep-linear-contracts}, under \Cref{ass:applications-regularity} these distributions are implemented by the affine contracts $w_\theta^*(x)=a(\theta)+\beta(\theta)x$ with $\beta(\theta)=h(\theta)/\Phi(\theta)$. Since $D_{KL}(\mu_\theta^*\|\nu)=m(\mu_\theta^*)/\Phi(\theta)-\log Z_\theta$, where $Z_\theta:=\sum_{j=1}^n \nu(x_j)\exp\bigl(x_j/\Phi(\theta)\bigr)$, the intercept simplifies to
\[
a(\theta)=V^*(\theta)-h(\theta)\log Z_\theta.
\]
Off path, a type $\theta$ that reports $\hat\theta$ again chooses an exponential tilt of $\nu$, with $\beta(\hat\theta)/h(\theta)$ in place of $1/\Phi(\theta)$ in \eqref{eq:KL-on-path}.

\Cref{fig:KL-corollary} illustrates Example~1 with $n=100$ equally spaced outputs $X=\{0.05,0.10,\ldots,5\}$, $\nu$ uniform on $X$, $\theta\sim U[0,1]$, and $h(\theta)=1.5-\theta$. In this case, $\beta(\theta)=(1.5-\theta)/(2.5-2\theta)$ is strictly increasing, so \Cref{ass:applications-regularity} holds. Panel~(a) plots the distributions $\mu^*_\theta$ and panel~(b) the associated affine wage schedules $w^*_\theta$, for $\theta\in\{0,3/4,1\}$.

This example is especially transparent because under the KL technology each type chooses an exponential tilt of the baseline distribution, and the corresponding implementing contract is linear in output. When $a(\theta)<0$, the contract can be interpreted as an upfront fee $|a(\theta)|$ from the agent to the principal which gives the right to a revenue share $\beta(\theta)$. Hidden information reduces the slope below the first-best benchmark of one, because the principal acts as if the relevant cost coefficient were the virtual term $\Phi(\theta)$ rather than the technological term $h(\theta)$.

\begin{figure}[H]
\centering

\begin{subfigure}[t]{0.48\textwidth}
\centering
\begin{tikzpicture}
\begin{axis}[
axis lines = center,
    width=\textwidth,
    height=0.62\textwidth,
    xmin=0, xmax=5,
    ymin=0, ymax=0.08,
    samples=100,
    domain=0.05:5,
    xlabel={$x$},
    ylabel={probability mass},
    xtick={0,5},
    xticklabels={$0$,$5$},
    xtick style={draw=none},
    legend style={at={(0.03,0.97)},anchor=north west,draw=none,fill=none},
    ylabel style={
        rotate=90,
        at={(axis description cs:-0.08,0.5)},
        anchor=south,
    },
]

\addplot[dotted, very thick]
    {0.01};
\addlegendentry{$\nu$}

\addplot[very thick,dotted, Green]
    {exp(0.4*x)/322.65798};
\addlegendentry{$\mu_0^*$}

\addplot[very thick,dotted, blue]
    {exp(x)/3022.58396};
\addlegendentry{$\mu_{3/4}^*$}

\addplot[very thick,dotted, red]
    {exp(2*x)/231450.90581};
\addlegendentry{$\mu_1^*$}

\end{axis}
\end{tikzpicture}
\caption{Probability masses of the baseline distribution $\nu$ and the on-path optimal KL-tilted distributions $\mu_0^*$, $\mu_{3/4}^*$, and $\mu_1^*$.}
\label{fig:KL-tilts-uniform-unit}
\end{subfigure}
\hfill
\begin{subfigure}[t]{0.48\textwidth}
\centering
\begin{tikzpicture}
\begin{axis}[
axis lines = center,
    width=\textwidth,
    height=0.7\textwidth,
    xmin=0, xmax=5,
    ymin=-3.6, ymax=2.0,
    samples=100,
    domain=0.05:5,
    xtick={0,5},
    xticklabels={$0$,$5$},
    xtick style={draw=none},
    xlabel={$x$},
    ylabel={$w(x)$},
    legend style={at={(0.03,0.99)},anchor=north west,draw=none,fill=none},
     ylabel style={
    rotate=90,
    at={(axis description cs:-0.08,0.5)},
    anchor=south,
},
]

\addplot[very thick,dotted, Green]
    {0.6*x - 1.75713};
\addlegendentry{$w_0^*$}

\addplot[very thick,dotted , blue]
    {0.75*x - 2.31645};
\addlegendentry{$w_{3/4}^*$}

\addplot[very thick,dotted, red]
    {x - 3.40102};
\addlegendentry{$w_1^*$}

\end{axis}
\end{tikzpicture}
\caption{On-path linear schedules $w_0^*$, $w_{3/4}^*$, and $w_1^*$. Higher types face steeper piece rates, consistent with the monotonicity condition.}
\label{fig:KL-contracts-uniform-unit}
\end{subfigure}
\caption{Optimal allocation and contracts under KL-divergence.}
\label{fig:KL-corollary}
\end{figure}

%%%%%%%%%%%%%%%%%%%%%%%

\paragraph{Example 2 (Moment-based costs).}\label{sec:ExampleMoment}
Let $X=[\underline{x},\overline{x}]$ and suppose
\[
\kappa(\mu)=K\!\left(Z(\mu)\right),
\qquad
Z(\mu):=\int_X z(x)\mu(\de x),
\]
where $K\colon\mathbb{R}_+\to\mathbb{R}_+$ is strictly increasing, convex, and differentiable, and $z\colon X\to\mathbb{R}_+$ is strictly increasing and twice
differentiable. The curvature of $z$ determines the cost of generating output dispersion. If $z$ is convex, a mean-preserving spread raises $Z(\mu)$ and therefore raises effort costs. If $z$ is concave, the same spread lowers effort costs. We also allow $z$ to be S-shaped, with a convex region followed by a concave region.

\paragraph{Optimal allocation.} Under Assumption~\ref{ass:applications-regularity}, Proposition~\ref{prop:multsep-linear-contracts} reduces the principal's search for an optimal on-path allocation to the pointwise problems
\[
\mu_\theta^*
\in
\arg\max_{\mu\in\Delta(X)}
\left\{m(\mu)-\Phi(\theta)K\!\left(Z(\mu)\right)\right\},
\qquad \theta\in\Theta,
\]
and guarantees that the resulting profile is implementable and optimal in the original contracting problem. For the moment-based technology, each distributional problem reduces to a one-dimensional choice. Let $\hat z$ denote the convex minorant of $z$,\footnote{The convex minorant of $z$ is the greatest convex function $\hat z\colon X\to\mathbb{R}$ satisfying $\hat z(x)\le z(x)$ for every $x\in X$. Its construction for the S-shaped specification is given in Appendix~\ref{app:moment-costs}.} and let $x_0$ be the point at which $\hat z$ leaves $z$ and becomes the chord connecting $(x_0,z(x_0))$ and $(\overline{x},z(\overline{x}))$; the limiting cases of full convexity and full concavity correspond to $x_0=\overline{x}$ and
$x_0=\underline{x}$, respectively. Since $\hat z^{-1}(Z)$ is the largest mean output attainable at moment $Z$, each type chooses
\begin{equation}\label{eq:moment-scalar-main}
Z_\theta^*
\in
\arg\max_{Z\in[z(\underline{x}),z(\overline{x})]}
\left\{\hat z^{-1}(Z)-\Phi(\theta)K(Z)\right\}.
\end{equation}
The associated optimal distribution is
\begin{equation}\label{eq:moment-distribution-main}
\mu_\theta^*
=
\begin{cases}
\delta_{z^{-1}(Z_\theta^*)},
& \text{ if } Z_\theta^*\le z(x_0),\\[7pt]
(1-p_\theta^*)\delta_{x_0}
+p_\theta^*\delta_{\overline{x}},
& \text{ if } Z_\theta^*>z(x_0),
\end{cases} \qquad \text{ where } \,
p_\theta^*
:=
\dfrac{Z_\theta^*-z(x_0)}{z(\overline{x})-z(x_0)}.
\end{equation}
The convex-minorant reduction and the formal verification are collected in Appendix~\ref{app:moment-costs}.

\begin{figure}[H]
  \centering
  % --- First plot ---
  \begin{subfigure}{0.3\textwidth}
    \centering
  \includegraphics[width=\textwidth]{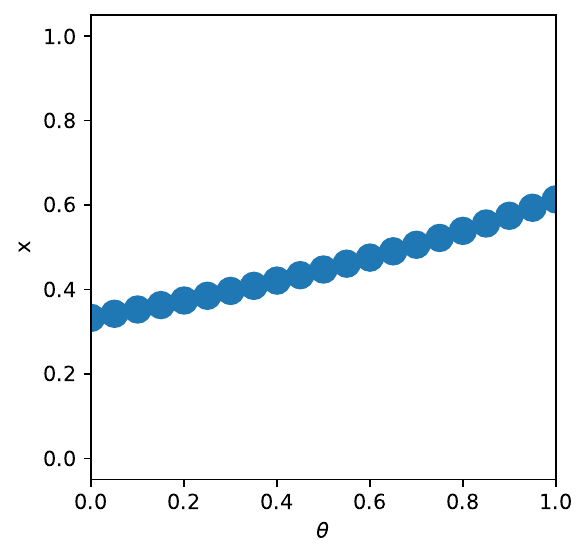}
    \caption{\small Convex $z(\cdot)$}
    % \label{fig:AS_MH_probs}
  \end{subfigure}
  \hfill
  % --- Second plot ---
  \begin{subfigure}{0.3\textwidth}
    \centering
  \includegraphics[width=\textwidth]{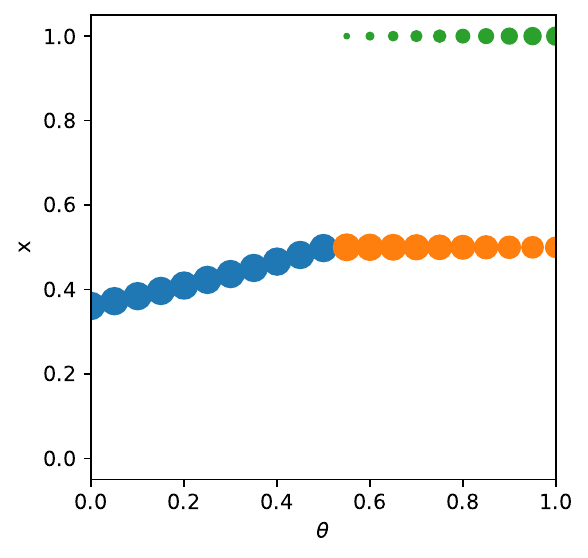}
    \caption{\small S-shaped $z(\cdot)$}
    % \label{fig:AS_MH_probs}
  \end{subfigure}
  \hfill
    % --- Third plot ---
   \begin{subfigure}{0.3\textwidth}
    \centering
  \includegraphics[width=\textwidth]{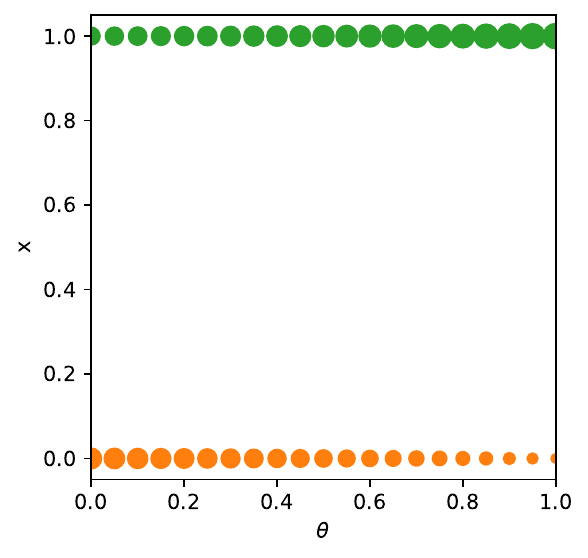}
    \caption{\small Concave $z(\cdot)$}
    % \label{fig:AS_MH_probs}
  \end{subfigure}
\caption{Optimal distribution profiles for different shapes of $z(\cdot)$. The horizontal axis is the type $\theta$, the vertical axis is output $x$, and the size of each dot represents the probability mass assigned to that output.}
  \label{fig:S-shaped}
\end{figure}

Equation~\eqref{eq:moment-distribution-main} produces the three distribution patterns displayed in Figure~\ref{fig:S-shaped}. The horizontal axis indexes the agent's type, the vertical axis records output, and the area of each dot represents the probability assigned to a support point.

The convex and concave specifications are the two polar cases. If $z$ is convex, as in Panel~(a), then $x_0=\overline{x}$ and the first line of
\eqref{eq:moment-distribution-main} applies to every type. The principal implements a deterministic output target that rises with type. If
$z$ is concave, as in Panel~(c), then $x_0=\underline{x}$ and every optimal distribution is binary with support on $\{\underline{x},\overline{x}\}$. More productive types place progressively more probability on the high output.

The S-shaped case in Panel~(b) combines these two patterns. Here $x_0\in(\underline{x},\overline{x})$. Types for which $Z_\theta^*\le z(x_0)$ choose a degenerate distribution, while types for which $Z_\theta^*>z(x_0)$ randomize between $x_0$ and $\overline{x}$. The structure of the allocation changes at the type for which the optimal moment crosses $z(x_0)$: lower types are recommended  degenerate distributions, whereas higher types are recommended binary distributions that place progressively more mass on
$\overline{x}$.

\paragraph{Implementation.} By Proposition~\ref{prop:multsep-linear-contracts}, the optimum above has an affine implementation
\[
w_\theta(x)=a(\theta)+\beta(\theta)x,
\qquad
\beta(\theta)=\frac{h(\theta)}{\Phi(\theta)}.
\]
Affine contracts are sufficient, but they are not necessary. The same allocation and rent schedule can be implemented by contracts that agree
with the affine schedule on the support of $\mu_\theta^*$ and lower payments elsewhere. Such a construction is given in Appendix~\ref{app:moment-costs}.

For the degenerate distributions in Panel~(a) and the lower region of Panel~(b), the alternative implementation is a target contract: it pays a fixed wage when output reaches the prescribed target and a punishment wage below it. Since the target is met with probability one on path, output
and pay are deterministic. This resembles an efficiency-wage arrangement in which a relatively high wage is paid conditional on satisfying a performance
standard.

For the binary distributions in Panel~(c) and the upper region of Panel~(b), the alternative is a single-bonus contract: it pays a baseline wage at the lower support point and a discrete bonus at $\overline{x}$, together with an off path punishment wage below the lowest point in the implemented distribution support. In summary, convex $z$ leads to target contracts, concave $z$ leads to bonus contracts, and S-shaped $z$ generates a screening menu that combines the two.

%%%%%%%%%%%%%%%%%%%%%%%%%%%% 

\subsubsection*{Risk-averse agent}

We retain the profit-maximizing principal and multiplicative costs, but let the agent be risk averse.

\paragraph{Example 3 (Log utility and KL costs).} Let $X=\{x_1,\ldots,x_n\}$, with $x_1<\cdots<x_n$, and let \(\nu\in\Delta(X)\) have full support. The agent has utility
\[
u(w)=\log w,
\qquad
w>0,
\]
and incurs cost
\[
C(\theta,\mu)
=
h(\theta)D_{\mathrm{KL}}(\mu\Vert\nu),
\qquad
h'(\theta)<0.
\]

It is convenient to express contracts in utils,
\[
q_\theta(x):=\log w_\theta(x).
\]
A type \(t\) facing a contract \(q\) chooses the distribution
\begin{equation}
\mu_t^q(x_i)
=
\frac{\nu_i\exp\!\left(q(x_i)/h(t)\right)}
{\sum_j\nu_j\exp\!\left(q(x_j)/h(t)\right)},
\qquad
\nu_i:=\nu(x_i).
\label{eq:log-kl-gibbs-short}
\end{equation}
Consequently, a full-support distribution \(\mu_\theta\) that delivers
utility \(V(\theta)\) is implemented by
\begin{equation}
\log w_\theta(x_i)
=
V(\theta)
+
h(\theta)
\log\!\left(\frac{\mu_\theta(x_i)}{\nu_i}\right).
\label{eq:log-kl-implementation-short}
\end{equation}

We follow a similar approach to Example 1, where we first solve a relaxed problem that imposes the envelope condition but drops
generalized integral monotonicity and later check that the solution satisfies all dropped constraints.

\paragraph{The relaxed problem.}
Define
\[
\mathcal P_h(\mu)
:=
\sum_i\nu_i
\left(\frac{\mu(x_i)}{\nu_i}\right)^{h+1}.
\]
Using \eqref{eq:log-kl-implementation-short}, the expected wage paid to
type \(\theta\) is
\[
\sum_i\mu_\theta(x_i)w_\theta(x_i)
=
e^{V(\theta)}\mathcal P_{h(\theta)}(\mu_\theta).
\]
The relaxed problem can be written as:\footnote{Note that \eqref{eq:log-kl-implementation-short} already incorporates on-path obedience. Off-path obedience plays a role only when verifying feasibility of the relaxed solution in the original problem.}
\begin{equation}
\begin{aligned}
\max_{\{\mu_\theta,V(\theta)\}}
\quad&
\int_{\underline\theta}^{\overline\theta}
f(\theta)
\left[
\sum_i x_i\mu_\theta(x_i)
-
e^{V(\theta)}\mathcal P_{h(\theta)}(\mu_\theta)
\right]d\theta,
\\
\text{s.t.}\quad&
V'(\theta)
=
-h'(\theta)D_{\mathrm{KL}}(\mu_\theta\Vert\nu) \; \text{for a.e } \th,
\qquad
V(\underline\theta)=0.
\end{aligned}
\label{eq:log-kl-relaxed-short}
\end{equation}

\paragraph{Solving the relaxed problem.}
Problem \eqref{eq:log-kl-relaxed-short} can be solved using standard optimal-control methods, with the promised utility \(V(\theta)\) as the state variable. Assigning a more costly distribution to type \(\theta\) raises the information rents of all more productive types. The co-state associated with \(V(\theta)\) captures this continuation wage cost.

Let \(L(\theta)\) denote the shadow cost of promised utility. Under log utility, it admits the representation
\begin{equation}
    L(\theta)
    :=
    \int_\theta^{\overline\theta}
    f(t)\mathbb E_{\mu_t}[w_t]\,dt.
    \label{eq:log-kl-shadow-cost}
\end{equation}
Thus, \(L(\theta)\) measures the marginal continuation wage cost of increasing the utility promised to type \(\theta\).
This cost is larger for lower types and vanishes at the highest type.

Define the \emph{rent-loading coefficient}
\begin{equation}
    \rho(\theta)
    :=
    \frac{-h'(\theta)}
         {f(\theta)h(\theta)}
    L(\theta).
    \label{eq:log-kl-rent-loading}
\end{equation}
The coefficient \(\rho(\theta)\) translates the shadow cost of promised utility into the local cost of providing stronger output
incentives. It is zero in the pure moral-hazard problem and at the highest type, but is generally positive when incentives provided to type \(\theta\) increase the information rents of more productive types.

Conditional on \(V(\theta)\) and \(L(\theta)\), the optimal wage schedule associated with report \(\theta\) is available in closed form. There exists a scalar \(\chi(\theta)\) such that, whenever \(\rho(\theta)>0\),
\begin{equation}
    w_\theta(x_i)
    =
    \frac{\rho(\theta)}{1+h(\theta)}
    W_0\left(
        \frac{1+h(\theta)}{\rho(\theta)}
        \exp\left\{
            \frac{x_i-\chi(\theta)}{\rho(\theta)}
        \right\}
    \right),
    \label{eq:log-kl-lambert-short}
\end{equation}
where \(W_0\) denotes the principal branch of the Lambert-\(W\) function.\footnote{If $z>0$, then $W_0(z)$ is the real number $w$ satisfying $w e^{w }=z$.} When \(\rho(\theta)=0\), the formula reduces to the affine
contract\footnote{This formula and its benchmark interpretations require an interior relaxed allocation with strictly positive wages. Assumption 5 ensures interiority in the joint problem; see Appendix~\ref{app:log-kl-exactness-proof}.}
\begin{equation}
    w_\theta(x_i)
    =
    \frac{x_i-\chi(\theta)}
         {1+h(\theta)}.
    \label{eq:log-kl-affine-short}
\end{equation}

The scalar \(\chi(\theta)\) determines the level of the contract and is uniquely chosen so that the induced probabilities sum to one:
\begin{equation}
    \sum_i
    \nu_i
    \left(
        e^{-V(\theta)}w_\theta(x_i)
    \right)^{1/h(\theta)}
    =
    1.
    \label{eq:log-kl-normalization-short}
\end{equation}
The corresponding output distribution is
\begin{equation}
    \mu_\theta(x_i)
    =
    \nu_i
    \left(
        e^{-V(\theta)}w_\theta(x_i)
    \right)^{1/h(\theta)}.
    \label{eq:log-kl-distribution-short}
\end{equation}
Thus, the optimal distribution in a solution to the relaxed problem is a power tilt of the baseline distribution: outputs carrying higher wages receive more probability, and types with lower effort costs respond more strongly to a given wage schedule.

The economic role of the rent-loading coefficient is particularly transparent from the pairwise first-order condition. Fix a report
\(\theta\), and consider shifting an infinitesimal amount of probability from \(x_i\) to \(x_j>x_i\). Per unit of probability shifted, the principal's expected output increases by \(x_j-x_i\). At an interior optimum, this marginal output gain must equal the associated increase in wage and information-rent costs. Subtracting the report-by-report first-order conditions for \(\mu_\theta(x_j)\) and \(\mu_\theta(x_i)\) gives
\begin{equation}
\begin{aligned}
    x_j-x_i
    ={}&
    [1+h(\theta)]
    \left[
        w_\theta(x_j)-w_\theta(x_i)
    \right]
    +
    \rho(\theta)
    \log\left(
        \frac{w_\theta(x_j)}
             {w_\theta(x_i)}
    \right).
\end{aligned}
\label{eq:log-kl-pairwise-solution}
\end{equation}
The first term on the right-hand side is the direct moral-hazard cost of shifting probability toward the higher output: it captures the direct increase in payments due to a higher output with a higher wage being more likely, i.e., $[w_{\th}(x_{j})-w_{\th}(x_{i})]$, as well as the change in wages required to induce the agent to choose a costlier distribution, i.e., $h(\th)[w_{\th}(x_{j})-w_{\th}(x_{i})]$. The second term is the additional information-rent cost associated with screening: stronger incentives induce a more costly distribution and thereby raise the information rents of more productive types.

When \(\rho(\theta)=0\), this screening cost disappears and the optimal wage is affine, with slope \(1/[1+h(\theta)]\). This is the pure moral-hazard solution and also applies to the highest type. 

When \(\rho(\theta)>0\), the schedule is no longer affine. The wage-payment cost depends on the difference in wages, whereas the information-rent cost depends on the difference in \emph{log} wages. Applying \eqref{eq:log-kl-pairwise-solution} to nearby outputs, so that \(\log\left(w_\theta(x_j)/w_\theta(x_i)\right)\) is approximately \(\left[w_\theta(x_j)-w_\theta(x_i)\right]/w\), the output gain required to justify an additional dollar of pay at wage level $w$ is
$
    [1+h(\theta)]
    +
    {\rho(\theta)}/ {w}$. 
The direct moral-hazard component, $[1+h(\theta)]$, is independent of the wage level, but the rent component, $\rho(\th)/w$, declines in \(w\): an additional dollar paid at a low wage represents a large proportional increase and therefore creates a large rent cost, while the same dollar paid at a high wage represents a smaller proportional increase and creates a smaller rent cost. Incentives are therefore cheapest, in rent terms, when concentrated at high outputs. As a result, the schedule becomes steeper as wages rise and, hence, each contract is strictly convex in output.

The promised-utility schedule \(V(\theta)\) and its shadow cost \(L(\theta)\) are determined jointly by a two-point boundary-value problem. The full optimal-control system and the derivation of \eqref{eq:log-kl-lambert-short}-\eqref{eq:log-kl-pairwise-solution} are provided in Appendix \ref{app:log-kl-relaxed-proof}.

\paragraph{Feasibility of the relaxed solution.} It remains to determine when the solution to the relaxed problem also satisfies the omitted generalized integral monotonicity constraint. We proceed in two steps. First, we identify an intuitive property of the wage menu that is sufficient for GIM. Second, we provide conditions on primitives under which every relaxed optimum satisfies this property.

A natural measure of incentive power is a wage ratio rather than a wage difference. We say that a menu satisfies \emph{increasing proportional incentives} if, for every \(x_j>x_i\),
\begin{equation}
\theta
\longmapsto
\frac{w_\theta(x_j)}{w_\theta(x_i)}
\quad\text{is nondecreasing.}
\label{eq:log-single-crossing-short}
\end{equation}
Thus, higher-type contracts provide a larger proportional reward for high output relative to low output.

\begin{lemma}\label{lem:log-single-crossing-short}
Suppose each wage schedule is increasing in output and the menu satisfies increasing proportional incentives. Then, for every true type \(t\),
\[
\hat \th
\longmapsto
D_{\mathrm{KL}}(\mu_{t,\hat \th}\Vert\nu)
\]
is nondecreasing, where \(\mu_{t,\hat \th}\) is type \(t\)'s optimal continuation distribution after reporting \(\hat \th\). Therefore, the
mechanism satisfies generalized integral monotonicity.
\end{lemma}

The lemma has a simple interpretation. A higher report exposes the agent to a contract with stronger proportional rewards for high output. Every type then chooses a distribution that is more tilted away from the baseline and therefore more costly in KL divergence. 

\noindent\emph{A primitive sufficient condition.} The following assumption guarantees that the contracts in the solution to the relaxed problem satisfy increasing proportional incentives.

\begin{assumption}
\label{ass:log-kl-primitives-short}
The primitives satisfy
\begin{equation}
\theta
\longmapsto
\frac{-h'(\theta)}{f(\theta)h(\theta)}
\quad\text{is nonincreasing,}
\label{eq:log-kl-rent-loading-short}
\end{equation}
and
\begin{equation}
\frac{x_n-x_1}{1+h(\overline\theta)}
\leq
1-e^{-1}.
\label{eq:log-kl-output-span-short}
\end{equation}
\end{assumption}

The first condition says that the rent burden generated by a marginal reduction in effort costs does not become more severe for higher types. The second condition bounds the amount of output variation that the contract must reward. In particular, it guarantees that the log-wage spread between the highest and lowest outputs is at most one, so the largest wage ratio is at most \(e\).

\begin{proposition}
\label{prop:log-kl-exact-short}
Suppose \(h\in C^2\), \(f\in C^1\) is strictly positive, \(\nu\) has full support, and Assumption \ref{ass:log-kl-primitives-short} holds. Then every relaxed optimum is interior and has increasing proportional incentives. It therefore satisfies all conditions in \Cref{thm:Characterization} and, thus, solves the original problem.
\end{proposition}

\begin{figure}[ht!]
\begin{subfigure}{0.48\textwidth}
        \centering
        \includegraphics[width=0.85\linewidth]{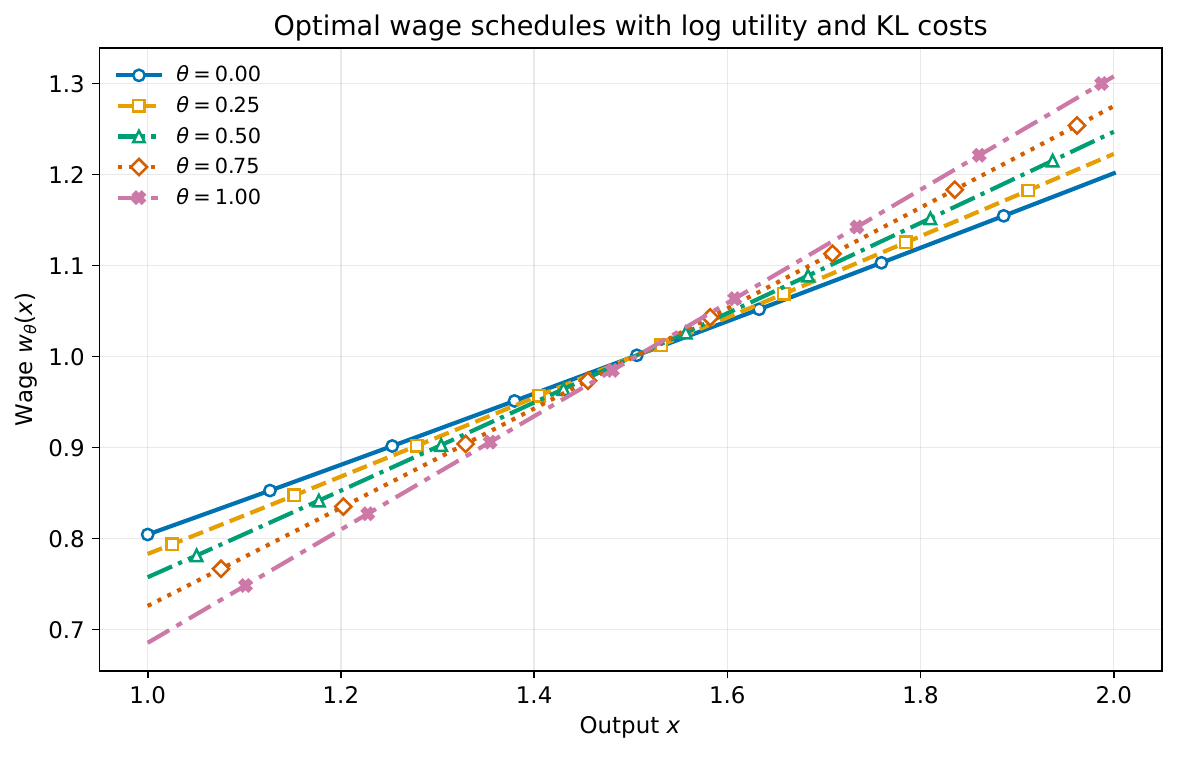}
        \caption{Higher-type contracts pay
        less at low outputs and more at high outputs, while the wage ratio
        between high and low outputs increases with type.}
        \label{fig:log-kl-short-wages}
  \end{subfigure}
  \begin{subfigure}{0.48\textwidth}
        \centering
        \includegraphics[width=0.85\linewidth]{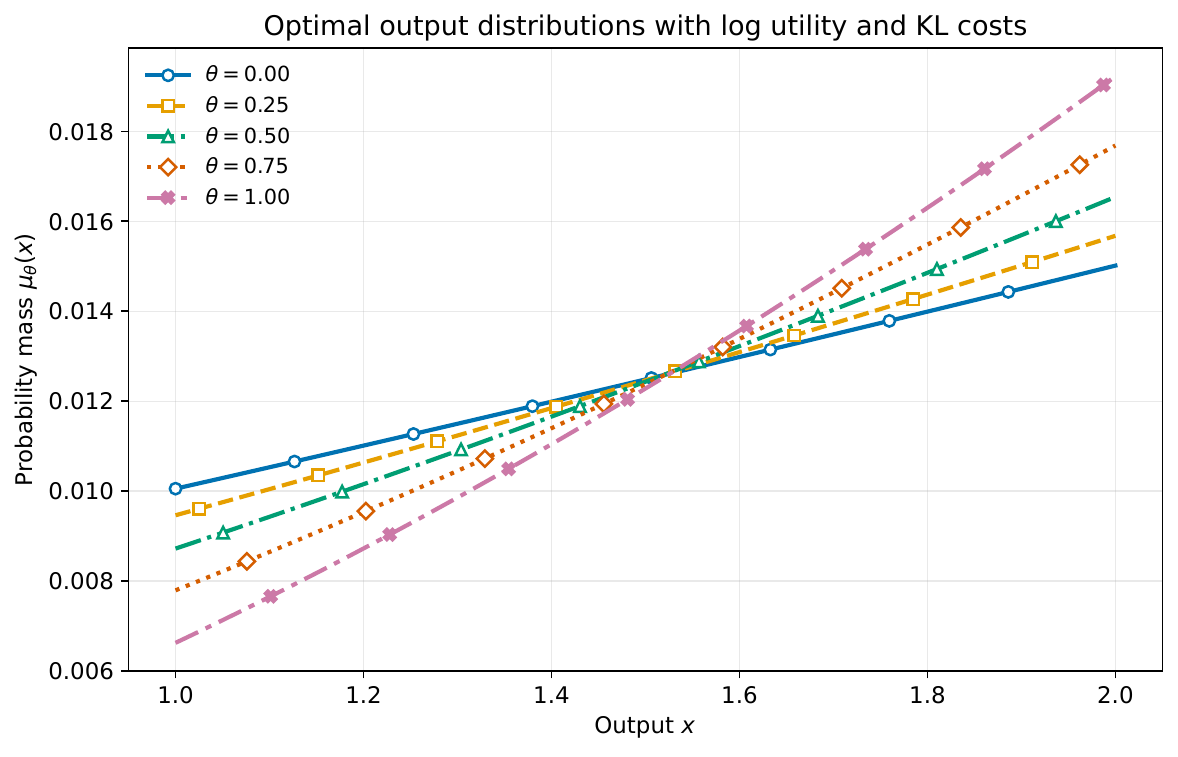}
        \caption{Optimal output distributions for the same five
        types. More productive types choose progressively stronger tilts
        toward high output.}
        \label{fig:log-kl-short-distributions}
    \end{subfigure}
    \caption{Optimal menu for risk averse agent}
\end{figure}

\paragraph{Illustration.} Let $\Theta=[0,1]$, $F(\theta)=\theta$, $h(\theta)=e^{-\theta/2}$, and let \(X\) contain 80 equally spaced points on \([1,2]\), with
\(\nu\) uniform. In this case,
\[
\frac{-h'(\theta)}{f(\theta)h(\theta)}
=
\frac12
\]
is constant, while
\[
\frac{x_n-x_1}{1+h(1)}
=
\frac{1}{1+e^{-1/2}}
<
1-e^{-1}.
\]
Hence, Proposition \ref{prop:log-kl-exact-short} applies. Figures
\ref{fig:log-kl-short-wages} and
\ref{fig:log-kl-short-distributions} plot the optimal menu for five
types. Higher types face stronger proportional incentives and choose
output distributions with more mass on high realizations.

\subsubsection{Binding Generalized Integral Monotonicity (GIM) Constraint}
Returning to a risk-neutral agent, we now turn to environments in which
GIM must be imposed directly because the solution to the relaxed problem
violates it. In Example~4, the regularity condition (Assumption~\ref{ass:applications-regularity}) is violated. Example~5 instead introduces a minimum-payment constraint (limited liability), which may cause (GIM) to bind even when Assumption~\ref{ass:applications-regularity} is satisfied.

\paragraph{Example 4 (Lack of regularity leads to binding GIM).}
Consider the moment-based technology of Example~2 with a binary outcome space $X=\{\underline{x},\bar{x}\}$. We identify a distribution recommended to type $\theta$ with the probability of the high outcome, $p:=\mu(\{\bar{x}\})$, and write $\underline{z}:=z(\underline{x})$ and $\bar{z}:=z(\bar{x})$, so that the cost function is
\[
C(\theta,p)=h(\theta)K\big((1-p)\underline{z}+p\bar{z}\big),
\]
with $h(\theta)=A-e^{\theta}$, $\theta\sim U([0,1])$, for a constant $A> e$.

In this case, the regularity condition in \Cref{ass:applications-regularity} is violated. The relaxed solution is infeasible because the wage differential $w_\th(\bar x) - w_\th(\ul x)$ is not increasing for all types. However, since the standard virtual cost term $h(\th) - h'(\th)\frac{1-F(\th)}{f(\th)} = A-\th e^{\th}$ is decreasing, the result of a pointwise maximization would be feasible (and optimal) under pure adverse selection, when actions are contractible.

For ease of exposition, define $\D x=\bar x - \ul x$, and assume $K(Z)=Z^2/2 $ and $\ul z=0$. To illustrate the effects of the combined moral hazard and adverse selection, we compute the optimal distribution in three cases: pure moral hazard, pure adverse selection, and the two frictions jointly.
 We use the superscript $MH$ for the pure moral-hazard case, $AS$ for the pure adverse-selection case, and ``$*$'' for the joint problem.

Since the agent is risk neutral and there is no limited liability constraint, under pure moral hazard the principal can implement the first-best distributions for each type $\theta$, i.e.,
\[p^{MH}_{\th} = \frac{\Delta x} {\bar{z}^2} \frac{1}{A - e^\th }  \wedge 1,\]
and the wages for each type are given by the on-path obedience constraints and the binding participation constraint. 

In the pure adverse-selection case, the principal distorts allocations to lower types downwards to reduce information rents. The optimal pure adverse-selection probabilities are
\begin{align*}
p^\text{AS}_\th =\frac{\D x}{\bar{z}^2 } \frac{1}{A - \th e^\th  }  \wedge 1. 
\end{align*}
Since $p_{\th}^{AS}$ is increasing, it is implementable when moral hazard is absent. In fact, when the principal can directly control effort, she can implement such distributions with wages that are constant in output:
\[w^{AS}_{\th}(x) = V^{AS}(\th)+C(\th,\mu_{\th}^{AS}) \quad \text{ for all } x\in X.\]

Once we introduce moral hazard on top of adverse selection, $p_{\th}$ increasing in $\theta$ is no longer sufficient. The generalized integral monotonicity constraint now requires the difference in pay between high and low outputs to be increasing. If one tries to implement $p_{\th}^{AS}$ using the on-path obedience constraints to recover the associated wages, this difference would be given by
\[ (A-e^{\th})\bar{z}^{2}p_{\th}^{AS} = \Delta x \frac{A-e^{\th}}{A-\th e^{\th}},\]
which violates generalized integral monotonicity for low $\th$. To find the optimal allocation with joint adverse selection and moral hazard, we must explicitly incorporate the constraint that $(A-e^{\th}) p_\th$ be nondecreasing whenever $p_\th$ is interior. We solve this optimal control problem in Appendix \ref{app:Example4} and find that the optimal allocation is 
\begin{align*}
    p^\ast_{\th} = p^\ast_0 \frac{A-1}{A-e^{\th}} \wedge 1,  
\end{align*}
for low enough $\th$. We always have $p^\ast_0 < p^\text{AS}_0$ unless $p^\ast_\th=p^\text{AS}_\th =1$ for all $\th$. Generalized integral monotonicity is then satisfied on the interior range since
\[w^{*}_{\th}(\bar{x})-w^{*}_{\th}(\ul x) = \bar{z}^{2}p^{*}_{0}[A-1].\]

\begin{figure}[h]
  \centering
  % --- First plot ---
  \begin{subfigure}{0.27\textwidth}
    \centering
    \begin{tikzpicture}
      \begin{axis}[
        axis lines = center,
        scale = 0.6,
        xtick={0.720},
        xticklabels={$\bar\th^\ast$},
        ytick={.474,1},
        yticklabels={$p^\ast_0$, $1$},
        xmin=0, xmax=0.8,
        ymin=0.45, ymax=1.1,
        xlabel={$\th$},
        ylabel={$p$},
        xlabel style={at={(current axis.right of origin)},anchor=west,yshift=-4pt},
        ylabel style={at={(current axis.above origin)},anchor=south,xshift=4pt},
      ]
        \addplot [dashed, lightgray, domain=0:0.8, samples=2] {1};
        \draw [dashed, lightgray] (axis cs:0.72,0) -- (axis cs:0.72,1);

        \addplot [blue, dashed, very thick, domain=0:0.8, samples=100]
          {min(1.6/(3 - exp(.8*x)*x),1)}
          node [above ,pos=0.3] {$p^\mathrm{AS}$};

        \addplot [Green, dotted, very thick, domain=0:0.8, samples=100]
          {min(1.5/(3 - exp(x)), 1)}
          node [above ,pos=0.3] {$p^\mathrm{MH}$};

        \addplot [red, dashdotted, very thick, domain=0:0.8, samples=100]
          {min(0.474 * (3-1)/(3 - exp(x)),1)}
          node [below right,pos=0.3] {$p^\ast$};

        \fill [blue] (axis cs:0,0.632120559) circle (0.5pt);
        \fill [red]  (axis cs:0,0.585)       circle (0.5pt);
      \end{axis}
    \end{tikzpicture}
    \caption{\small Effort.}
    \label{fig:AS_MH_probs}
  \end{subfigure}
  \hfill
  % --- Second plot ---
  \begin{subfigure}{0.27\textwidth}
    \centering
    \begin{tikzpicture}
      \begin{axis}[
        axis lines = center,
        scale = 0.6,
        xtick=\empty,
        ytick=\empty,
        xmin=0, xmax=.85,
        ymin=1.5, ymax=3.25,
        xlabel={$\th$},
        ylabel={$c_{\mu_\th}(\th, \bar x)- c_{\mu_\th}(\th,\ul x)$},
        xlabel style={at={(current axis.right of origin)},anchor=west,yshift=-4pt},
        ylabel style={at={(current axis.above origin)},anchor=south,xshift=25pt},
      ]
        \addplot [blue, dashed, very thick, domain=0:0.9, samples=100]
        {2*(3-exp(x))*min(1.6/(3 - exp(.8*x)*x),1)}
          node [above ,pos=0.1] {$AS$};

        \addplot [Green, dotted, very thick, domain=0:0.9, samples=100]
          {2*(3-exp(x))*min(1.5/(3 - exp(x)), 1)}
          node [below ,pos=0.1] {$MH$};

        \addplot [red, dashdotted, very thick, domain=0:0.9, samples=100]
          {2*(3-exp(x))*min(0.474 * (3-1)/(3 - exp(x)),1)}
          node [below,pos=0.1] {$Both$};

        \fill [blue] (axis cs:0,0.632120559) circle (0.5pt);
        \fill [red]  (axis cs:0,0.585)       circle (0.5pt);
      \end{axis}
    \end{tikzpicture}
    \caption{\small $c_{\mu_{\th}}$-spread.}
    \label{fig:AS_MH_costs}
  \end{subfigure}
  \hfill
    % --- Third plot ---
  \begin{subfigure}{0.27\textwidth}
    \centering
    \begin{tikzpicture}
      \begin{axis}[
        axis lines = center,
        scale = 0.6,
        xtick=\empty,
        ytick=\empty,
        xmin=0, xmax=.85,
        ymin=1.5, ymax=3.25,
        xlabel={$\th$},
        ylabel={$w_{\th}(\bar x)- w_\th(\ul x)$},
        xlabel style={at={(current axis.right of origin)},anchor=west,yshift=-4pt},
        ylabel style={at={(current axis.above origin)},anchor=south,xshift=20pt},
      ]
        \addplot [blue, dashed, very thick, domain=0:0.9, samples=100]
        {1.515}
          node [above ,pos=0.7] {$AS$};

        \addplot [Green, dotted, very thick, domain=0:0.9, samples=100]
          {2*(3-exp(x))*min(1.5/(3 - exp(x)), 1)}
          node [below ,pos=0.1] {$MH$};

          \addplot [red, dashdotted, very thick, domain=0:0.9, samples=100]
          {2*0.474*(3-1)}
          node [below,pos=0.1] {$Both$};

        \fill [blue] (axis cs:0,0.632120559) circle (0.5pt);
        \fill [red]  (axis cs:0,0.585)       circle (0.5pt);
      \end{axis}
    \end{tikzpicture}
    \caption{\small Bonus for high output.}
    \label{fig:AS_MH_wages}
  \end{subfigure}

  \caption{Effort, Gâteaux derivative spread and wage spread under AS, MH, and both.}
  \label{fig:ASvMH}
\end{figure}

Figure \ref{fig:ASvMH} illustrates the three scenarios. Panel (a) plots the probability of the high output, panel (b) the spread in the cost Gâteaux derivatives, and panel (c) the wage spread between the high and the low output.

In the pure moral-hazard case, the wage spread must exactly match the spread in the cost Gâteaux derivatives, so panels (b) and (c) coincide. In contrast, under pure adverse selection the principal directly observes $p$, eliminating the need to induce payment dispersion; consequently, the wage spread for all types is zero. As panel (b) shows, $p^{AS}$ does not increase fast enough to produce a nondecreasing spread in the cost Gâteaux derivatives, which is why it violates generalized integral monotonicity. The implemented probability $p^\ast$ must therefore be steeper than $p^{AS}$: it starts below $p^{AS}$, crosses it from below, and rises rapidly enough to ensure that the difference in the cost Gâteaux derivatives is nondecreasing in the agent's type for any interior probability. The eventual decline in $[
c_{\mu_\theta}(\theta,\bar x)- c_{\mu_\theta}(\theta,\underline x)]$ occurs only for types who choose the high output with probability one.  

In terms of wages, when both moral hazard and adverse selection are present, all types receive the same contract: the bonus exactly equals the cost Gâteaux derivative spread for types mixing over outputs, while the highest types have a strict incentive to choose the high output with probability one. Comparing $p^{AS}$ and $p^\ast$ with $p^{MH}$ (equal to the first-best here), we see that introducing moral hazard into the adverse-selection problem increases the distortion for low types but reduces it for higher types.

Our characterization is closely related to that in \cite{castro2024disentangling}. When output is binary, every distribution is summarized by a single number---the probability of the high-output realization---so the flexible moral-hazard problem collapses to the standard one-dimensional effort formulation. In that case, on-path obedience identifies the utility gap between high and low output with the marginal cost of raising the success probability, so generalized integral monotonicity reduces exactly to the increasing marginal cost condition (IMC) of \cite{castro2024disentangling}. As they already note, in the binary case IMC is sufficient, and indeed necessary, once envelope and on-path obedience are imposed. Our characterization shows that, with binary output, IMC is precisely the specialization of GIM and therefore captures the double-deviation constraints.

\paragraph{Example~5 (Limited liability leads to binding GIM).}\label{sec:ExampleKL_LL}
We return to the environment of Example~1, but now suppose that the agent is cash constrained and cannot make payments to the principal. Thus,
\[
w_\theta(x_i)\geq 0
\qquad
\text{for every }\theta\in \Th \text{ and }x_i\in X.
\]
As in Example~1, the agent's cost function is 
$
C(\theta,\mu)
=
h(\theta)D_{KL}(\mu\Vert\nu)$, for some finite $X=\{x_1,\ldots,x_n\}$ with a full-support baseline distribution $\nu$. Write $\nu_i:=\nu(x_i)$. We show that limited liability can make GIM bind even when the regularity condition in \Cref{ass:applications-regularity} holds. The optimal contracts can then feature upward distortions in the incentive power of wages above the limited liability bound.

As before, on-path obedience identifies wages with the log likelihood ratio of the implemented distribution. If $\mu_\theta$ is implemented and type $\theta$ receives utility $V(\theta)$, then
\begin{equation}\label{eq:KLLLWageFromDensity}
w_\theta(x_i)
=
V(\theta)
+
h(\theta)\log\!\left(\frac{\mu_\theta(x_i)}{\nu_i}\right).
\end{equation}
Limited liability can therefore be written directly as the likelihood-ratio floor
\begin{equation}\label{eq:KLLLDensityFloor}
\mu_\theta(x_i)
\geq
\nu_i\exp\!\left(-\frac{V(\theta)}{h(\theta)}\right)
\qquad
\text{for every }i.
\end{equation}
Whenever this constraint binds, the wage at that output is zero. 

The KL specification gives the off-path distribution in closed form. If true type $\theta$ selects the contract intended for report $r$, then
\begin{equation*}%\label{eq:KL_LL_offpath_distribution}
\mu_{\theta,r}(x_i)
=
\frac{
\nu_i\left(\mu_r(x_i)/\nu_i\right)^{h(r)/h(\theta)}
}{
\sum_{j=1}^n
\nu_j\left(\mu_r(x_j)/\nu_j\right)^{h(r)/h(\theta)}
}.
\end{equation*}
Consequently, the utility loss from reporting $r$ rather than truthfully is
\begin{equation}\label{eq:KL_LL_reporting_gap}
G(\theta,r)
:=
V(\theta)-V(r)
-
h(\theta)
\log\!\left[
\sum_{i=1}^n
\nu_i
\left(\frac{\mu_r(x_i)}{\nu_i}\right)^{h(r)/h(\theta)}
\right].
\end{equation}
Incentive compatibility is equivalent to
$
G(\theta,r)\geq0$ 
for every $(\theta,r)\in\Th^2$.

When solving for a profit-maximizing mechanism for the principal, limited liability changes the solution method relative to Example~1. Without limited liability, the relaxed pointwise problem selects an exponential tilt, and monotonicity of the resulting piece rate $\b$ is sufficient for GIM. 
Here, the floor in \eqref{eq:KLLLDensityFloor} depends on the endogenous rent  $V(\th)$.  Through the envelope condition, changing the distribution of one type changes the rents, and hence the limited-liability constraints, of all higher types.
Moreover, with limited liability, some GIM constraints will typically bind at the optimal mechanism. This introduces another global effect since changing the distribution attached to any report $r$ changes the payoff of every true type $\th$ who considers selecting contract $w_r$. Unlike in Example~1, these constraints prevent the problem from decomposing immediately into separate type-by-type maximization problems.

To solve the problem, we adjoin the limited-liability and GIM constraints with multipliers. \Cref{prop:KL_LL_verification} in Appendix~\ref{app:Example5} shows that the problem then decomposes into type-by-type problems analogous to \eqref{eq:multsep-relaxed}, with two changes: the virtual cost $\Phi(\th)$ is adjusted for the shadow cost of the limited-liability and reporting constraints, and an additional term penalizes distributions that make it attractive for other types to report $\th$ and then adjust their distribution. A mechanism is optimal if its on-path distributions solve these problems for multipliers that satisfy complementary slackness. We apply the sufficient conditions in \Cref{prop:KL_LL_verification} to prove optimality for the following numerical illustration.

\paragraph{Illustration.}
Consider a uniform baseline distribution $\nu$ and the parameters
\[
X=\{0,12,13\},\qquad
\theta\sim U[0,1],\qquad
h(\theta)=4-\theta.
\]

\begin{figure}[H]
\centering

\begin{subfigure}[t]{0.49\textwidth}
\centering
\begin{tikzpicture}
\begin{axis}[
    width=\textwidth,
    height=0.67\textwidth,
    xmin=0, xmax=13,
    ymin=-5, ymax=2.3,
    xlabel={output $x$},
    ylabel={wage},
    axis lines=left,
    xtick={0,12,13},
    ymajorgrids=true,
    grid style={gray!20},
    legend style={
        at={(0.5,1.03)},
        anchor=south,
        draw=none,
        fill=none,
        font=\scriptsize,
        legend columns=2
    }
]

% Highest type NO LL
\addplot[ thick,red,dashed]%,mark=square*]
coordinates {(0,-11.022847) (12,0.977153) (13,1.977153)};
\addlegendentry{$\theta=1$ no LL}

% Lowest type NO LL
\addplot[ thick,blue,dashed]%,mark=*]
coordinates {(0,-8.558211) (12,1.041789) (13,1.841789)};
\addlegendentry{$\theta=0$ no LL}

% Lowest type
\addplot[ thick,blue]%,mark=*]
coordinates {(0,0) (12,0.125) (13,1.125)};
\addlegendentry{$\theta=0$ with LL}

% Highest type
\addplot[ thick,red]%,mark=square*]
coordinates {(0,0) (12,0.05) (13,1.25)};
\addlegendentry{$\theta=1$ with LL}

\end{axis}
\end{tikzpicture}

\caption{Wage schedules for the lowest and highest type.}
\label{fig:KL_LL_013_left}
\end{subfigure}
\hfill
\begin{subfigure}[t]{0.49\textwidth}
\centering\begin{tikzpicture}
\begin{axis}[
    width=\textwidth,
    height=0.67\textwidth,
    xmin=0, xmax=1,
    ymin=0.8, ymax=1.025,
    xlabel={type $\theta$},
    ylabel={$w(13)-w(12)$},
    axis lines=left,
    xtick={0,0.297211,1},
    xticklabels={$0$,$\theta^P$,$1$},
    ymajorgrids=true,
    grid style={gray!50},
    legend style={
        at={(0.5,1.03)},
        anchor=south,
        draw=none,
        fill=none,
        font=\scriptsize,
        legend columns=1
    },
    clip=false
]

% LL, x=12
\addplot[ thick,blue]
coordinates {
(0.000000,1.120143- 0.120143)
(0.027019,1.120476-0.119716)
(0.054038,1.120798-0.119302)
(0.081058,1.121107-0.118902)
(0.108077,1.121404-0.118518)
(0.135096,1.121689-0.118150)
(0.162115,1.121960-0.117798)
(0.189134,1.122217-0.117463)
(0.216153,1.122460-0.117147)
(0.243173,1.122688-0.116849)
(0.270192,1.122901-0.116571)
(0.297211,1.123098-0.116313)
(1.000000,1.123098-0.116313)
};
\addlegendentry{$[w_\th(13) - w_\theta(12)]$ with LL}

No LL, x=12
\addplot[ thick,blue,dashed,domain=0:1,samples=200]
{
13*(4-x)/(5-2*x)
 -1.5*ln((1+exp(12/(5-2*x))+exp(13/(5-2*x)))/3)
 -2.5*ln((1+exp(12/5)+exp(13/5))/3)
- (12*(4-x)/(5-2*x)
 -1.5*ln((1+exp(12/(5-2*x))+exp(13/(5-2*x)))/3)
 -2.5*ln((1+exp(12/5)+exp(13/5))/3))};
\addlegendentry{$[w_\th(13) - w_\theta(12)]$ no LL}

\addplot[black,densely dotted] coordinates {(0.297211,.8) (0.297211,1.025)};
\end{axis}

\end{tikzpicture}
\caption{Wage spread $[w_\theta(13)-w_\theta(12)]$ across types. Pooling begins at $\theta^P$.}
\label{fig:KL_LL_013_right}
\end{subfigure}

\caption{Optimal menus with and without agent-side limited liability; dashed lines depict the unrestricted affine benchmark.}
\label{fig:KL_LL_013}
\end{figure}

Figure~\ref{fig:KL_LL_013} depicts the profit-maximizing menu under agent-side limited liability and without limited liability (dashed).
Without limited liability, regularity (\Cref{ass:applications-regularity}) holds since $
\beta(\theta)
%= \frac{h(\theta)}{h(\theta)-h'(\theta)(1-F(\theta))/f(\theta)}
=
\frac{4-\theta}{5-2\theta}$ 
is strictly increasing, so \Cref{prop:multsep-linear-contracts} implies that the optimal mechanism is the strictly separating menu of affine contracts.

With limited liability, the solution has a different structure. The wage for output \(x=0\) is pinned at the floor, \(w_\theta(0)=0\) for all types $\th$, and the principal screens only through the two upper wages. At the lowest type, $w_{\ul \th}(13)-w_{\ul \th}(12)=13-12=1$: the lowest type's contract has the efficient slope of one on the range of strictly positive wages. The reason is that raising the rent $V(\ul \th)$ of the lowest type raises the rents of all types by the same amount and thereby relaxes every limited-liability constraint. At the bottom, this benefit exactly offsets the usual motive to reduce information rents, so the lowest type's incentives are undistorted, much like those of the highest type without limited liability, where $w_{\bar \th}(13)-w_{\bar \th}(12)=\beta(\bar \th)=1$.

This is also why GIM binds even though \Cref{ass:applications-regularity} holds. If the principal ignored GIM, the highest type would also receive slope one, and the relaxed optimum would feature decreasing incentive power on a range of types, which GIM rules out. Instead, starting from slope one at the bottom, the lower-output reward \(w_\theta(12)\) decreases in $\th$ while the spread \(w_\theta(13)-w_\theta(12)\) increases, strictly so on a separating arc $[\ul \th, \th^P)$. Separation stops at $\theta^P\simeq0.297$, after which all higher types are pooled on the common contract
\[
w_\theta(12)\simeq0.116,
\qquad
w_\theta(13)\simeq1.123,
\qquad
\theta\in[\theta^P,1].
\]
Thus the slope $w_\th(13)-w_\th(12)$ is strictly above one for all $\th>\ul \th$: conditional on paying positive wages, incentive power is inefficiently high. Appendix~\ref{app:KL_LL_013} verifies that this menu is optimal.

%%%%%%%

\subsection{Optimal Taxation}
In the previous examples, we considered a principal who maximizes expected profits. Theorem \ref{thm:Characterization} characterizes incentive compatibility more generally, regardless of the principal's objective. The next example showcases how our  characterization can be applied to a Mirrlees-style optimal-taxation problem (see \citealp{M1971Restud}) with flexible moral hazard.

\paragraph{Example 6.} \label{sec:ExampleMirrlees}
Suppose $X= [\ul x, \bar x]$. Let \( u(\cdot) \) be strictly concave, and define the cost function as
\[
C(\theta, \mu) = h(\theta)\int_X z(x)\mu(\de x),
\]
where \( h \) is strictly decreasing and \( z \) is differentiable, strictly increasing, and strictly convex. Each agent of type \( \theta \) chooses a distribution \( \mu \) over gross incomes \( x \) at cost \( C(\theta, \mu) \). The principal designs a menu of tax schedules \( T_{\theta}:\mathbb{R} \to \mathbb{R} \), so that an agent who reports type $\theta$ and earns income \( x \) pays \( T_{\theta}(x) \) in taxes and thus retains net income \( w_{\theta}(x) := x - T_{\theta}(x) \). Given a menu \( \{T_{\theta}\}_{\theta\in \Th} \), each agent attains utility \( U(\theta, w_{\theta}, \mu_{\theta}) \).

The principal's objective is to maximize weighted social welfare:
\[
SW := \int U(\theta, w_{\theta}, \mu_{\theta}) G(\theta) \, \de F(\theta),
\]
where \( G \) assigns Pareto weights to different types. 
The principal designs a profile of outcome distributions $\mu_\th$ and net incomes \( w_{\theta} \) (or equivalently, tax schedules \( T_{\theta} \)) to maximize welfare  \( SW \), subject to  incentive compatibility and 
a minimum revenue requirement for \( R \ge 0 \): 
\begin{align}
\int_{\underline{\theta}}^{\overline{\theta}} \int_X [x - w_{\theta}(x)] \mu_{\theta}(\de x) \, \de F(\theta) \ge R.
\tag{Min-Rev} \label{eq:Min-Rev}
\end{align}

\begin{proposition}\label{prop:MirrleesExample}
    In the above specification, the optimal mechanism can be implemented by a single, type-independent tax function $T:\mathbb{R}\rightarrow\mathbb{R}$ and for all $\th\colon$ $\mu_{\th} = \d_{\{x_\th\}}$ for some $x_\th \in X$.  
\end{proposition}

Note that once each type $\th$ is asked to implement a degenerate distribution with probability one on a single outcome $x_\th$, the agent's action choice becomes essentially observable. Proposition \ref{prop:MirrleesExample} shows that optimal-taxation problems with flexible moral hazard where mean-preserving income spreads are costly endogenously result in the familiar hidden-type-only problem sometimes referred to as featuring false moral hazard.

Proposition \ref{prop:MirrleesExample} establishes a strong version of the Taxation Principle: the optimal mechanism can be implemented using a single tax schedule $T:\mathbb{R}\rightarrow\mathbb{R}$, where agents report only their realized gross income---not their type. This simplifies implementation by eliminating the need for communication prior to income realization; agents need not disclose their type or expected income beforehand (see \citealp{poggi2025} for a discussion of the taxation principle).
Crucially, the possibility of implementing the optimal mechanism via a single tax schedule does not follow directly from the use of degenerate distributions. In a direct revelation mechanism, the principal can freely adjust net incomes for gross incomes lying outside the support of the distribution recommended to a given type, thereby deterring agents from selecting distributions with positive mass outside that support after misreporting their type. 
In contrast, with a single tax schedule, the net income corresponding to each gross income is determined by the on-path recommendation for each type. 
This limits the principal's ability to penalize off-support choices, as agents retain the option to select distributions spanning the union of supports across type-specific recommendations.

This stylized example shows conditions under which the optimal taxation problem with hidden types and hidden effort is solved by distribution choices that eliminate the moral-hazard component. We hope that the tractability of our framework will be useful for future analyses of optimal taxation problems where both frictions play a central role.

%%%%%%%%%%%%%%%%%%%%%%%%%%%

\section{Discussion and concluding remarks}

This paper provides a necessary-and-sufficient characterization of incentive compatibility in environments with hidden types and flexible moral hazard. Theorem~1 delivers a tractable representation of the relevant constraints by combining an envelope condition for information rents, first-order (on- and off-path) obedience conditions, and a generalized integral monotonicity requirement that captures incentives against double deviations.

The characterization is also constructive and lends itself to two complementary numerical approaches. One can search over candidate profiles of on-path distributions, using Theorem~1 to check implementability and to construct the implementing contracts along the way. Alternatively, one can solve a relaxed problem that drops generalized integral monotonicity and then verify whether its solution satisfies the omitted condition, as Section \ref{sec:applications} illustrates.

While our baseline model allows the agent to choose any distribution over outputs, our implementability logic only requires \emph{local} flexibility. The tractability of the flexible approach is that, once an on-path profile of distributions is fixed, the contract menu is pinned down: on-path obedience ties down $u\circ w_\theta$ on $\supp(\mu_\theta)$ up to an additive, type-dependent constant,
and the screening restrictions then determine this constant. A key step is the first-order implementability logic in \cite{georgiadis2024flexible}. As they emphasize, this version of the first-order approach does not require full distributional freedom: it continues to hold as long as the agent can make small reallocations of probability mass around the target distribution and the cost is smooth in a neighborhood of that target. Moreover, their discussion shows that the same reasoning can extend (with minor reformulations) to some settings without local flexibility, such as when feasible distributions impose lower bounds on event probabilities.\footnote{One can treat the lower bounds as fixed and reframe the agent's choice as allocating the residual probability mass above those bounds, rewriting payoffs accordingly.}

Finally, we assume that  $X \subset \real$ is one-dimensional. More generally, consider a convex and compact $X \subset \real^d$ for some finite positive integer $d$. For example, the agent could be a worker who controls different output measures that the principal can observe separately.
 Since  \Cref{thm:Characterization} focuses on implementability and abstracts from the principal's preferences, this characterization result extends directly to this case if we again restrict the principal to propose upper-semicontinuous wage contracts. 
  Allowing a multidimensional $X$ and a principal payoff function $v\colon X \to \real$ would permit the analysis of screening with multitasking. Multi-agent and dynamic environments are additional promising directions for future research.

 \newpage

\appendix

\section*{APPENDIX} 
	\setcounter{section}{0}

\section{Main Proofs} \label{sec:proofs}

\subsection{Proof of \texorpdfstring{\Cref{lm:SemiContWages}}{Lemma \ref{lm:SemiContWages}}}   \label{sec:proof_SemiContWages}
We show that  $\argmax_\mu U(\th, w, \mu) $ is nonempty.  Suppose that for a sequence of distributions $(\mu_n)_{n\ge 1}$, the utility $U(\theta,w,\mu_n)$ converges to $\sup_{\nu\in\D(X)}U(\theta,w,\nu)$. The space $\D(X)$ is compact by assumption, so there is a converging subsequence $n_k$ and $\mu\in\D(X)$ with  $\mu_{n_k} \overset{\text{weak}^*}{\longrightarrow} \mu$.
 Since $w$ is upper-semicontinuous and $C(\th,\cdot)$ is weak$^*$-continuous, the Portmanteau Theorem \citep[see][Theorem 2.1. and Problem 2.6.]{billingsley2013convergence} implies that any sequence satisfies the inequality below: 
 \begin{align*} \sup_{\nu\in \D(X)}U(\th,w,\nu) &=\underset{k\to\infty}\limsup \Paren{\int_X u\circ w(x)\mu_{n_k}(\de x)  - C(\th, \mu_{n_k})} \\ & \le \int_X u\circ w(x)\mu(\de x)  - C(\th, \mu) =  U(\th, w, \mu). \end{align*}
Hence, the limit distribution $\mu \in \D(X)$ is a maximizer.  \qed

\subsection{Proof of \texorpdfstring{\Cref{lm:MonotoneWages}}{Lemma \ref{lm:MonotoneWages}}}   \label{sec:proof_MonotoneWages}
Define $ \tilde w_\th(x)  \; := \sup_{ y \le x } w_\th(y)$. If $X=[\ul x,\bar x]$, set $\overline{w}_{\theta}(x)\; := \underset{\e \searrow 0}{\lim} \ \tilde w_{\theta} (x+\epsilon)$. If $X$ is finite, set $\overline w_\theta=\widetilde w_\theta$.
Note that each $\tilde{w}_{\theta}$ is nondecreasing, while each $\overline{w}_{\theta}$ is nondecreasing and right-continuous. Moreover, since every $w_{\theta}$ is feasible---that is, $w_{\theta}\in \mathcal{W}$---it follows that both $\tilde{w}_{\th}$ and $\overline{w}_{\th}$ also belong to $\mathcal{W}$.

First, we prove that $\{ \tilde{w}_\th,  \mu_\th\}_{\theta \in \Th}$ is incentive compatible and  $ \mu_\theta \Paren{ \{x \colon  \tilde w_\th(x) \ne w_\th(x) \}} = 0$ for all $\th\in \Th$. 
Then, we show for the right-continuous modifications that  $\{ \overline{w}_\th,  \mu_\th\}_{\theta \in \Th}$ is incentive compatible and  $ \mu_\theta \Paren{ \{x \colon  \overline{w}_\th(x) \ne w_\th(x) \}} = 0$ for all $\th\in \Th$. 
For the first part, we apply a result from \citetalias{georgiadis2024flexible}:
\begin{lemma*}[Lemma 5, \citetalias{georgiadis2024flexible}:]
For every $\nu\in \D(X)$ and every  $\e >0$, a measurable function $g : X \rightarrow X$ exists such that $g(x)\leq x$ for all $x$, and $w\circ g(x) \geq [\tilde w(x)-\e] \ \nu$-almost surely.
\end{lemma*}

Define
\begin{align*}
    V(\theta,\hat{\theta}) := & \underset{\mu\in \D(X)}{\sup}\ \left\{\int u\circ w_{\hat \theta}(x)\mu(\de x)- C(\theta,\mu) \right\}, \\
        \tilde V(\theta,\hat{\theta}) := & \underset{\mu\in \D(X)}{\sup}\ \left\{\int u\circ \tilde w_{\hat \theta}(x)\mu(\de x)- C(\theta,\mu) \right\},\\
     \bar V(\theta,\hat{\theta}) := & \underset{\mu\in \D(X)}{\sup}\ \left\{\int u\circ \overline{w}_{\hat{\theta}}(x)\mu(\de x)- C(\theta,\mu) \right\}. 
\end{align*}

\begin{claim}\label{claim:WageTilde}
    $\tilde V(\theta,\hat{\theta}) = V(\theta,\hat{\theta})$ for all $\theta$, $\hat{\theta}$. Moreover, $\mu_\theta \Paren{ \{x \colon  \tilde w_\th(x) \ne w_\th(x) \}} = 0$ for all $\th\in \Th$ and
    \[\mu_{\theta} \in \underset{\mu\in \D(X)}{\argmax} \ \left\{\int u\circ\tilde{w}_{\th}(x)\mu(\de x)- C(\theta,\mu)\right\}.\]
\end{claim}
\noindent
\begin{proof}[Proof of \Cref{claim:WageTilde}]
    As $\tilde{w}_{\hat{\th}}(x)\geq w_{\hat{\th}}(x)$ for all $x$, $\tilde V(\theta,\hat{\theta}) \geq V(\theta,\hat{\theta})$ for all $\theta$, $\hat{\theta}$. It remains to show that $\tilde V(\theta,\hat{\theta}) \leq V(\theta,\hat{\theta})$. Suppose, by way of contradiction, that there exist $\theta$ and $\hat{\theta}$ such that $\tilde{V}(\theta,\hat{\theta})>V(\theta,\hat{\theta})$. Then, there must exist $\nu\in\D(X)$ and $\e>0$ such that
    \begin{align*}
        \int u\circ \tilde w_{\hat{\theta}}(x) \nu (\de x) - C(\theta,\nu) > V(\theta,\hat{\theta})+\e.
    \end{align*}

    Let $g:X\rightarrow X$ be such that $g(x)\leq x$ for all $x$ and $u\circ w_{\hat{\theta}} \circ g\geq [u\circ \tilde w_{\hat{\theta}} - \e] \ \nu$-almost surely (such a $g$ exists by Lemma 5 in \citetalias{georgiadis2024flexible} and monotonicity of $u$). Define $\bar{\nu} = \nu \circ g^{-1}$, that is, $\bar{\nu} (Y) = \nu \circ g^{-1}(Y)$ for all Borel $Y$. We claim that $\bar{\nu}$ delivers the agent strictly higher utility under $w_{\hat{\theta}}$ than $V(\theta,\hat{\theta})$, thereby contradicting the definition of $V(\theta,\hat{\theta})$. This claim is implied by the following inequality chain:
    \begin{align*}
        \int u\circ w_{\hat{\theta}}(x)\bar{\nu}(\de x) - C(\th, \bar{\nu}) &\geq \int [u\circ \tilde w_{\hat{\theta}}(x)-\e]\nu(\de x)-C(\theta, \bar{\nu})\\
        &\geq \int u\circ \tilde{w}_{\hat{\theta}}(x)\nu(\de x)- C(\th, \nu)-\e >V(\theta,\hat{\theta}),
    \end{align*}
    where the first inequality comes from $u\circ w_{\hat{\theta}}\circ g\geq [u\circ\tilde w_{\hat{\theta}}-\e]$ holding $\nu$-almost surely, the second inequality from $\nu$ first-order stochastically dominating $\bar{\nu}$, and the third inequality from the contradiction assumption. Hence, $V(\theta,\hat{\theta}) = \tilde{V}(\theta,\hat{\theta})$.

    Applying the chain of inequalities above with $\theta=\hat{\theta}$, it implies that $\mu_{\theta}$ satisfies on-path obedience under $\tilde{w}_{\theta}$, which then implies  $\mu_\theta \Paren{ \{x \colon  \tilde w_\th(x) \ne w_\th(x) \}} = 0$ for all $\th\in \Th$.
\end{proof}

If $X$ is finite, $\tilde w_\th$ is right-continuous and the proof is complete. To complete the proof of \Cref{lm:MonotoneWages} for the remaining case with $X=[\ul x, \bar x]$, we prove an analogous claim to \Cref{claim:WageTilde} for $\overline{w}$: 
\begin{claim}\label{claim:WageBar}
    $\bar V(\theta,\hat{\theta}) =\tilde  V(\theta,\hat{\theta})$ for all $\theta$, $\hat{\theta}$. Moreover, $\mu_\theta \Paren{ \{x \colon  \overline{w}_\th(x) \ne \tilde w_\th(x) \}} = 0$ for all $\th\in \Th$ and all 
    \[\mu_{\theta} \in \underset{\mu\in \D(X)}{\argmax} \ \left\{\int u\circ\overline{w}_{\th}(x)\mu(\de x)- C(\th, \mu)\right\}.\]
\end{claim}
\noindent
\begin{proof}[Proof of \Cref{claim:WageBar}]
First, as $\tilde{w}_{\th}$ is nondecreasing, we have $\overline{w}_{\hat{\theta}}(x)\geq \tilde{w}_{\hat{\theta}}(x)$ for all $x$, $\theta$, $\hat{\theta}$. Hence, $\bar{V}(\theta,\hat{\theta}) \geq \tilde{V}(\theta,\hat{\theta}) $. It remains to prove  $\bar{V}(\theta,\hat{\theta}) \leq \tilde{V}(\theta,\hat{\theta}) $. Suppose, toward a contradiction, that there exists $(\theta,\hat{\th})$ such that $\bar{V}(\theta,\hat{\theta})>\tilde{V}(\theta,\hat{\theta})$. Then, there must exist $\nu$ and $\e>0$ such that
\begin{align*}
    \int u\circ \overline{w}_{\hat{\theta}}(x)\nu(dx)- C(\theta,\nu) > \tilde{V}(\theta,\hat{\theta})+\e.
\end{align*}
Let $g:X \rightarrow X$ be a measurable function such that $g\Paren{\max X} = \max X$, and for all $x<\max X\colon$  $x<g(x)\leq x+\delta$, where $\delta>0$. Define $\nu_{\delta} = \nu \circ g^{-1}$. We claim that for a small enough $\delta$, $\nu_{\delta}$ delivers the agent strictly higher utility under $\tilde{w}_{\hat{\theta}}$ than $\tilde{V}(\theta,\hat{\theta})$, thereby reaching a contradiction. This claim is implied by the following chain of inequalities:
\begin{align*}
           \int u\circ \tilde{w}_{\hat{\theta}}(x)\nu_\delta(\de x) - C(\th, \nu_{\delta}) &\geq \int u\circ \overline{w}_{\hat{\theta}}(x)\nu(\de x)-C(\th, \nu_{\delta})\\
        &\geq \int u\circ \overline{w}_{\hat{\theta}}(x)\nu(\de x)- C(\th, \nu)-\e > \tilde{V}(\theta,\hat{\theta}), 
\end{align*}
where the first inequality comes from $u\circ \tilde{w}_{\hat{\theta}}\circ g\geq u\circ\overline{w}_{\hat{\theta}}$, the second inequality from $\delta$ being sufficiently small and $C(\theta,\cdot)$ continuous, and the third inequality from the contradiction assumption. 

 Applying the chain of inequalities above with $\theta=\hat{\theta}$, it implies that $\mu_{\theta}$ satisfies on-path obedience under $\overline{w}_{\theta}$, which, together with $\overline{w} \ge w$,  implies  $\mu_\theta \Paren{ \{x \colon  \overline{w}_\th(x) \ne w_\th(x) \}} = 0$ for all $\th\in \Th$.
\end{proof}

\subsection{Proof of \texorpdfstring{\Cref{thm:Characterization}}{Theorem \ref{thm:Characterization}}}   \label{sec:proof_Characterization}

Note that $U(\cdot , w_{\hat{\theta}},\mu )$ is continuously differentiable for all  $ (w_{\hat{\theta}},\mu)$  and $\partial U(\th, w_{\hat{\theta}},\mu)/ \partial \th  = - \partial C(\th, \mu )/\partial \th$, which by Assumption \ref{Ass:Diff} is bounded by an integrable function $b$.
Hence, by Theorem 2 of \cite{milgrom2002envelope}, any mechanism that induces truthtelling must satisfy
$$V(\theta) = V(\underline{\theta})-\int_{\underline{\theta}}^{\theta} \frac{\partial}{\partial \th} C(t, \mu_{t})\de t.$$

We have shown that \cref{it:Envelope} is necessary for truthtelling. Items \ref{it:On-path-OB} and \ref{it:Off-path-OB} follow from \citetalias{georgiadis2024flexible} as follows: Lemma 1 in \citetalias{georgiadis2024flexible} immediately gives \cref{it:Off-path-OB}. For \cref{it:On-path-OB}, note that we need to ensure that the agent of type $\th$  gets utility $V(\theta)$, instead of the minimum-payment constraint in \citetalias{georgiadis2024flexible}. Here,  $\Big[ V(\theta)+C(\th, \mu_{\theta})-\int c_{\mu_{\theta}}(\th, x)\mu_{\theta}(\de x)\Big]$ is the constant $m$ in Proposition 2 in \citetalias{georgiadis2024flexible}, which guarantees that the agent gets utility $V(\theta)$.\footnote{GRS assume Gâteaux differentiability at every $\mu\in\Delta(X)$, but the proofs of their Lemma 1 and Proposition 2 only use it at the distribution being characterized. For necessity, this distribution is a maximizer, so Assumption~\ref{Ass:Gat} applies; for sufficiency, differentiability is part of items 2 and 3.}

Now, we show that given that items  \ref{it:Envelope}-\ref{it:Off-path-OB} hold, item \ref{it:GIM} is necessary and sufficient for incentive compatibility. 
Define $V(\theta, \hat \th) := \sup_{\nu \in \D(X)}U(\th, w_{\hat\th}, \nu)=U(\th, w_{\hat\th}, \mu_{\th,\hat \th})$ and $g(\theta,\hat{\theta}) := V(\theta)-V(\th ,\hat{\theta})$ as $\th$'s loss from misreporting $\hat \th$. 
The mechanism $\{w_\th,\mu_\th\}_{\th\in\Th}$ satisfies \eqref{eq:IC} if and only if $g(\theta,\hat \th)\geq 0$ for all $\theta,\hat{\theta}$. Note that  $g(\hat{\theta},\hat{\theta}) = V(\hat{\theta})-V(\hat{\theta},\hat{\theta})=0$. Hence, $\{w_\th,\mu_\th\}_{\th\in\Th}$ satisfies \eqref{eq:IC} if and only if
\begin{equation*}
    g(\th, \hat{\theta}) =  g(\th, \hat{\theta})-g(\hat{\theta},\hat{\theta}) 
    = \int_{\hat{\theta}}^{\theta}\frac{\partial g(t,\hat{\theta})}{\partial\theta}\de t 
    % = \int_{\hat{\theta}}^{\theta}\big[V^{\prime}(t)-\frac{\partial V(t,\hat \th)}{\partial \theta}\big]\de t 
    =\int_{\hat{\theta}}^{\theta}\Big[-\frac{\partial C(t, \mu_{t})}{\partial \th}+\frac{\partial C(t, \mu_{t, \hat \th })}{\partial \th }\Big]\de t \geq 0.
\end{equation*}
\qed

\subsection{Proof of Corollary \ref{cor:GenMon-MultSeparableCost}} \label{sec:ProofSufficientCorol}

Suppose the single-crossing property in the Corollary statement holds. Monotonicity of $\hat \th \to \k(\mu_{\th, \hat \th})$ implies the following property.
\begin{itemize}
    \item \( \kappa(\mu_{\theta,\hat{\theta}}) \geq \kappa(\mu_{\theta}) \) for all \( \hat{\theta} \geq \theta \);
    \item \( \kappa(\mu_{\theta,\hat{\theta}}) \leq \kappa(\mu_{\theta}) \) for all \( \hat{\theta} \leq \theta \).
\end{itemize}
 Then,
\begin{align*}
   \Big[- \frac{\partial C(t,\mu_{t})}{\partial \th }+ \frac{\partial C(t,\mu_{t,\hat{\theta}})}{\partial \th }\Big]= -h^{\prime}(t)\Big[\kappa(\mu_{t})-\kappa(\mu_{t,\hat{\th}})\Big] \begin{cases}\leq 0 \quad \text{for all } t\leq \hat{\theta}\\
   \geq 0 \quad \text{for all } t \geq \hat{\theta}.
   \end{cases}
\end{align*}
Integrating from $\hat{\theta}$ to $\theta$ delivers the result. \qed

\section{Proofs for Applications}\label{sec:proofsApplications}

\subsection{Proof of Proposition \ref{prop:multsep-linear-contracts}}
Let $\kappa_\mu$ denote the G\^ateaux derivative of $\kappa$ at $\mu$, so that for every $\mu,\nu\in\Delta(X)$,
\[
\lim_{\varepsilon\searrow 0}
\frac{\kappa((1-\varepsilon)\mu+\varepsilon\nu)-\kappa(\mu)}{\varepsilon}
=
\int_X \kappa_\mu(x)\,[\nu-\mu](\de x).
\]

For each type $\theta$, define
\[
J_\theta(\mu):=m(\mu)-\Phi(\theta)\kappa(\mu).
\]
Since $V(\underline\theta)$ enters the principal's objective in \eqref{eq:multsep-virtual-profit} with coefficient $-1$ and individual rationality requires $V(\underline\theta)\ge 0$, any solution to the relaxed problem sets $V^*(\underline\theta)=0$. Hence, once we verify incentive compatibility, optimality in the original problem will follow directly from \eqref{eq:multsep-virtual-profit} and the fact that $\mu_\theta^*$ maximizes $J_\theta$ pointwise.

We therefore verify the four conditions in \Cref{thm:Characterization}.

\paragraph{Envelope.}
Under the multiplicatively separable specification,
\[
\frac{\partial C(\theta,\mu)}{\partial \theta}=h'(\theta)\kappa(\mu).
\]
By the definition of $V^*$,
\[
V^*(\theta)
=
\int_{\underline\theta}^{\theta}[-h'(t)]\kappa(\mu_t^*)\,\de t
=
V^*(\underline\theta)-\int_{\underline\theta}^{\theta}\frac{\partial C(t,\mu_t^*)}{\partial t}\,\de t,
\]
with $V^*(\underline\theta)=0$. Thus item~\ref{it:Envelope} of \Cref{thm:Characterization} holds.

\paragraph{On-path obedience and linearity of wages.}
Fix $\theta\in\Theta$. Since $\mu_\theta^*$ solves \eqref{eq:multsep-relaxed} and $J_\theta$ is concave, the directional derivative of $J_\theta$ at $\mu_\theta^*$ in the direction of any $\nu\in\Delta(X)$ must be non-positive:
\[
0
\ge
\lim_{\varepsilon\searrow 0}
\frac{J_\theta((1-\varepsilon)\mu_\theta^*+\varepsilon\nu)-J_\theta(\mu_\theta^*)}{\varepsilon}.
\]
Using the G\^ateaux derivative of $\kappa$, this becomes
\[
m(\nu)-m(\mu_\theta^*)
-
\Phi(\theta)
\int_X \kappa_{\mu_\theta^*}(x)\,[\nu-\mu_\theta^*](\de x)
\le 0.
\]
Equivalently,
\[
\int_X \bigl[x-\Phi(\theta)\kappa_{\mu_\theta^*}(x)\bigr] \,\nu(\de x)
\le
m(\mu_\theta^*)-\Phi(\theta)\int_X \kappa_{\mu_\theta^*}(x)\,\mu_\theta^*(\de x)
\qquad \text{for all } \nu\in\Delta(X).
\]
Because $\nu$ ranges over all probability measures on $X$, the last equation implies
\begin{equation}\label{eq:proof-multsep-support-ineq}
x-\Phi(\theta)\kappa_{\mu_\theta^*}(x)
\le
m(\mu_\theta^*)-\Phi(\theta)\int_X \kappa_{\mu_\theta^*}(y)\,\mu_\theta^*(\de y)
\qquad \text{for all } x\in X,
\end{equation}
with equality for $\mu_\theta^*$-almost every $x$.

Now define
\[
a(\theta):=V^*(\theta)+h(\theta)\kappa(\mu_\theta^*)-\beta(\theta)m(\mu_\theta^*),
\qquad
w_\theta^*(x):=a(\theta)+\beta(\theta)x.
\]
Since $\theta\mapsto \mu_\theta^*$ is measurable and both $m$ and $\kappa$ are weak$^*$-continuous, the map $a$ is Borel.

Using $\beta(\theta)\Phi(\theta)=h(\theta)$ and \eqref{eq:proof-multsep-support-ineq}, we obtain for every $x\in X$,
\begin{align*}
w_\theta^*(x)
&=
V^*(\theta)+h(\theta)\kappa(\mu_\theta^*)-\beta(\theta)m(\mu_\theta^*)+\beta(\theta)x \\
&\le
V^*(\theta)+h(\theta)\kappa(\mu_\theta^*)
-h(\theta)\int_X \kappa_{\mu_\theta^*}(y)\,\mu_\theta^*(\de y)
+h(\theta)\kappa_{\mu_\theta^*}(x),
\end{align*}
with equality for $\mu_\theta^*$-almost every $x$. Since
\[
c_{\mu_\theta^*}(\theta,x)=h(\theta)\kappa_{\mu_\theta^*}(x),
\qquad
C(\theta,\mu_\theta^*)=h(\theta)\kappa(\mu_\theta^*),
\]
this is exactly item~\ref{it:On-path-OB} of \Cref{thm:Characterization}. In particular, $w_\theta^*$ implements $\mu_\theta^*$ on path. Integrating $w_\theta^*$ against $\mu_\theta^*$ gives
\[
\int_X w_\theta^*(x)\,\mu_\theta^*(\de x)
=
a(\theta)+\beta(\theta)m(\mu_\theta^*)
=
V^*(\theta)+h(\theta)\kappa(\mu_\theta^*),
\]
so the utility delivered on path is indeed $V^*(\theta)$.

\paragraph{Off-path obedience.}
Fix $(t,\hat\theta)\in\Theta^2$ and let
\[
\tilde\mu:=\mu_{t,\hat\theta}^*,
\qquad
\hat\beta:=\beta(\hat\theta).
\]
By construction, $\tilde\mu$ solves
\[
\max_{\nu\in\Delta(X)}\bigl\{\hat\beta m(\nu)-h(t)\kappa(\nu)\bigr\}.
\]
The objective is concave in $\nu$, so the same directional-derivative argument implies that for every $\nu\in\Delta(X)$,
\[
\hat\beta\bigl[m(\nu)-m(\tilde\mu)\bigr]
-
h(t)\int_X \kappa_{\tilde\mu}(x)\,[\nu-\tilde\mu](\de x)
\le 0.
\]
Equivalently,
\[
\int_X \bigl[\hat\beta x-h(t)\kappa_{\tilde\mu}(x)\bigr]\,\nu(\de x)
\le
\int_X \bigl[\hat\beta x-h(t)\kappa_{\tilde\mu}(x)\bigr]\,\tilde\mu(\de x)
\qquad \text{for all } \nu\in\Delta(X).
\]
Adding the constant $a(\hat\theta)$ to the integrand and recalling that
\[
w_{\hat\theta}^*(x)=a(\hat\theta)+\beta(\hat\theta)x,
\qquad
c_{\tilde\mu}(t,x)=h(t)\kappa_{\tilde\mu}(x),
\]
we obtain
\[
\tilde\mu
\in
\arg\max_{\nu\in\Delta(X)}
\int_X \bigl[w_{\hat\theta}^*(x)-c_{\tilde\mu}(t,x)\bigr] \,\nu(\de x).
\]
Thus item~\ref{it:Off-path-OB} of \Cref{thm:Characterization} holds.

\paragraph{Generalized integral monotonicity.}
Fix $t\in\Theta$ and take two reports $\hat\theta_1<\hat\theta_2$. Write
\[
\mu_i:=\mu_{t,\hat\theta_i}^*,
\qquad
\beta_i:=\beta(\hat\theta_i),
\qquad i=1,2.
\]
Since $\mu_i$ maximizes $\beta_i m(\nu)-h(t)\kappa(\nu)$, we have
\[
\beta_1 m(\mu_1)-h(t)\kappa(\mu_1)
\ge
\beta_1 m(\mu_2)-h(t)\kappa(\mu_2)
\]
and
\[
\beta_2 m(\mu_2)-h(t)\kappa(\mu_2)
\ge
\beta_2 m(\mu_1)-h(t)\kappa(\mu_1).
\]
Adding these inequalities yields
\[
(\beta_2-\beta_1)\bigl[m(\mu_2)-m(\mu_1)\bigr]\ge 0.
\]
Under \Cref{ass:applications-regularity}, $\beta_2\ge\beta_1$.
If $\beta_2>\beta_1$, the preceding inequality implies
$m(\mu_2)\ge m(\mu_1)$. If $\beta_2=\beta_1$, the maximization problems at the two reports $\hat{\th}_1$ and $\hat{\th}_2$ coincide. We select the same off-path best response across reports with the same value of $\beta$.\footnote{Retaining the
prescribed on-path response whenever
$\beta(\hat\theta)=\beta(t)$.} Thus $\mu_2=\mu_1$ in this case,
and $m(\mu_2)\ge m(\mu_1)$ holds in both cases.
 Substituting this back into the first optimality inequality gives
\[
h(t)\bigl[\kappa(\mu_2)-\kappa(\mu_1)\bigr]
\ge
\beta_1\bigl[m(\mu_2)-m(\mu_1)\bigr]
\ge 0.
\]
Therefore, for each fixed true type $t$, the map $\hat\theta\mapsto \kappa(\mu_{t,\hat\theta}^*)$ is weakly increasing. By \Cref{cor:GenMon-MultSeparableCost}, item~\ref{it:GIM} of \Cref{thm:Characterization} follows.

We have verified all four conditions in \Cref{thm:Characterization}, so the extended mechanism
\[
\bigl\{w_\theta^*,\mu_{t,\hat\theta}^*\bigr\}_{(t,\hat\theta)\in\Theta^2}
\]
is incentive compatible. Finally, because $V^*(\underline\theta)=0$ and $\mu_\theta^*$ maximizes $J_\theta$ pointwise, \eqref{eq:multsep-virtual-profit} implies that no incentive-compatible mechanism can yield a higher value than this one. Hence the relaxed solution is optimal in the original problem and can be implemented by a menu of affine contracts. \qed

\subsection{Formal derivations and proofs for Example~2}
\label{app:moment-costs}

Let $\hat z$ be the convex minorant of $z$. In the fully convex case take $x_0=\overline x$ and define $\hat z=z$ on all of $X$. Otherwise, there is a unique $x_0\in[\underline{x},\overline{x})$ such that \begin{equation}\label{eq:moment-convex-minorant}
\hat z(x)
=
\begin{cases}
z(x), & x<x_0,\\[4pt]
z(x_0)+(x-x_0)
\dfrac{z(\overline{x})-z(x_0)}{\overline{x}-x_0},
& x\in[x_0,\overline{x}].
\end{cases}
\end{equation}

\begin{lemma}\label{Lem-S-shaped}
For each $Z\in[z(\underline{x}),z(\overline{x})]$, define
\[
\widetilde X(Z)
:=
\max_{\mu\in\Delta(X)}
\left\{
\int_X x\,\mu(\de x)
\;\middle|\;
\int_X z(x)\mu(\de x)=Z
\right\}.
\]
Then $\widetilde X(Z)=\hat z^{-1}(Z)$. Moreover, an optimizer is
\begin{equation}\label{eq:moment-canonical-distribution}
\mu^Z
=
\begin{cases}
\delta_{z^{-1}(Z)},
& Z\le z(x_0),\\[7pt]
\dfrac{z(\overline{x})-Z}{z(\overline{x})-z(x_0)}\delta_{x_0}
+
\dfrac{Z-z(x_0)}{z(\overline{x})-z(x_0)}\delta_{\overline{x}},
& Z> z(x_0).
\end{cases}
\end{equation}
\end{lemma}

\begin{proof}
Fix a feasible $Z$ and a distribution $\mu$ satisfying $\int_X z(x)\mu(\de x)=Z$. Since $\hat z$ is convex and $\hat z\le z$, Jensen's inequality gives
\[
\hat z\!\left(\int_X x\,\mu(\de x)\right)
\le
\int_X \hat z(x)\mu(\de x)
\le
\int_X z(x)\mu(\de x)
=Z.
\]
Because $\hat z$ is increasing,
\[
\int_X x\,\mu(\de x)\le \hat z^{-1}(Z).
\]
If $Z\le z(x_0)$, the degenerate distribution $\delta_{z^{-1}(Z)}$ attains the bound. If $Z\ge z(x_0)$, let
\[
p(Z):=\frac{Z-z(x_0)}{z(\overline{x})-z(x_0)}.
\]
The distribution $(1-p(Z))\delta_{x_0}+p(Z)\delta_{\overline{x}}$ has $z$-moment $Z$ and mean
\[
(1-p(Z))x_0+p(Z)\overline{x}=\hat z^{-1}(Z),
\]
where the equality uses the fact that $\hat z$ is affine on $[x_0,\overline{x}]$. Hence the upper bound is attained in both regions.
\end{proof}

The next result records both the application of Proposition~\ref{prop:multsep-linear-contracts} and a non-affine implementation.

\begin{proposition}\label{Prop-S-shaped}
Suppose Assumption~\ref{ass:applications-regularity} holds. Let $Z_\theta^*$ solve \eqref{eq:moment-scalar-main} and let $\mu_\theta^*$ be the corresponding
distribution in \eqref{eq:moment-canonical-distribution}. Define
\[
V^*(\underline\theta)=0,
\qquad
V^*(\theta)
=
\int_{\underline\theta}^{\theta}[-h'(t)]K(Z_t^*)\,\de t,
\qquad
Y_\theta^*:=\hat z^{-1}(Z_\theta^*),
\]
and
\[
\beta(\theta):=\frac{h(\theta)}{\Phi(\theta)},
\qquad
H_\theta^*:=V^*(\theta)+h(\theta)K(Z_\theta^*).
\]
For each $\theta$, choose $\ell_\theta$ such that
\begin{equation}\label{eq:moment-punishment-bound}
\ell_\theta
\le
H_\theta^*-
\beta(\theta)(Y_\theta^*-\underline{x}).
\end{equation}

If $Z_\theta^*\le z(x_0)$, write $x_\theta^*:=z^{-1}(Z_\theta^*)$ and set
\begin{equation}\label{eq:moment-target-contract}
w_\theta^*(x)
=
\begin{cases}
\ell_\theta, & x<x_\theta^*,\\[4pt]
H_\theta^*, & x\ge x_\theta^*.
\end{cases}
\end{equation}

If $Z_\theta^*>z(x_0)$, write
\[
p_\theta^*
:=
\frac{Z_\theta^*-z(x_0)}{z(\overline{x})-z(x_0)}
\]
and set
\begin{equation}\label{eq:moment-single-value-bonus}
w_\theta^*(x)
=
\begin{cases}
\ell_\theta,
& x<x_0,\\[4pt]
H_\theta^*-
\beta(\theta)p_\theta^*(\overline{x}-x_0),
& x\in[x_0,\overline{x}),\\[4pt]
H_\theta^*+
\beta(\theta)(1-p_\theta^*)(\overline{x}-x_0),
& x=\overline{x}.
\end{cases}
\end{equation}
Then the step-contract menu $\{w_\theta^*,\mu_\theta^*\}_{\theta\in\Theta}$ implements the optimal allocation and rent schedule. In particular, there is
an extended recommendation schedule for which the mechanism satisfies all four conditions in Theorem~\ref{thm:Characterization}.
\end{proposition}

\begin{proof}
By Lemma~\ref{Lem-S-shaped}, fixing the moment $Z$ and maximizing mean output yields $\hat z^{-1}(Z)$. Hence the relaxed problem in
\eqref{eq:multsep-relaxed} is exactly \eqref{eq:moment-scalar-main}, and the canonical distribution associated with $Z_\theta^*$ solves that relaxed problem for type $\theta$.

Assumption~\ref{ass:applications-regularity} is the condition required by Proposition~\ref{prop:multsep-linear-contracts}. Proposition~\ref{prop:multsep-linear-contracts} therefore implies that the distribution profile is optimal in the original problem and that it admits an affine implementation. In terms of the notation above, the affine contract for report $\theta$ can be written as 
\begin{equation}\label{eq:moment-affine-benchmark} \overline w_\theta(x)
=
H_\theta^*+\beta(\theta)(x-Y_\theta^*).
\end{equation}
Its expected payment under $\mu_\theta^*$ is $H_\theta^*$, so it delivers rent $V^*(\theta)$.

We show that the step contract in the statement agrees with \eqref{eq:moment-affine-benchmark} on the intended support and is weakly below
it elsewhere. First suppose $Z_\theta^*\le z(x_0)$. Then $\mu_\theta^*=\delta_{x_\theta^*}$ and $Y_\theta^*=x_\theta^*$. Both contracts pay $H_\theta^*$ at $x_\theta^*$. For $x\ge x_\theta^*$, the affine contract is weakly above $H_\theta^*$; for $x<x_\theta^*$, \eqref{eq:moment-punishment-bound} and the monotonicity of the affine contract give $\ell_\theta\le\overline w_\theta(x)$.

Now suppose $Z_\theta^*>z(x_0)$. Then
\[
\mu_\theta^*
=(1-p_\theta^*)\delta_{x_0}
+p_\theta^*\delta_{\overline{x}},
\qquad
Y_\theta^*
=(1-p_\theta^*)x_0+p_\theta^*\overline{x}.
\]
Consequently,
\[
\overline w_\theta(x_0)
=H_\theta^*-\beta(\theta)p_\theta^*(\overline{x}-x_0)
\]
and
\[
\overline w_\theta(\overline{x})
=H_\theta^*+\beta(\theta)(1-p_\theta^*)(\overline{x}-x_0),
\]
which are exactly the two on-support payments in \eqref{eq:moment-single-value-bonus}. Between $x_0$ and $\overline{x}$, the step contract remains at its payment at $x_0$, while the affine contract is increasing. Below $x_0$, \eqref{eq:moment-punishment-bound} again places the step contract weakly below the affine contract. Thus, in both cases, 
\begin{equation}\label{eq:moment-step-dominance} w_\theta^*(x)\le\overline w_\theta(x)
\quad\text{for every }x\in X,
\qquad
w_\theta^*(x)=\overline w_\theta(x)
\quad\text{on }\supp(\mu_\theta^*).
\end{equation}

Fix a true type $t$, a report $\hat\theta$, and a distribution $\nu\in\Delta(X)$. Equation~\eqref{eq:moment-step-dominance} implies
\[
\int_X w_{\hat\theta}^*(x)\nu(\de x)
-h(t)K\!\left(\int_X z(x)\nu(\de x)\right)
\le
\int_X \overline w_{\hat\theta}(x)\nu(\de x)
-h(t)K\!\left(\int_X z(x)\nu(\de x)\right).
\]
Every report-action deviation under the step menu therefore yields no more utility than the same deviation under the incentive-compatible affine menu. At the truthful allocation, the two menus agree on the support of $\mu_\theta^*$, so truthful utility and expected wages are unchanged. The step menu is consequently incentive compatible and gives the principal the same payoff as the affine optimum. Selecting a best response after every off-path report produces the required extended recommendation schedule.
\end{proof}

\subsection{Formal derivations and proofs for Example~3}\label{app:log-kl-relaxed-proof}
\subsubsection{Derivation of the relaxed solution}

Let \(\lambda(\theta)\) denote the co-state associated with the state variable \(V(\theta)\) in \eqref{eq:log-kl-relaxed-short}. The
Hamiltonian is
\[
\begin{aligned}
\mathcal H(\theta,V,\mu,\lambda)
={}&
f(\theta)
\left[
\sum_i x_i\mu(x_i)
-
e^V\mathcal P_{h(\theta)}(\mu)
\right]
+
\lambda[-h'(\theta)]
D_{\mathrm{KL}}(\mu\Vert\nu).
\end{aligned}
\]
The co-state equation and terminal condition are
\[
\lambda'(\theta)
=
f(\theta)e^{V(\theta)}
\mathcal P_{h(\theta)}(\mu_\theta),
\qquad
\lambda(\overline\theta)=0.
\]
By \eqref{eq:log-kl-implementation-short}, the term on the right-hand side is the expected wage paid to type \(\theta\). Defining
\(L(\theta):=-\lambda(\theta)\) and integrating backward therefore gives \eqref{eq:log-kl-shadow-cost}; equivalently,
\[
L'(\theta)
=
-f(\theta)\mathbb E_{\mu_\theta}[w_\theta],
\qquad
L(\overline\theta)=0.
\]

At an interior solution, the first-order condition for
\(\mu_\theta(x_i)\) is
\[
x_i
-
[1+h(\theta)]e^{V(\theta)}
\left(\frac{\mu_\theta(x_i)}{\nu_i}\right)^{h(\theta)}
-
\frac{[-h'(\theta)]L(\theta)}{f(\theta)}
\left[
\log\left(\frac{\mu_\theta(x_i)}{\nu_i}\right)+1
\right]
=
\xi(\theta),
\]
where \(\xi(\theta)\) is the multiplier on the probability-normalization constraint. Using \eqref{eq:log-kl-implementation-short} and
\eqref{eq:log-kl-rent-loading}, this condition can be written as
\begin{equation}
x_i
=
\chi(\theta)
+
[1+h(\theta)]w_\theta(x_i)
+
\rho(\theta)\log w_\theta(x_i),
\label{eq:log-kl-implicit-wage-appendix}
\end{equation}
where
\[
\chi(\theta)
:=
\xi(\theta)
+
\frac{[-h'(\theta)]L(\theta)}{f(\theta)}
-
\rho(\theta)V(\theta).
\]
Solving \eqref{eq:log-kl-implicit-wage-appendix} for the wage gives \eqref{eq:log-kl-lambert-short} when \(\rho(\theta)>0\) and
\eqref{eq:log-kl-affine-short} when \(\rho(\theta)=0\). Subtracting \eqref{eq:log-kl-implicit-wage-appendix} at outputs \(x_j\) and
\(x_i\) gives the pairwise condition \eqref{eq:log-kl-pairwise-solution}. Finally, \eqref{eq:log-kl-normalization-short} and
\eqref{eq:log-kl-distribution-short} follow directly from \eqref{eq:log-kl-implementation-short} and probability normalization.\footnote{These conditions are necessary for a relaxed optimum, which is all that the proof of Proposition~\ref{prop:log-kl-exact-short} uses. Online Appendix~\ref{app:log-kl-verification} gives a sufficient condition under which a solution of these conditions solves the relaxed problem \eqref{eq:log-kl-relaxed-short} and shows that it holds in the illustration of Example~3.}

\subsubsection{Proof of Lemma \ref{lem:log-single-crossing-short}}
\label{app:log-kl-single-crossing-proof}

Write \(q_r(x_i):=\log w_r(x_i)\). Fix a true type \(t\) and two reports \(r_2>r_1\), and define
\[
q_i^m:=q_{r_m}(x_i),
\qquad
\Delta q_i:=q_i^2-q_i^1,
\qquad
m\in\{1,2\}.
\]
Increasing proportional incentives imply that \(\Delta q_i\) is nondecreasing in \(i\). Moreover, because each wage schedule is
increasing in output, both \(q^1\) and \(q^2\) are nondecreasing in \(i\).

For \(s\in[0,1]\), define
\[
q_i^s:=(1-s)q_i^1+s q_i^2
\]
and the corresponding distribution
\[
\mu_i^s
:=
\frac{
    \nu_i\exp\left(q_i^s/h(t)\right)
}{
    \displaystyle\sum_k\nu_k\exp\left(q_k^s/h(t)\right)
}.
\]
This auxiliary path is used only to compare the continuation distributions induced by the two reports. Let
\[
D(s):=D_{\mathrm{KL}}(\mu^s\Vert\nu).
\]
Differentiating it with respect to $s$ gives
\[
D'(s)
=
\frac{1}{h(t)^2}
\operatorname{Cov}_{\mu^s}(q^s,\Delta q).
\]
Both \(q^s\) and \(\Delta q\) are nondecreasing in output, so their covariance is nonnegative. Indeed, for any probability vector \(p\),
\[
\operatorname{Cov}_p(a,b)
=
\frac12\sum_{i,j}p_ip_j(a_i-a_j)(b_i-b_j).
\]
Thus \(D'(s)\geq0\), and integration over \([0,1]\) yields
\[
D_{\mathrm{KL}}(\mu_{t,r_2}\Vert\nu)
\geq
D_{\mathrm{KL}}(\mu_{t,r_1}\Vert\nu).
\]
Since \(t\), \(r_1\), and \(r_2\) were arbitrary, the off-path KL divergence is nondecreasing in the report for every true type.
Corollary~2 therefore implies generalized integral monotonicity.
\qed

\subsubsection{Proof of Proposition \ref{prop:log-kl-exact-short}}
\label{app:log-kl-exactness-proof}

The proof proceeds in three steps. First, we show that every relaxed optimum is interior and use the output-span restriction in
\eqref{eq:log-kl-output-span-short} to obtain a uniform bound on the log-wage spread. Second, we use the monotonicity condition in
\eqref{eq:log-kl-rent-loading-short} to show that the rent-loading coefficient \(\rho(\theta)\) is nonincreasing. Third, we combine these two properties with the report-by-report first-order condition to show that the relaxed wage menu satisfies increasing proportional incentives. Lemma~\ref{lem:log-single-crossing-short} then implies generalized integral monotonicity and optimality of the relaxed solution in the original problem.

Let the solution to the relaxed problem be
\[
q_i(\theta):=\log w_\theta(x_i).
\]
The envelope condition and \(V(\underline\theta)=0\) imply
\begin{equation}
    V(\theta)\geq0
    \qquad
    \text{for every }\theta.
    \label{eq:log-kl-V-positive-proof}
\end{equation}

\medskip
\noindent
\textbf{Step 1: Interiority and the log-wage spread.}

For every \(\theta<\overline\theta\), one has \(L(\theta)>0\). The KL term in the report-by-report problem then rules out zero
probabilities, so \(\mu_\theta\) has full support.

At \(\theta=\overline\theta\), one has \(L(\overline\theta)=0\) and hence \(\rho(\overline\theta)=0\). If the top-type distribution did not have full support, its active set would be an upper set and \(\chi(\overline\theta)\geq x_1\). Every active wage would therefore satisfy
\[
w_{\overline\theta}(x_i)
=
\frac{x_i-\chi(\overline\theta)}{1+h(\overline\theta)}
\leq
\frac{x_n-x_1}{1+h(\overline\theta)}
<1.
\]
Together with \eqref{eq:log-kl-V-positive-proof}, this contradicts the normalization condition \eqref{eq:log-kl-normalization-short}. Thus the relaxed optimum is interior for every type.

Equation \eqref{eq:log-kl-implicit-wage-appendix} shows that wages are strictly increasing in output. Set
\[
d:=\frac{x_n-x_1}{1+h(\overline\theta)}.
\]
Applying \eqref{eq:log-kl-pairwise-solution} to \(x_n\) and \(x_1\)
gives
\[
w_\theta(x_n)-w_\theta(x_1)\leq d.
\]
The normalization condition implies
\[
w_\theta(x_1)
\leq
e^{V(\theta)}
\leq
w_\theta(x_n).
\]
Since \(V(\theta)\geq0\), it follows that \(w_\theta(x_1)\geq1-d\). Consequently,
\begin{equation}
q_n(\theta)-q_1(\theta)
=
\log\frac{w_\theta(x_n)}{w_\theta(x_1)}
\leq
-\log(1-d)
\leq1,
\label{eq:log-kl-log-range-proof}
\end{equation}
where the final inequality follows from \eqref{eq:log-kl-output-span-short}.

\medskip
\noindent
\textbf{Step 2: The rent-loading coefficient is nonincreasing.}

Let
\[
\alpha(\theta)
:=
\frac{-h'(\theta)}{f(\theta)h(\theta)}.
\]
By \eqref{eq:log-kl-rent-loading}, $\rho(\theta)=\alpha(\theta)L(\theta)$. Condition \eqref{eq:log-kl-rent-loading-short} gives
\(\alpha'(\theta)\leq0\), while \eqref{eq:log-kl-shadow-cost} gives
\[
L'(\theta)
=
-f(\theta)\mathbb E_{\mu_\theta}[w_\theta]
<0.
\]
Hence, 
\begin{equation}
\rho'(\theta)\leq0.
\label{eq:log-kl-rho-decreasing-proof}
\end{equation}

\medskip
\noindent
\textbf{Step 3: Increasing proportional incentives.}

Fix a type \(\theta\) at which the relaxed solution is differentiable; the following argument therefore applies almost everywhere.
Differentiating \eqref{eq:log-kl-implicit-wage-appendix} with respect to \(\theta\) gives
\begin{equation}
\begin{aligned}
&\left(
    [1+h(\theta)]w_\theta(x_i)+\rho(\theta)
\right)
\frac{\partial q_\theta(x_i)}{\partial\theta}=
-\chi'(\theta)
-h'(\theta)w_\theta(x_i)
-\rho'(\theta)q_\theta(x_i).
\end{aligned}
\label{eq:log-kl-qtheta-direct-proof}
\end{equation}

Differentiating \eqref{eq:log-kl-implementation-short}, multiplying by \(\mu_\theta(x_i)\), and summing over outputs gives
\begin{equation}
    \sum_i
    \mu_\theta(x_i)
    \frac{\partial q_\theta(x_i)}{\partial\theta}
    =
    0,
\label{eq:log-kl-average-qtheta-direct-proof}
\end{equation}
where we have used the envelope condition and \(\sum_i\mu_\theta(x_i)=1\).

Combining \eqref{eq:log-kl-qtheta-direct-proof} and \eqref{eq:log-kl-average-qtheta-direct-proof} yields
\begin{equation}
\chi'(\theta)
=
\frac{
\displaystyle
\sum_k
\frac{
    \mu_\theta(x_k)
}{
    [1+h(\theta)]w_\theta(x_k)+\rho(\theta)
}\left[
        -h'(\theta)w_\theta(x_k)
        -
        \rho'(\theta)q_\theta(x_k)
    \right]
}{
\displaystyle
\sum_k
\frac{
    \mu_\theta(x_k)
}{
    [1+h(\theta)]w_\theta(x_k)+\rho(\theta)
}
}.
\label{eq:log-kl-chi-weighted-average-proof}
\end{equation}
Thus, \(\chi'(\theta)\) is a weighted average of $\left[ -h'(\theta)w_\theta(x_k) - \rho'(\theta)q_\theta(x_k)
\right]$. Because wages are increasing in output, \(q_\theta(x)=\log w_\theta(x)\) is also increasing in output.
Moreover,
\[
-h'(\theta)>0
\qquad\text{and}\qquad
-\rho'(\theta)\geq0
\]
by \eqref{eq:log-kl-rho-decreasing-proof}. It follows that
\begin{equation}
    \chi'(\theta)
    \geq
    -h'(\theta)w_\theta(x_1)
    -
    \rho'(\theta)q_\theta(x_1).
\label{eq:log-kl-chi-lower-bound-proof}
\end{equation}

For fixed \(\theta\), the right-hand side of \eqref{eq:log-kl-qtheta-direct-proof}, after dividing by its positive coefficient, can be viewed as a function of the wage \(w\). Its derivative is
\begin{equation}
\begin{aligned}
&\frac{\partial}{\partial w}
\left[
\frac{
    -\chi'(\theta)
    -
    h'(\theta)w
    -
    \rho'(\theta)\log w
}{
    [1+h(\theta)]w+\rho(\theta)
}
\right]
\\
&=
\frac{
    [1+h(\theta)]\chi'(\theta)
    -
    h'(\theta)\rho(\theta)
    -
    \rho'(\theta)
    \left[
        [1+h(\theta)](1-\log w)
        +
        \dfrac{\rho(\theta)}{w}
    \right]
}{
    \left(
        [1+h(\theta)]w+\rho(\theta)
    \right)^2
}.
\end{aligned}
\label{eq:log-kl-qtheta-wage-derivative-proof}
\end{equation}

Using \eqref{eq:log-kl-chi-lower-bound-proof}, the numerator in \eqref{eq:log-kl-qtheta-wage-derivative-proof} is bounded below by
\begin{align}
-h'(\theta)
\left(
    [1+h(\theta)]w_\theta(x_1)+\rho(\theta)
\right)
-
\rho'(\theta)
\left[
    [1+h(\theta)]
    \left(
        1+q_\theta(x_1)-\log w
    \right)
    +
    \frac{\rho(\theta)}{w}
\right].
\label{eq:log-kl-qtheta-derivative-lower-bound-proof}
\end{align}

For every $w\in \left[ w_\theta(x_1),w_\theta(x_n) \right]$, the log-wage bound \eqref{eq:log-kl-log-range-proof} implies
\[
1+q_\theta(x_1)-\log w
\geq
1-
\left[
    q_\theta(x_n)-q_\theta(x_1)
\right]
\geq0.
\]
Because
\[
-h'(\theta)>0,
\qquad
-\rho'(\theta)\geq0,
\qquad
\rho(\theta)\geq0,
\]
the expression in \eqref{eq:log-kl-qtheta-derivative-lower-bound-proof} is nonnegative.
Consequently, the right-hand side of \eqref{eq:log-kl-qtheta-direct-proof} is nondecreasing in the wage.

Since \(x_j>x_i\) implies
\[
w_\theta(x_j)>w_\theta(x_i),
\]
we obtain
\[
\frac{\partial q_\theta(x_j)}{\partial\theta}
\geq
\frac{\partial q_\theta(x_i)}{\partial\theta}
\]
for almost every \(\theta\). Therefore,
\begin{equation}
    \frac{\partial}{\partial\theta}
    \left[
        q_\theta(x_j)-q_\theta(x_i)
    \right]
    \geq0
    \qquad
    \text{for almost every }\theta.
\label{eq:log-kl-increasing-differences-direct-proof}
\end{equation}

The regularity assumptions and interiority imply that the difference in \eqref{eq:log-kl-increasing-differences-direct-proof} is absolutely continuous in \(\theta\). Hence
\[
\theta
\longmapsto
q_\theta(x_j)-q_\theta(x_i)
=
\log
\frac{w_\theta(x_j)}{w_\theta(x_i)}
\]
is nondecreasing for every \(x_j>x_i\). Thus, the relaxed menu satisfies increasing proportional incentives.

Lemma~\ref{lem:log-single-crossing-short} then implies generalized integral monotonicity. The envelope condition is imposed in the relaxed problem, while on-path and off-path obedience follow from the wage and continuation-distribution formulas. Hence all four conditions in Theorem~1 hold, and the relaxed solution solves the original problem.
\qed 

%%%%%%%%%%%%%%%%%%%%%%%%%%%%

\subsection{Formal derivations and proofs for Example~4}\label{app:Example4}

Solving for the optimal allocation with the monotonicity constraint explicitly taken into account results in the following control problem, where we write $\bar z = z(\bar x)$ and define $\varphi(\th) = (A - e^\th \th)\bar z^2$ for the virtual cost term. 

We first justify the regularity implicit in the control formulation.
Write
\[
 h(\theta)=A-e^\theta,\qquad
 L(\theta,p)=\Delta x\,p-\frac12\varphi(\theta)p^2,
 \qquad
 r(\theta)=\frac{\Delta x\,h(\theta)}{\varphi(\theta)}.
\]
By binary GIM, implementable allocations admit the representation
$p_\theta=\min\{q_\theta/h(\theta),1\}$ with a nonnegative,
nondecreasing normalized bonus $q$; after saturation, $q$ may be
chosen constant. Conversely, such a bonus implements the associated
allocation together with the envelope utility. We choose endpoint
values by one-sided limits.

The function $r$ is strictly positive, with its value at $\theta=1$
defined as $\Delta x/(2\bar z^2)$ when $A=e$. For sufficiently small
$\varepsilon>0$, replacing $q$ by $\max\{q,\varepsilon\}$ moves every
changed probability towards its pointwise maximizer
$\min\{\Delta x/\varphi(\theta),1\}$. This strictly increases profit
if the right-continuous representative has $q_0=0$. Hence an optimum
has $p_0>0$ and $\underline\theta=0$. When $A=e$, it also reaches
$p=1$ strictly before $\theta=1$.

An optimal allocation cannot jump. At a jump at $t$, let
$p_-<p_+\le1$ be its one-sided limits. Strict concavity of $L(t,\cdot)$
gives $p_c\in(p_-,p_+)$ such that
\[
 L(t,p_c)>\min\{L(t,p_-),L(t,p_+)\}.
\]
Replacing $q$ by $h(t)p_c$ on a sufficiently short interval on
the corresponding side of the jump preserves monotonicity and
strictly increases profit. This argument also excludes a jump
directly to $p=1$.

Before saturation, moreover, $q=r$ on the support of the measure
$dq$. Otherwise, on a sufficiently small interval around a point
of this support, flattening $q$ to its right-endpoint value when
$q<r$, or its left-endpoint value when $q>r$, strictly increases
profit. Outside this support, $q$ is locally constant. Trimming
constant pieces from the ends of any interval therefore gives
\[
 0\le q(t)-q(s)\le \operatorname{Lip}(r)(t-s).
\]
Thus $q$ is Lipschitz. Since $h$ is bounded away
from zero on this interval, $p=q/h$ is absolutely continuous there,
and its continuous extension by $p=1$ remains absolutely continuous.
The control formulation therefore loses no potentially optimal
allocation.

\begin{align*}
    \max_{\ul \th, \bar \th, p_{\ul \th}, p_{\bar \th} , u} \;\; \int_{\ul \th}^{\bar \th} \ \Bigl( p_\th \D x -\varphi(\th) \frac12 p_\th^2
     \Bigr) \de \th + (1-\bar \th)\Paren{\D x - (A -e^{\bar \th})\frac{1}{2}(\bar z)^2 }, 
\end{align*}
subject to 
\begin{align*}
p_{\ul \th} \ge 0, \quad 1-p_{\bar \th}\ge 0,\qquad \quad  \ul \th \ge 0, \quad  1- \bar \th \ge 0,
\\
\dot p_\th = u_\th,    \qquad\quad  \text{ and } \qquad\quad  u_\th \ge p_\th \frac{e^{\th}}{A - e^{\th}}.
\end{align*}
Let $p_\th$ be the state variable and $u_\th$ the control. Let $\psi_\th$ denote the co-state. For the endpoint constraints, let $\g_{\ul \th}$ be the multiplier associated with the constraint $\ul \th \ge  0$, and $\g_{\bar \th}$ with $1 - \bar \th\ge 0$. Let $\g_{\ul p}$ be the multiplier associated with the constraint  $p_{\ul \th } \ge 0 $ and $\g_{\bar p} $  for $1-p_{\bar \th} \ge 0 $. We must have $\g_i\ge0$ for all $i\in\{\ul \th, \bar\th, \ul p,\bar p\}$. 
Finally, we let $\tau_\th \ge 0$ be the multiplier for the constraint $u_\th -   p_\th \frac{e^{\th} }{A- e^{\th}}  \ge 0$.

The extended Hamiltonian is given by 
\begin{align*}
    \HH_\th = \psi^\circ \Brac{p_\th \D x - \varphi(\th) \frac12 p_\th^2} + \psi_\th u_\th  + \tau_\th \Paren{u_\th -    p_\th\frac{ e^{\th}}{A - e^{\th}} }
\end{align*}
Denote the scrap value by 
\begin{align*}
    S(\bar \th) =   (1-\bar \th)\Paren{\D x - (A-e^{\bar \th})\frac{1}{2}(\bar z)^2}. 
\end{align*}
We maximize the integral
\begin{align*}
    \int_{\ul \th}^{\bar \th} \big(\HH_\theta-\psi_\theta\dot p_\theta \big) \de \th \, + \,  \psi^\circ S(\bar \th) \, +\g_{\ul p} p_{\ul \th }  + \g_{\bar p}(1-p_{\bar\th}) + \g_{\ul \th} \ul \th + \g_{\bar \th} (1-\bar \th), 
\end{align*}
where $\psi^\circ \in \{0,1\}$ is $0$ in case the problem degenerates. 
% \begin{align*}
%     \LL_{\th} &= \HH_\th + 
%     \LL_{\th} &= \HH_\th + 
%     \\
%     & = \psi^\circ \Brac{p_\th \D x -  \h(\th)\frac12 p_\th^2} + \psi_\th \dot p_\th+ \mu_\th \Paren{  \frac{p_\th }{e^{1-\th}-1} - \dot p_\th }
% \end{align*}
We first solve for an optimal policy with $\ul\th < \bar \th$ and return to the bang-bang case $\ul\th=\bar \th$ below. 
Note first that the first-order condition on $\HH$ with respect to $u$ gives $$\psi_\th + \tau_\th = 0.$$ 

The necessary condition for optimality of the endpoint $\bar \th$ is  
$
  \HH_{\bar \th} + \psi^\circ  \partial_{\bar \th} S(\bar \th) -\g_{\bar \th} = 0
$, or after substitution,
\begin{align}\label{eq:FOC_theta_O}
    &\psi^\circ\Brac{p_{\bar \th} \D x -\varphi(\bar\th) \frac12 p_{\bar \th}^2- \D x +\varphi(\bar\th)\frac{1}{2}} -\tau_{\bar \th}p_{\bar \th}\frac{ e^{\bar \th}}{A - e^{\bar \th}} -\g_{\bar\th} = 0.
\end{align}
At the initial point, an optimum has $\ul\th=0$ and $p_0>0$ by the perturbation argument above. The transversality condition for the initial state is $\psi_{\ul\th}=-\g_{\ul p}$, and complementary slackness gives $\g_{\ul p}\,p_{\ul\th}=0$. Hence $\psi_0=0$. 

The problem is normal. Indeed, if $\psi^\circ=0$, the adjoint
equation becomes
\[
 \dot\psi_\theta
 =-\frac{e^\theta}{A-e^\theta}\psi_\theta.
\]
Together with $\psi_0=0$, this gives $\psi\equiv0$. Stationarity
and the endpoint conditions then force all remaining multipliers
to vanish, contradicting nontriviality. Hence $\psi^\circ=1$.

For an optimla policy, the monotonicity constraint must bind for all $\th \le \bar r$, where $\th=\bar r$ solves $A \th = e^{\th}$.\footnote{That is, when $A=e$, then $\bar r=1$ and when $A>e$, then $0<\bar r<1$} For $\th\ge \bar r$, the solution from the relaxed problem without the monotonicity constraint (i.e. $p^\text{AS}$) is sufficiently increasing so that $p^\text{AS}_\th\, h(\th) = \frac{\D x}{\bar z^2}\frac{A-e^\th}{A-\th e^{\th}}$ is nondecreasing. 

Define $$
\th^B = \inf\Set{\th \in [\ul \th, \bar \th] \colon u_\th > p_\th \frac{e^\th}{A-e^\th}},$$
i.e. the lowest type at which the monotonicity constraint is slack. 

Next, we show that if an optimal policy has a slack interior segment, so that $\th^B< \bar \th$, then $\ge \bar r$. 
Using the identity $\psi_\th = -\tau_\th$, the evolution of the co-state is given by 
\begin{align}\label{eq:CoStateODE}
    \dot \psi_\th = - \partial_{p_{\th}} \HH_\th = - \D x +  \varphi(\th) p_\th - \psi_\th  \frac{ e^{\th}}{A - e^{\th}}.
\end{align} 
Suppose there is an interval of types $[s,t]\ne \varnothing$ such that the constraint $u_{\th}- p_\th\frac{ e^{\th}}{A - e^{\th}} \ge 0$ is slack. Then we must have $\tau_\th=0$ and, a fortiori, $\psi_\th = \dot \psi_\th = 0$ for all $\th \in [s,t]$.
This implies that $p_\th = \D x / \varphi(\th)$, which is exactly the solution without moral hazard $p_\th^\text{AS}$. Thus, for $\th< \bar r$, this candidate violates the monotonicity constraint. We conclude that $\th^B \ge \bar r$ whenever $\th^{B}<\bar\th$. 
 Given an initial value $p_{\ul \th}$, the ODE representing the binding monotonicity constraint  has the unique solution 
\begin{align}\label{eq:pSolved}
    p_{\th} = p_{\ul \th} \frac{A- e^{\ul{\th}}}{A-e^{\th}}.
\end{align}

\paragraph{Case 1: monotonicity constraint becomes slack.} If $\th^B < \bar \th$, then complementary slackness of $\tau_{\th^B}$ together with the FOC for $u$ implies that $\psi_{\th^B} = 0$ and $p_{\th^B} = \frac{\D x}{\varphi(\th^B)}>0$. The latter requires $p_{\ul\th}>0$ and thereby $\psi_{\ul \th} = 0$. Since the monotonicity constraint is binding for all lower types, \eqref{eq:CoStateODE} gives for $\th\in[0, \th^B]$:
\begin{align*}
      \psi_{\th} = \int_{\th}^{\th^B} \frac{A-e^{\th}}{A-e^{t}} \Paren{\D x - p_{t}\,\varphi(t)}\de t =  \int_\th^{\th^B}\frac{p_t}{p_{\th}}\Paren{\D x -p_t \, \varphi(t)}\de t. 
\end{align*}
We need that $\psi_{0} = \psi_{\th^B}=0$ and $\psi_\th = - \tau_\th \le 0$ for all $\th \in [0, \th^B]$. 
Note that the term in parentheses, $\D x - p_t\, \varphi(t)$, is strictly decreasing for $t < \bar r$ and  strictly increasing for $t>\bar r$. 
For $\psi_\th$ to be initially decreasing we must have $\D x - p_{0} \,\varphi(0) > 0 $, or, equivalently, $p_{0 } <  p^\text{AS}_{0}$.  
Conversely, for $\psi_\th$ to increase toward $\psi_{\th^B} = 0$, we need $\D x - p_t \varphi (t) < 0$ for all $t$ in some interval $[\th^B-\e, \th^B)$. Hence, as $\th \to \th^B$, we have $p_\th  - p^\text{AS}_\th >0$ and (because $\th^B<\bar \th$) $p_{\th^B} - p^\text{AS}_{\th^B} =0$. Therefore, $\th^B$ must be strictly greater than $\bar r$ (which is possible only if $A>e$).  

For this candidate solution, the end points are determined by the pair $(p_0, \th^B)$ that satisfies $\th^B > \bar r$ and solves the system of equations 
\begin{align*}
    &\D x - p_0 \frac{A-1}{A-e^{\th^B}}(A-\th^B e^{\th^B})\bar z^2   = 0,
    \\
    &    \int_{0}^{\th^B}\frac{A-1}{A-e^{t}}\Paren{\D x - p_0 \frac{A-1}{A-e^t} (A- t e^t) \bar z ^2 }\de t =0.
    \end{align*}

For  $\th \in [\th^B, \bar \th]$, we have $p_\th = \frac{\D x}{\varphi(\th)}$ for all $\th \in [\th^B, \bar \th]$ as the alternative (binding monotonicity constraint) would require $\psi_\th  > 0$, which is incompatible with $\tau_\th \ge 0$. 
Therefore, the boundary $\bar \th$ in this case is determined by the behavior of $p^\text{AS}$, with $\bar \th = 1$ if $p^\text{AS}_{1-\e} < 1$ and $\bar \th = \inf\Set{ \th \in [0,1] \colon p^\text{AS}_\th = 1 }$ otherwise.

\paragraph{Case 2: monotonicity binds throughout.} Next, we consider the case where $\th^B = \bar \th$. In this case $p_\th = p_0 \frac{A-1}{A-e^\th}$ for all $\th\in[0,\bar\th]$. 
The optimality condition for the boundary $\bar \th$ \eqref{eq:FOC_theta_O} shows immediately that 
$\tau_{\bar \th} = 0$: either $p_{\bar \th} = 1$, in which the term in square brackets is 0 and $\tau_{\bar\th}\ge 0$ and $\g_{\bar \th}\ge 0$ implies that both are 0; or $p_{\bar \th}<1$, and $\psi_{\bar \th} = -\tau_{\bar\th} = 0$ by complementary slackness. 

Hence, we get 
\begin{align*}
      \psi_{\th} = \int_{\th}^{\bar\th} \frac{A-e^{\th}}{A-e^{t}} \Paren{\D x - p_{t}\,\varphi(t)}\de t =  \int_\th^{\bar \th}\frac{p_t}{p_{\th}}\Paren{\D x -p_t \, \varphi(t)}\de t. 
\end{align*}
As in the previous case, keeping the co-state negative requires $p_0 < p^\text{AS}_0$ as well as $p_{\bar \th} > p^\text{AS}_{\bar \th}$. 
For the case $\bar\th<1$, we must have $p_{\bar \th}=1$. Thus, the end points $p_0$ and $\bar \th$ must satisfy the system 
\begin{align*}
    &p_0 = \frac{A- e^{\bar \th}}{A-1},
    \\
    &\int_{0}^{\bar \th} \frac{A-1}{A-e^t}\Paren{\D x - p_0 \frac{A-1}{A-e^t}\varphi(t)}\de t = 0.
\end{align*}
In this case $p_0 < p^{AS}_0$ is equivalent to $\D x / \bar z^2 > \frac{A}{A-1}( A - e^{\bar \th})$, whereas $p_{\bar \th} > p^\text{AS}_{\bar \th}$ is equivalent to $\D x /\bar z ^2 < A - \bar \th e^{\bar \th }$.

The remaining possibility $\theta^B=\bar\theta=1$ cannot be optimal.
For $A=e$, this already follows from $q_0>0$, which implies saturation
before $1$. Suppose therefore that $A>e$, and write
$q=(A-1)p_0>0$ for the constant normalized bonus. The endpoint
conditions give $\psi_0=\psi_1=0$. Integrating the adjoint equation
with integrating factor $1/h$ yields
\[
 q
 =\Delta x\,
   \frac{\int_0^1 h(t)^{-1}\,dt}
        {\int_0^1\varphi(t)h(t)^{-2}\,dt}
 <\frac{\Delta x}{\bar z^2},
\]
because
$\varphi(t)-\bar z^2h(t)=\bar z^2(1-t)e^t>0$ for $t<1$.
Consequently,
\[
 \dot\psi_{1-}
 =-\Delta x+\varphi(1)p_1
 =-\Delta x+\bar z^2q<0.
\]
Since $\psi_1=0$, this requires $\psi_\theta>0$ immediately below
$1$, contradicting $\psi=-\tau\le0$.

Finally, we have to consider the case $\bar \th = \ul \th = 0$. In this case $p_\th=1$ for all $\th \in[0,1]$. This is optimal if and only if $ \frac{\partial }{\partial p} \Paren{ \D x p - \varphi(0) p^2/2 }\lvert_{p=1} \ge 0$, which is equivalent to $A \le \Delta x/\bar z ^2$, which is also the condition under which $p^\text{AS}_0$ is equal to 1.

An optimum exists: normalized bonuses may be restricted to
nondecreasing functions taking values in $[0,h(0)]$. Helly's
selection theorem and dominated convergence therefore give
attainment of the objective.

The preceding cases exhaust the possible maximizers. Indeed,
$r$ decreases up to $\bar r$ and increases thereafter. Together
with continuity and $q=r$ on the support of $dq$, this permits
only an initial constant bonus followed, if at all, by tracking
$r$. An intervening constant segment cannot reconnect to strictly
increasing $r$. A final constant segment after tracking would have
$q<r$ and could be improved by slightly raising its bonus,
allowing earlier saturation.

Thus an optimum belongs to Case 1, Case 2 with $\bar\theta<1$,
or the allocation $p\equiv1$. Comparing the objective over all
feasible solutions of these cases therefore identifies a global
optimum. \qed

\subsection{Formal derivations and proofs for Example~5}\label{app:Example5}
\subsubsection{Sufficient conditions for optimality}
We first state sufficient conditions for a mechanism to solve the principal's problem in Example~5.

To capture the limited liability constraints, let $\eta_i(\th)\geq0$ denote the multiplier on the inequality $w_\th(x_i) \ge 0$.
To capture the GIM constraints, suppose there exists a kernel
$
r\mapsto\G_r
$
of finite nonnegative measures on $\Th$ such that
\[
\int_\Th \G_r(\Th)\,\de F(r)<\infty.
\]
Let $\G$ denote the induced measure on $\Th^2$, defined by
\[
\G(A\times B)
=
\int_B \G_r(A)\,\de F(r)
\]
for all Borel sets $A,B\subseteq\Th$.
 Here $\G_r$ assigns multiplier weight only to true types for which the reporting constraint associated with report $r$ is relevant.\footnote{This disintegrated formulation rules out a singular component of the multiplier in the report dimension. Allowing such components is straightforward by stating the below objective with its dual variables in integrated form, but this would obscure the report-by-report representation. In the optimal contract derived below we use a multiplier which is indeed absolutely continuous with respect to $F$.} Denote the true-type and report marginals of $\G$ by $\G_T$ and $\G_R$,
respectively.\footnote{That is, $\G_T(\de \th) = \G(\de \th \times \Th)$ and $\G_R(\de r) = \G(\Th \times \de r)$.}

For any pair of such multipliers, define the cumulative shadow cost by 
\begin{equation}\label{eq:KL_LL_multiplier_stock}
\L_\th
:=
\int_\th^{\bar\th}
\sum_{i=1}^n\eta_i(t)\,\de F(t)
+
\G_T([\th,\bar\th])
-
\G_R([\th ,\bar\th]).
\end{equation}
The first term captures the effect of the limited-liability constraints
of higher types. The remaining terms capture the effect of the global
reporting constraints on information rents.\footnote{We have
$
\G_T([\th,\bar\th])-\G_R([\th,\bar\th])
=
\G\bigl(\{(t,r):r<\th \leq t \}\bigr)
-
\G\bigl(\{(t,r):t<\th \leq r\}\bigr)$. 
This difference is the net shadow cost of reporting constraints that cross type $\th$: downward misreports across $
\th$ enter positively, whereas upward misreports across $\th$ enter negatively.}
Define, for $\L\in\real$, the adjusted virtual cost
\begin{equation}\label{eq:KL_LL_adjusted_virtual_cost}
\widetilde\Phi(\th,\L)
:=
h(\th)
-
h'(\th)
\frac{1-F(\th)-\L}{f(\th)}.
\end{equation}
When limited liability and all reporting constraints are slack, $\L=0$ and $\widetilde\Phi$ equals the virtual cost $\Phi$ of the relaxed problem \eqref{eq:multsep-relaxed}. 

The GIM multiplier also has a second direct effect. For a fixed report
$r$, the expression
\[
h(\th)
\log\!\left[
\sum_{i=1}^n
\nu_i
\left(
\frac{\mu(x_i)}{\nu_i}
\right)^{h(r)/h(\th)}
\right]
\]
is the optimized continuation payoff of true type $\th$ from reporting $r$, net of that report's promised utility (see equation \ref{eq:KL_LL_reporting_gap}).
The multiplier $\G_r$ directly penalizes distributions associated with report $r$ that make this double deviation attractive. The adjusted virtual cost $\tilde \Phi$ captures the effect of GIM on the rent schedule, while this direct effect on off-path behavior will be captured in the final term in the type-by-type objective below.

\begin{proposition}
\label{prop:KL_LL_verification}
Let
$\left\{
w_\th,V(\th),\mu_{\th,\hat \th}
\right\}_{(\th,\hat \th)\in\Th^2}$
satisfy the conditions in \Cref{thm:Characterization} and agent-side limited liability.
Denote by $G$ the reporting loss in \eqref{eq:KL_LL_reporting_gap}.
Suppose there exist integrable functions $
\{\eta_i\}_{
i=1}^n$ from $\Th$ to $\real_+$
and a finite nonnegative measure
\[
\G(\de\th,\de r)
=
\G_r(\de\th)\,\de F(r),
\]
such that, for $F$-a.e.\ type $\th$
\begin{align}
\mu_\theta
\in
\arg\max_{\mu\in\Delta(X)}
\Bigg\{
& m(\mu)-\widetilde\Phi(\th,\L_\th)
D_{KL}(\mu\Vert\nu) \notag
\\
&+ \sum_{i=1}^n
\eta_i(\th)h(\th)
\log\!\left(\frac{\mu(x_i)}{\nu_i}\right)
-
\int_\Th
h(t)
\log\!\left[
\sum_{i=1}^n
\nu_i
\left(
\frac{\mu(x_i)}{\nu_i}
\right)^{h(\th)/h(t)}
\right]
\G_\th(\de t)
\Bigg\} 
\label{eq:KL_LL_pointwise_suff}
\end{align}

Finally, suppose the multipliers and the cumulative shadow cost $\L$ defined in
\eqref{eq:KL_LL_multiplier_stock} satisfy complementary slackness:
\begin{align}
\eta_i(\th)w_\th (x_i)
&=0
\quad
\text{for every }i
\text{ and for $F$-a.e.\ }\th,
\label{eq:KL_LL_CS_LL}
\\
\int_{\Th^2}
G(\th,r)\,
\G(\de\th,\de r)
&=0,
\label{eq:KL_LL_CS_GIM}
\\
\L_{\ul\th}\leq1,
\qquad
\left[1-\L_{\ul\th}\right]V(\ul\th)
&=0.
\label{eq:KL_LL_CS_IR}
\end{align}

Then the mechanism maximizes the principal's expected profit among all individually rational and incentive-compatible mechanisms satisfying
agent-side limited liability.
\end{proposition}

The proposition has the same basic logic as \Cref{prop:multsep-linear-contracts}: the multipliers transform the principal's objective into a collection of type-by-type maximization problems, with $\L_\th$ adjusting the virtual cost and $\G_\th$ penalizing the double deviations that $\mu_\th$ makes attractive to other types.
\subsubsection{Proof of Proposition \ref{prop:KL_LL_verification}}

We add stars to the objects of the candidate mechanism and leave an
arbitrary feasible mechanism unstarred. For the latter, write
\[
D_\th:=D_{KL}(\mu_\th\Vert\nu),
\qquad
S(\th):=\sum_{i=1}^n\eta_i(\th),
\]
and let \(G\) denote its reporting loss as defined in
\eqref{eq:KL_LL_reporting_gap}. Denote by
\(\mathcal Q_\th(\mu)\) the expression in braces in
\eqref{eq:KL_LL_pointwise_suff}.

We first note that the candidate is individually rational. Indeed,
on-path obedience and agent-side limited liability imply
\[
\begin{aligned}
V^*(\th)
&=
\max_{\mu\in\Delta(X)}
\left\{
\sum_{i=1}^n w_\th^*(x_i)\mu(x_i)
-
h(\th)D_{KL}(\mu\Vert\nu)
\right\}
\geq
\sum_{i=1}^n\nu_iw_\th^*(x_i)
\geq0.
\end{aligned}
\]

Because \(X\) is finite, \(\nu\) has full support, and \(h\) is positive
and continuous, \(D_{KL}(\mu\Vert\nu)\) is bounded over
\(\Delta(X)\), as is the power-log expression appearing in
\eqref{eq:KL_LL_pointwise_suff}. The envelope condition therefore
bounds \(V\) on \(\Th\). Together with limited liability,
\eqref{eq:KLLLWageFromDensity}, the integrability of the functions
\(\eta_i\), and the finiteness of \(\G\), this justifies the changes in
the order of integration below.

\paragraph{Adjoining the constraints.}
Fix an arbitrary individually rational and incentive-compatible mechanism satisfying agent-side limited liability. 
The expected profit of the principal if the allocation is $\mu = \{\mu_\th\}_{\th \in \Th}$ and the agent's value function $V$ can be written as
\begin{equation}
\Pi(\mu,V)
=
\int_\Th
\left[
m(\mu_\th)-h(\th)D_\th-V(\th)
\right]\de F(\th).
\label{eq:KL_LL_verification_profit}
\end{equation}

Adjoin the limited-liability and GIM inequalities and define the Lagrangian 
\begin{align}
\mathcal L_{\eta,\G}(\mu,V)
:={}&
\Pi(\mu,V)
+
\int_\Th
\sum_{i=1}^n
\eta_i(\th)w_\th(x_i)\,\de F(\th)
+
\int_{\Th^2}
G(t,r)\,\G(\de t,\de r).
\label{eq:KL_LL_verification_lagrangian}
\end{align}
Since
\[
w_\th(x_i)\geq0,
\qquad
G(t,r)\geq0,
\qquad
\eta_i(\th)\geq0,
\qquad
\G\geq0,
\]
we have
\begin{equation}
\Pi(\mu,V)
\leq
\mathcal L_{\eta,\G}(\mu,V).
\label{eq:KL_LL_verification_first_bound}
\end{equation}

Substituting for the wages $w$ and the misreporting losses $G$ using \eqref{eq:KLLLWageFromDensity} and
\eqref{eq:KL_LL_reporting_gap} into the Lagrangian 
\eqref{eq:KL_LL_verification_lagrangian} gives
\begin{align}
\mathcal L_{\eta,\G}(\mu,V)
={}&
\int_\Th
\Bigg\{
m(\mu_\th)-h(\th)D_\th
+
\sum_{i=1}^n
\eta_i(\th)h(\th)
\log\left(\frac{\mu_\th(x_i)}{\nu_i}\right)
\Bigg\}\de F(\th)
\notag\\
&-
\int_{\Th^2}
h(t)
\log\left[
\sum_{i=1}^n
\nu_i
\left(
\frac{\mu_r(x_i)}{\nu_i}
\right)^{h(r)/h(t)}
\right]
\G(\de t,\de r)
\notag\\
&+
\int_\Th
[S(\th)-1]V(\th)\,\de F(\th)
+
\int_\Th V(\th)\,\G_T(\de\th)
-
\int_\Th V(\th)\,\G_R(\de\th).
\label{eq:KL_LL_verification_expanded}
\end{align}

\paragraph{Eliminating the rent schedule.}
The envelope condition gives
\begin{equation}
V(\th)
=
V(\ul\th)
+
\int_{\ul\th}^{\th}
[-h'(s)]D_s\,\de s.
\label{eq:KL_LL_verification_envelope}
\end{equation}
Substituting \eqref{eq:KL_LL_verification_envelope} into the last line of
\eqref{eq:KL_LL_verification_expanded} and changing the order of
integration yields
\begin{align}
&
\int_\Th
[S(\th)-1]V(\th)\,\de F(\th)
+
\int_\Th V(\th)\,\G_T(\de\th)
-
\int_\Th V(\th)\,\G_R(\de\th)
\notag\\
={}&
\left[
\int_\Th[S(\th)-1]\,\de F(\th)
+
\G_T(\Th)-\G_R(\Th)
\right]V(\ul\th)
\notag\\
&+
\int_{\ul\th}^{\bar\th}
[-h'(s)]D_s
\Bigg[
\int_s^{\bar\th}
[S(\th)-1]\,\de F(\th)
+
\G_T([s,\bar\th])
-
\G_R([s,\bar\th])
\Bigg]\de s.
\label{eq:KL_LL_verification_rent_terms}
\end{align}
Naturally, the two marginals of \(\G\) have the same total mass:
$
\G_T(\Th)=\G_R(\Th)=\G(\Th \times \Th )$. 
Moreover, by the definition of $\L$ in \eqref{eq:KL_LL_multiplier_stock}, 
$ 
\L_{\ul\th}
=
\int_\Th S(\th)\,\de F(\th)$ ,

and, thus, for every \(s\in\Th\),
\[
\begin{aligned}
&
\int_s^{\bar\th}
[S(\th)-1]\,\de F(\th)
+
\G_T([s,\bar\th])
-
\G_R([s,\bar\th])
%\\ & \qquad
\; = \; 
\L_s-[1-F(s)].
\end{aligned}
\]
Equation \eqref{eq:KL_LL_verification_rent_terms} therefore becomes
\begin{align}
&
\int_\Th
[S(\th)-1]V(\th)\,\de F(\th)
+
\int_\Th V(\th)\,\G_T(\de\th)
-
\int_\Th V(\th)\,\G_R(\de\th)
\notag\\
={}&
-\left[1-\L_{\ul\th}\right]V(\ul\th)
+
\int_{\ul\th}^{\bar\th}
[-h'(\th)]D_\th
\left[\L_\th-(1-F(\th))\right]\de\th
\notag\\
={}&
-\left[1-\L_{\ul\th}\right]V(\ul\th)
+
\int_\Th
\frac{h'(\th)}{f(\th)}
\left[1-F(\th)-\L_\th\right]
D_\th\,\de F(\th).
\label{eq:KL_LL_verification_rent_identity}
\end{align}
Inserting 
\eqref{eq:KL_LL_verification_rent_identity} for the last line in  \eqref{eq:KL_LL_verification_expanded}, and using $
\G(\de t,\de r)=\G_r(\de t)\,\de F(r)$,  
gives the central dual identity
\begin{equation}
\mathcal L_{\eta,\G}(\mu,V)
=
\int_\Th
\mathcal Q_\th(\mu_\th)\,\de F(\th)
-\left[1-\L_{\ul\th}\right]V(\ul\th).
\label{eq:KL_LL_verification_dual_identity}
\end{equation}
Indeed, the coefficient on \(D_\th\) in $Q_\th$ is 
$
-\widetilde\Phi(\th,\L_\th)$, 
and the disintegration of the remaining term in the second line of 
\eqref{eq:KL_LL_verification_expanded} gives
\[
\begin{aligned}
&
\int_{\Th^2}
h(t)
\log\left[
\sum_{i=1}^n
\nu_i
\left(
\frac{\mu_r(x_i)}{\nu_i}
\right)^{h(r)/h(t)}
\right]
\G(\de t,\de r)
\\
&\qquad
=
\int_\Th
\left\{
\int_\Th
h(t)
\log\left[
\sum_{i=1}^n
\nu_i
\left(
\frac{\mu_r(x_i)}{\nu_i}
\right)^{h(r)/h(t)}
\right]
\G_r(\de t)
\right\}
\de F(r).
\end{aligned}
\]

\paragraph{The dual bound.}
Individual rationality gives \(V(\ul\th)\geq0\). Since
\(\L_{\ul\th}\leq1\), the boundary term in
\eqref{eq:KL_LL_verification_dual_identity} is nonpositive. Hence
\[
\mathcal L_{\eta,\G}(\mu,V)
\leq
\int_\Th
\mathcal Q_\th(\mu_\th)\,\de F(\th).
\]
By the pointwise-maximization condition
\eqref{eq:KL_LL_pointwise_suff},
\[
\mathcal Q_\th(\mu_\th)
\leq
\mathcal Q_\th(\mu_\th^*)
\qquad
\text{for \(F\)-a.e.\ }\th.
\]
Combining these inequalities with
\eqref{eq:KL_LL_verification_first_bound} gives
\begin{equation}
\Pi(\mu,V)
\leq
\int_\Th
\mathcal Q_\th(\mu_\th^*)\,\de F(\th)
\label{eq:KL_LL_verification_upper_bound}
\end{equation}
for every individually rational, incentive-compatible mechanism
satisfying agent-side limited liability.

\paragraph{Attainment by the candidate.}
Apply \eqref{eq:KL_LL_verification_lagrangian} to the candidate. Complementary slackness ensures that 
\[
\int_\Th
\sum_{i=1}^n
\eta_i(\th)w_\th^*(x_i)\,\de F(\th)
=0,
\quad
\text{ and }
\quad
\int_{\Th^2}
G^*(t,r)\,\G(\de t,\de r)
=0.
\]
Consequently,
\begin{equation}
\mathcal L_{\eta,\G}(\mu^*,V^*)
=
\Pi(\mu^*,V^*).
\label{eq:KL_LL_verification_candidate_L}
\end{equation}

On the other hand, the dual identity
\eqref{eq:KL_LL_verification_dual_identity} and
\eqref{eq:KL_LL_CS_IR} imply
\[
\begin{aligned}
\mathcal L_{\eta,\G}(\mu^*,V^*)
&=
-\left[1-\L_{\ul\th}\right]V^*(\ul\th)
+
\int_\Th
\mathcal Q_\th(\mu_\th^*)\,\de F(\th)
\\
&=
\int_\Th
\mathcal Q_\th(\mu_\th^*)\,\de F(\th).
\end{aligned}
\]
Together with \eqref{eq:KL_LL_verification_candidate_L}, this gives
\begin{equation}
\Pi(\mu^*,V^*)
=
\int_\Th
\mathcal Q_\th(\mu_\th^*)\,\de F(\th).
\label{eq:KL_LL_verification_candidate_value}
\end{equation}

Finally, combining
\eqref{eq:KL_LL_verification_upper_bound} and
\eqref{eq:KL_LL_verification_candidate_value} yields
\[
\Pi(\mu,V)
\leq
\Pi(\mu^*,V^*)
\]
for every individually rational and incentive-compatible mechanism
satisfying agent-side limited liability. The candidate mechanism is
therefore optimal.
\qed

\subsubsection{Derivation of numerical illustration in Example~5}
\label{app:KL_LL_013}

We construct the candidate directly and verify the sufficient conditions in Proposition~\ref{prop:KL_LL_verification}. In the candidate, the limited-liability constraints bind at output \(0\) for all types, so the associated multipliers are positive, whereas the multipliers on the upper wages are zero. In particular, $\L_{\ul \th}=1$, so the adjusted virtual cost at the bottom equals the actual cost, $\widetilde\Phi(\ul\th,\L_{\ul\th})=h(\ul\th)$. On the separating region, the GIM multiplier is zero. On the pooled region, all GIM constraints hold with equality, and the highest type's downward constraint has a positive GIM multiplier for each report in the pooled region. 
Write $\Delta w_\theta:=w_\theta(13)-w_\theta(12)$, or sometimes $d$ for short.
For a contract $(w(0),w(12),w(13))=(0,b,b+d)$, let
\begin{equation}\label{eq:KL_LL_013_conditional_profit}
\pi(\theta,b,d)
:=
\frac{(12-b)e^{b/h(\theta)}+(13-b-d)e^{(b+d)/h(\theta)}}
{1+e^{b/h(\theta)}+e^{(b+d)/h(\theta)}}.
\end{equation}
Here $b$ and $d$ are the level and spread of the two positive wages;
subscripts $b$ and $d$ on $\pi$ denote partial derivatives in these
coordinates. The chosen distribution under a contract $w_\theta$ with $w_\theta(0)=0$ is
\begin{equation}\label{eq:KL_LL_013_distribution}
\mu_\theta(x)=\frac{e^{w_\theta(x)/h(\theta)}}{Z_\theta},
\qquad
Z_\theta=1+e^{w_\theta(12)/h(\theta)}+e^{w_\theta(13)/h(\theta)}.
\end{equation}
Define $P_\theta:=\mu_\theta(12)+\mu_\theta(13)$ and
$s_\theta:=\mu_\theta(13)/P_\theta
=[1+e^{-\Delta w_\theta/h(\theta)}]^{-1}$.

\paragraph{The candidate.}
Set $w_\theta(0)=0$ in every contract.
For an initial payment $w_{\ul \th}(12)=z$, consider the system of differential equations 
\begin{align}
(\Delta w_\theta)'
&=\frac{\Delta w_\theta^2}{h(\theta)}
\left[2-\frac1{\Delta w_\theta}
-\frac{1-P_\theta}{h(\theta)}
\left(12-\frac{w_\theta(12)}{\Delta w_\theta}\right)\right],
\label{eq:KL_LL_013_separating_ODE}\\
w_\theta'(12)&=-s_\theta(\Delta w_\theta)',
\label{eq:KL_LL_013_local_IC}\\
w_{\ul \th}(12)&=z,\qquad \Delta w_{\ul \th}=1.
\label{eq:KL_LL_013_lower_boundary}
\end{align}
For a proposed switching type $T \in \Th$, set
$w^P=w_T$, $\Delta w^P:=\Delta w_T$, and hold this contract fixed for all types above $T$.
The two equations determining the boundary values $(z,T)$ are
\begin{align}
\int_T^1\pi_b(t,w^P(12),\Delta w^P)\,\de t
&=h(T)P_T\frac{\Delta w^P-1}{\Delta w^P},
\label{eq:KL_LL_013_FB_level}\\
\int_T^1\bigl[\pi_d(t,w^P(12),\Delta w^P)
-s_T\pi_b(t,w^P(12),\Delta w^P)\bigr]\,\de t
&=0.
\label{eq:KL_LL_013_FB_spread}
\end{align}
The following lemma establishes existence and the bounds used below.
Its proof, including the quantitative bounds, is given
in Supplementary Appendix~\ref{app:KL_LL_013_supplement}.

\begin{lemma}[Existence and bounds]\label{lem:KL_LL_013_existence}
Equations~\eqref{eq:KL_LL_013_separating_ODE}--\eqref{eq:KL_LL_013_FB_spread}
have a solution with
\[
\frac{119}{1000}<w_0(12)<\frac{121}{1000},
\qquad \frac{5}{20}<\theta^P:=T<\frac7{20}.
\]
Every solution in this box satisfies, on its separating interval $\th \in [0,T]$,
\begin{equation}\label{eq:KL_LL_013_separating_bounds}
\frac1{10}<w_\theta(12)<\frac{61}{500},
\qquad 1\leq\Delta w_\theta<\frac{51}{50},
\qquad \frac1{200}<(\Delta w_\theta)'<\frac1{20}.
\end{equation}
At the switching type $T$, the derivative is taken from the left.
In particular, $w_\theta(12)$ decreases strictly,
$w_\theta(13)$ increases strictly, and $\Delta w_\theta>1$ for
$0<\theta\leq\theta^P$.
\end{lemma}

We also use the following curvature property; its proof is in the same
supplementary appendix.
\begin{lemma}[Convexity of the power-log term]
\label{lem:KL_LL_power_log_convexity}
For a full-support distribution $\nu$ on a finite set with at least two
outputs, define
\[
H_q(\mu):=\log\left[\sum_{x\in X}\nu(x)
\left(\frac{\mu(x)}{\nu(x)}\right)^q\right],
\qquad \nu_{\min}:=\min_{x\in X}\nu(x).
\]
Then $H_q$ is convex on $\Delta(X)$ whenever
$1\leq q\leq(1-\nu_{\min})^{-1}$.
\end{lemma}

\paragraph{The multiplier certificate.}
Fix any solution in the bounds of Lemma~\ref{lem:KL_LL_013_existence}.
On the separating interval below $\th^P$,
set $\Gamma_r=0$, $\eta_{12}(r)=\eta_{13}(r)=0$, and
\begin{equation}\label{eq:KL_LL_013_L_separating}
\Lambda_r=1-r+h(r)\left(1-\frac1{\Delta w_r}\right),
\qquad
\eta_0(r)=\frac{\mu_r(0)}{h(r)}
\left(12-\frac{w_r(12)}{\Delta w_r}\right).
\end{equation}
Then $\Lambda_0=1$, $\Lambda_r'=-\eta_0(r)$, and
\[
\widetilde\Phi(r,\Lambda_r)=\frac{h(r)}{\Delta w_r}>0,
\qquad \eta_0(r)>0.
\]
These identities follow from the differential equations and the bounds
in Lemma~\ref{lem:KL_LL_013_existence}.

On the pool above $\th^P$, write $H=h(r)$, $H_P=h(\theta^P)$, $b=w^P(12)$, and
$d=\Delta w^P$. For $3\leq H\leq H_P$, define
\[
\mu_x^P(H):=
\frac{e^{w^P(x)/H}}{Z^P(H)},\qquad
Z^P(H):=1+e^{b/H}+e^{(b+d)/H},
\qquad R_x(H):=\frac{\mu_x^P(3)}{\mu_x^P(H)}.
\]
Let $N$ be the bounded solution on $(3,H_P]$ of
\begin{equation}\label{eq:KL_LL_013_N_ODE}
N'(H)=A(H)+B(H)N(H),
\end{equation}
where
\begin{align}
A(H)&=\frac{(12d-b)\mu_0^P(H)}{H^2}-\frac{2d-1}{H},
\label{eq:KL_LL_013_A}\\
B(H)&=-\frac1H+\frac{d\mu_0^P(H)}{H^2}
\left[\frac bd+\frac{R_0(H)-R_{12}(H)}{R_{13}(H)-R_{12}(H)}\right]
-\frac{d}{H^2[R_{13}(H)-R_{12}(H)]}.
\label{eq:KL_LL_013_B}
\end{align}
Define, with $H=h(r)$,
\begin{align}
\Lambda_r&=2H-3-\frac H d[1-N(H)],
\label{eq:KL_LL_013_L_pool}\\
\gamma(r)&=\frac{N(H)}{H[R_{13}(H)-R_{12}(H)]},
\qquad \Gamma_r(\de t)=\gamma(r)\delta_{\{1\}}(\de t),
\label{eq:KL_LL_013_gamma_pool}\\
\eta_0(r)&=\frac{\mu_0^P(H)}H
\left[12-\frac{b[1-N(H)]}{d}
+N(H)\frac{R_0(H)-R_{12}(H)}{R_{13}(H)-R_{12}(H)}\right],
\qquad \eta_{12}(r)=\eta_{13}(r)=0.
\label{eq:KL_LL_013_eta_pool}
\end{align}
The next lemma justifies these definitions, including the endpoint
$r=1$. The proof in the Supplementary Appendix \ref{app:KL_LL_013_supplement} establishes the boundary identities
before using them to determine the signs.

\begin{lemma}\label{lem:KL_LL_013_pool_certificate}
The bounded solution of \eqref{eq:KL_LL_013_N_ODE} is unique and
extends continuously differentiably to $H=3$, with
\[
N(3)=0,\qquad N'(3)=\frac{A(3)}2,\qquad N(H_P)=0,
\qquad 0<N(H)<1\quad(3<H<H_P).
\]
The multiplier densities extend continuously to the endpoints,
$\eta_0>0$ on $[\theta^P,1]$, $\gamma(\theta^P)=0$, and
$\gamma>0$ on $(\theta^P,1]$. Moreover,
\begin{equation}\label{eq:KL_LL_013_dual_normalization}
\int_{\theta^P}^1\gamma(r)\,\de r
=\Lambda_1=3\left(1-\frac1{\Delta w^P}\right)>0.
\end{equation}
The separating and pooled definitions of $\Lambda$ match at
$\theta^P$ and jointly satisfy \eqref{eq:KL_LL_multiplier_stock}.
\end{lemma}

\begin{proposition}
\label{prop:KL_LL_013_solution}
For $X=\{0,12,13\}$, uniform $\nu$, $\theta\sim U[0,1]$, and
$h(\theta)=4-\theta$, the candidate above maximizes expected profit
under agent-side limited liability. It has $w_\theta(0)=0$ for every
type, separates strictly below $\theta^P$, and offers a common contract
above $\theta^P$. The solution of the defining equations in the box
of Lemma~\ref{lem:KL_LL_013_existence} is unique.
Numerically,
\[
\theta^P\simeq0.2972110,\quad w_0(12)\simeq0.1201433,
\quad w^P(12)\simeq0.1163128,\quad w^P(13)\simeq1.1230975.
\]
\end{proposition}

\begin{proof}
\emph{Incentive compatibility.}
Fix a solution furnished by Lemma~\ref{lem:KL_LL_013_existence}.
A true type $t$ facing report $r$ obtains
\[
U(t,r)=h(t)\log\left[
\frac{1+e^{w_r(12)/h(t)}+e^{w_r(13)/h(t)}}3\right].
\]
For a separating report, let
$s_{t,r}:=[1+e^{-\Delta w_r/h(t)}]^{-1}$.
Equations~\eqref{eq:KL_LL_013_separating_ODE}--\eqref{eq:KL_LL_013_lower_boundary}
give
\begin{equation}\label{eq:KL_LL_013_report_derivative}
U_r(t,r)=[\mu_{t,r}(12)+\mu_{t,r}(13)]
(\Delta w_r)'(s_{t,r}-s_r).
\end{equation}
Since $h$ decreases and $\Delta w_r>0$, this derivative has the sign
of $t-r$. Thus each type below $\theta^P$ uniquely prefers his own
contract, while each type at or above $\theta^P$ strictly prefers the
pooled contract to every separating contract. Reports within the pool
are equivalent. Obedience follows from the exponential response
formula, and differentiating $U(\theta,\theta)$ gives the envelope
condition. Nonnegative wages imply participation by allowing the agent
to choose $\nu$ at zero cost.

\medskip
\emph{Global pointwise maximization.}
Denote the objective in \eqref{eq:KL_LL_pointwise_suff} by
$\mathcal Q_r$. Under the proposed multipliers it is
\begin{equation}\label{eq:KL_LL_013_pool_objective}
\mathcal Q_r(\mu)=m(\mu)
-\widetilde\Phi(r,\Lambda_r)D_{KL}(\mu\Vert\nu)
+h(r)\eta_0(r)\log[3\mu(0)]
-3\gamma(r)H_{h(r)/3}(\mu),
\end{equation}
where $\gamma=0$ below $\theta^P$.
On the pool,
$\widetilde\Phi(r,\Lambda_r)=h(r)[1-N(h(r))]/d>0$.
Since $1\leq h(r)/3\leq4/3<3/2$ and $\nu_{\min}=1/3$,
Lemma~\ref{lem:KL_LL_power_log_convexity} makes the last term
concave. The logarithm is concave, and the negative KL term is strictly
concave. Hence $\mathcal Q_r$ is strictly concave on its effective
domain. At $\mu(0)=0$ its value is $-\infty$; other boundary
distributions are included by continuity.

For an interior candidate, the simplex first-order conditions are
\begin{equation}\label{eq:KL_LL_013_certificate_FOC}
x-\widetilde\Phi(r,\Lambda_r)[\log(3\mu_r(x))+1]
+\frac{h(r)\eta_x(r)}{\mu_r(x)}
-h(r)\gamma(r)\frac{\mu_{1,r}(x)}{\mu_r(x)}=\xi_r.
\end{equation}
On the pool, the difference between the conditions at $13$ and $12$ is
\[
1-\frac{\widetilde\Phi(r,\Lambda_r)}{h(r)}d
-h(r)\gamma(r)[R_{13}(h(r))-R_{12}(h(r))]=0.
\]
This follows from the definitions of $\Lambda$ and $\gamma$.
The $12$ versus $0$ difference is zero by the definition of $\eta_0$.
The same two differences below $\theta^P$ vanish by
\eqref{eq:KL_LL_013_L_separating}. Thus $\mu_r$ is the unique
global maximizer required by \eqref{eq:KL_LL_pointwise_suff}.

\medskip
\emph{Complementary slackness and optimality.}
We have $\Lambda_0=1$. The only positive floor multiplier is attached
to $w_r(0)=0$. Finally, $\Gamma$ is supported on $(1,r)$ with
$r\geq\theta^P$, where $G(1,r)=0$ because both reports offer $w^P$.
Lemma~\ref{lem:KL_LL_013_pool_certificate} establishes integrability
and the multiplier-stock identity. All the conditions of
Proposition~\ref{prop:KL_LL_verification} are therefore satisfied.

Strict concavity also implies that every optimal mechanism has the
same on-path distributions almost everywhere: otherwise the pointwise
dual bound is strict on a set of positive measure. Equality in the
dual bound and $\eta_0>0$ then imply $w_\theta(0)=0$ almost
everywhere, which pins down rents and all remaining wages. In
particular, any two roots in the box of
Lemma~\ref{lem:KL_LL_013_existence} yield the same continuous menu.
Its strictly increasing spread below the switch identifies both the
initial wage and the switching type. This proves uniqueness of that
root.
\end{proof}

\subsection{Proof for Optimal Taxation Example~6}
\subsubsection{Proof of Proposition \ref{prop:MirrleesExample}}
The proof has two steps. Step 1 shows that the principal may restrict attention to mechanisms implementing degenerate gross-income distributions. Given an arbitrary incentive-compatible mechanism, we replace each recommended distribution \(\mu_\theta\) by the degenerate distribution at the income \(\tilde x_\theta\) that carries the same effort cost, and construct an accompanying net-income schedule that delivers each type the same utility as before. Because the effort cost is unchanged type by type, the envelope condition continues to hold with the same rent schedule; obedience on and off path follows from the shape of the constructed schedule; and incentive compatibility follows from the incentive constraints of the original mechanism. The construction leaves the objective unchanged, while concavity of \(u\) and convexity of \(z\) imply that it weakly raises revenue.

Step 2 then shows that a menu of type-contingent schedules can be replaced by the single schedule \(\overline{w}(x)=\sup_\theta w_\theta(x)\). The upper envelope weakly raises every agent's payoff, while the incentive constraints of the original mechanism imply that no type can gain from it at any income. The two bounds together leave utilities and on-path net incomes unchanged, so both the objective and revenue are unaffected, and \(\bar T(x):=x-\overline{w}(x)\) is the desired tax schedule.

Throughout, write \(Z(\mu):=\int_X z(x)\mu(\de x)\), so that \(C(\theta,\mu)=h(\theta)Z(\mu)\). Since the cost is linear in \(\mu\), its  G\^ateaux derivative is \(c_\mu(\theta,x)=h(\theta)z(x)\), independent of \(\mu\), and \(C(\theta,\mu)=\int_X c_\mu(\theta,x)\mu(\de x)\). Hence the additive constant in item 2 of Theorem \ref{thm:Characterization} equals \(V(\theta)\), and on-path obedience reads
\begin{equation}
u\circ w_\theta(x)\;\le\; V(\theta)+h(\theta)z(x)\quad\text{for all }x\in X,
\qquad\text{with equality }\mu_\theta\text{-a.s.}
\label{eq:mirrlees-obedience}
\end{equation}

\paragraph{Step 1: reduction to degenerate distributions.}
Consider any mechanism \((\mu,V,w)\) satisfying \eqref{eq:IC} and \eqref{eq:Min-Rev}. Since \(z\) is continuous and strictly increasing and \(Z(\mu_\theta)\in[z(\ul x),z(\bar x)]\), there is a unique \(\tilde x_\theta\in X\) with
\[
z(\tilde x_\theta)=Z(\mu_\theta),
\qquad\text{that is,}\qquad
Z(\delta_{\tilde x_\theta})=Z(\mu_\theta).
\]
Set \(\tilde V:=V\) and \(\tilde\mu_\theta:=\delta_{\tilde x_\theta}\). Let \(M:=\min_{t\in\Theta}\{V(t)+h(t)z(\ul x)\}\), which is attained since \(V\) and \(h\) are continuous on the compact set \(\Theta\); let \(\theta^\ast\) attain it and put \(\ell:=w_{\theta^\ast}(\ul x)\). By \eqref{eq:mirrlees-obedience} applied to the original mechanism at \(x=\ul x\),
\[
u(\ell)\;\le\; V(\theta^\ast)+h(\theta^\ast)z(\ul x)\;=\;M .
\]
Define
\[
\tilde w_\theta(x):=
\begin{cases}
\ell, & x<\tilde x_\theta,\\[2pt]
u^{-1}\!\left(V(\theta)+h(\theta)z(\tilde x_\theta)\right), & x\ge \tilde x_\theta .
\end{cases}
\]
The upper branch is well defined: \(V(\theta)+h(\theta)z(\tilde x_\theta)=V(\theta)+h(\theta)Z(\mu_\theta)=\int_X u\circ w_\theta(x)\,\mu_\theta(\de x)\) lies in the range of \(u\), being an average of values of \(u\circ w_\theta\).

\emph{Envelope.} Since \(\partial_\theta C(t,\mu)=h'(t)Z(\mu)\) and \(Z(\tilde\mu_t)=Z(\mu_t)\) for every \(t\), the function \(\tilde V=V\) satisfies item 1 of Theorem \ref{thm:Characterization} for the new profile.

\emph{On-path obedience.} As \(z\) is increasing, \(u\circ\tilde w_\theta(x)=V(\theta)+h(\theta)z(\tilde x_\theta)\le V(\theta)+h(\theta)z(x)\) for \(x\ge\tilde x_\theta\), with equality at \(\tilde x_\theta\); and \(u(\ell)\le M\le V(\theta)+h(\theta)z(x)\) for \(x<\tilde x_\theta\). Thus \eqref{eq:mirrlees-obedience} holds for \(\tilde w_\theta\) with equality at \(\tilde x_\theta\), so \(\tilde\mu_\theta\) is obedient and delivers utility exactly \(V(\theta)\).

\emph{Off-path obedience.} Fix \(\theta\) and a report \(\hat\theta\). Since \(\int_X[u\circ\tilde w_{\hat\theta}(x)-h(\theta)z(x)]\mu(\de x)\) is linear in \(\mu\), it suffices to maximize the integrand pointwise. On \([\tilde x_{\hat\theta},\bar x]\) the payment is constant while \(z\) is increasing, so \(\tilde x_{\hat\theta}\) is optimal there; on \([\ul x,\tilde x_{\hat\theta})\) the payment is constant at \(\ell\), so \(\ul x\) is optimal there. Hence a best response exists and
\[
\max_{\mu\in\Delta(X)} U(\theta,\tilde w_{\hat\theta},\mu)
=\max\Big\{\,V(\hat\theta)+\big[h(\hat\theta)-h(\theta)\big]z(\tilde x_{\hat\theta}),\;\;u(\ell)-h(\theta)z(\ul x)\Big\}.
\]

\emph{Incentive compatibility.} Both terms in the display are bounded by \(V(\theta)\). For the first, \eqref{eq:IC} for the original mechanism---applied to the deviation in which type \(\theta\) reports \(\hat\theta\) and chooses \(\mu_{\hat\theta}\)---gives
\[
V(\theta)\;\ge\;\int_X u\circ w_{\hat\theta}(x)\,\mu_{\hat\theta}(\de x)-h(\theta)Z(\mu_{\hat\theta})
\;=\;V(\hat\theta)+\big[h(\hat\theta)-h(\theta)\big]Z(\mu_{\hat\theta}),
\]
and \(Z(\mu_{\hat\theta})=z(\tilde x_{\hat\theta})\). For the second, \(u(\ell)\le M\le V(\theta)+h(\theta)z(\ul x)\). Hence \((\tilde\mu,V,\tilde w)\) satisfies \eqref{eq:IC}.

\emph{Payoffs.} Agents' utilities are unchanged, so the objective \(SW\) is unchanged. For revenue, \eqref{eq:mirrlees-obedience} gives
\[
u\big(\tilde w_\theta(\tilde x_\theta)\big)
= V(\theta)+h(\theta)Z(\mu_\theta)
= \int_X u\circ w_\theta(x)\,\mu_\theta(\de x)
\;\le\; u\!\left(\int_X w_\theta(x)\,\mu_\theta(\de x)\right),
\]
by concavity of \(u\), so \(\tilde w_\theta(\tilde x_\theta)\le\int_X w_\theta(x)\mu_\theta(\de x)\). By convexity of \(z\),
\[
z(\tilde x_\theta)=\int_X z(x)\,\mu_\theta(\de x)\;\ge\; z\!\left(\int_X x\,\mu_\theta(\de x)\right),
\]
so \(\tilde x_\theta\ge\int_X x\,\mu_\theta(\de x)\). Therefore \(\tilde x_\theta-\tilde w_\theta(\tilde x_\theta)\ge\int_X[x-w_\theta(x)]\mu_\theta(\de x)\) for every \(\theta\), and integrating against \(F\) shows that \eqref{eq:Min-Rev} remains satisfied. The alternative mechanism thus leaves the principal's payoff unchanged while relaxing the minimum revenue constraint, so the principal can restrict attention to mechanisms that implement degenerate gross-income distributions.

\emph{GIM.} Finally, \(\tilde x_\theta\) is weakly increasing. Adding the two inequalities from \eqref{eq:IC} obtained by letting \(\theta\) report \(t\) and \(t\) report \(\theta\), each while choosing the other's recommended distribution, yields
\[
\big[h(t)-h(\theta)\big]\big[Z(\mu_t)-Z(\mu_\theta)\big]\le 0 .
\]
For \(\theta>t\) we have \(h(t)>h(\theta)\), hence \(Z(\mu_t)\le Z(\mu_\theta)\) and therefore \(\tilde x_t\le\tilde x_\theta\).

\paragraph{Step 2: implementation by a single tax schedule.}
It remains to show that the optimum can be achieved with a single type-independent net income function \(\overline{w}:\mathbb{R}\rightarrow\mathbb{R}\) (or equivalently, a single tax schedule \(\bar{T}:\mathbb{R}\rightarrow\mathbb{R}\)). By Step 1, we may consider any mechanism \((\mu,V,w)\) that satisfies \eqref{eq:IC} and \eqref{eq:Min-Rev} and for which \(\mu_{\theta}=\delta_{\{x_{\theta}\}}\) for all \(\theta\), with \(x_{\theta}\) weakly increasing in \(\theta\). Let
\[
\overline{w}(x):=\sup_{\theta\in\Theta}\ w_{\theta}(x).
\]
Each \(w_\theta\) is nondecreasing, so \(\overline{w}\) is nondecreasing and hence Borel; it is also bounded above and satisfies any payment bounds satisfied by every \(w_\theta\), so \(\overline{w}\in\mathcal{W}\). We show that \((\mu,V,\overline{w})\) satisfies \eqref{eq:IC} and \eqref{eq:Min-Rev} and generates the same utilities for all agents, the same tax revenue, and the same payoff for the principal.

First, \(\overline{w}(x_{\theta})=w_{\theta}(x_{\theta})\) for every \(\theta\). If instead \(\overline{w}(x_\theta)>w_\theta(x_\theta)\), there would exist \(\tilde\theta\) with \(w_{\tilde{\theta}}(x_{\theta})>w_{\theta}(x_{\theta})\), and in the original mechanism type \(\theta\) would be better off reporting \(\tilde\theta\) while still choosing \(\delta_{\{x_{\theta}\}}\), contradicting \eqref{eq:IC}. Since \(\overline{w}\ge w_\theta\) pointwise, replacing \(w\) by \(\overline{w}\) weakly increases every agent's payoff: for all \(\theta\),
\begin{align*}
 \underset{\mu\in\Delta(X)}{\max} \ \left\{\int_{X} u(\overline{w}(x))\mu(\de x)-h(\theta)\int_{X}z(x)\mu(\de x)\right\}\geq V(\theta).
\end{align*}

Conversely, \((\mu,V,w)\) satisfying \eqref{eq:IC} implies \(V(\theta)\geq u(w_{\tilde{\theta}}(x))-h(\theta)z(x)\) for all \(x\) and \(\tilde\theta\). Hence
\begin{align*}
   & V(\theta) \geq \underset{\tilde{\theta}}{\sup} \ \left\{u(w_{\tilde{\theta}}(x))-h(\theta)z(x)\right\} =u(\overline{w}(x))-h(\theta)z(x) \ \text{ for all } x,
\end{align*}
where the equality uses continuity and monotonicity of \(u\). Taking expectations with respect to any \(\mu \in \Delta(X)\),
\begin{align*}
    V(\theta) \geq \int_{X}u(\overline{w}(x))\mu(\de x)-h(\theta)\int_{X}z(x)\mu(\de x) \quad \text{ for all } \mu \in \Delta(X).
\end{align*}
Combining the two displays, each type attains exactly \(V(\theta)\) under \(\overline{w}\), and since \(\overline{w}(x_{\theta})=w_{\theta}(x_{\theta})\), the choice \(\delta_{\{x_{\theta}\}}\) remains optimal. Thus \((\mu,V,\overline{w})\) satisfies \eqref{eq:IC} and delivers the same utilities as the original mechanism, so the principal's objective is unchanged. On-path net incomes are also unchanged, so tax revenue is unchanged and \eqref{eq:Min-Rev} continues to hold. Setting \(\bar{T}(x):=x-\overline{w}(x)\) yields the desired single tax schedule. \qed

% \newpage
\bibliographystyle{ecta}
\bibliography{References.bib}

\newpage 
\section*{ONLINE APPENDIX} 

\section{Formal Results and Proofs for the Separate Problems}\label{sec:proofsSP}

\subsection{Pure Moral Hazard}\label{sec:proofsMH}

\begin{proposition}\label{propIRMH}
    A mechanism $\{w_\th,\mu_\th\}_{\th\in\Th}$ satisfies obedience and participation if and only if for each $\th\in\Th$ there exists $m_{\th}\in \mathbb{R}$ such that the following two conditions hold:
\begin{enumerate}
\item \textbf{Obedience:}
\begin{align*}
u\circ w_{\theta}(x) \begin{cases}
       \le c_{\mu_{\theta}}(\th, x)+m_{\th}
    \quad \text{for all } x \in X,
    \\
 =  c_{\mu_{\theta}}(\th, x)+m_{\th}   \quad \mu_{\theta}\text{-almost all } x\in X;
\end{cases}
\end{align*}
    
    \item \textbf{Participation:} 
       \begin{align*}
 m_{\th} \geq C(\th,\mu_{\th})-\int c_{\mu_{\th}}(\th,x)\mu_{\th}(\de x). 
\end{align*}
\end{enumerate}
\end{proposition}
\begin{proof}
Item 1 follows directly from Proposition 2 in \citetalias{georgiadis2024flexible}. For item 2, substituting the expression for $w_{\th}$ from item 1 into (IR) yields
\[U(\th, w_{\th},\mu_{\th})= \int u\circ w_{\th}(x)\mu_{\th}(\de x)-C(\th,\mu_{\th}) = m_{\th}+\int c_{\mu_{\th}}(\th,x)\mu_{\th}(\de x) -C(\th,\mu_{\th})\geq 0.\]
\end{proof}
\begin{corollary}
    A distribution $\mu_{\th} \in \Delta(X)$ is implementable for type $\theta$ by some wage function $w_{\th}$ if and only if
    \begin{align*}
    \max \bigg\{C(\th,\mu_{\th})-\int c_{\mu_{\th}}(\th,x)\mu_{\th}(\de x), \ & \sup\left\{[u(\ul w) - c_{\mu_{\th}}(\th, x)]:x\in X\right\}\bigg\}  \\ & \leq  \inf\bigg\{[u(\overline{w} +x)-c_{\mu_{\th}}(\th,x)] :x \in \supp(\mu_{\th})\bigg\}.
    \end{align*}
\end{corollary}
\begin{proof}
By item 1 in Proposition \ref{propIRMH},we have pinned down all candidate wage functions $w_{\th}$, up to a constant $m_{\th}$. It remains to check if there exists $m_{\th}$ for which limited liability is satisfied, or equivalently, if $u(\ul w)\leq u\circ w_{\th}(x)\leq u(x+\overline{w})$ for all $x\in X$. The limited liability on the agent's side requires that
\begin{align*}
    u(\ul w) \leq c_{\mu_{\th}}(\th,x)+m_{\th} \quad \text{ for all } \ x \in X. 
\end{align*}
Combining this inequality with Item 2 of Proposition \ref{propIRMH} gives the lower bound for $m_{\th}$.
The limited liability on the principal's side requires that
\begin{align*}
    u(w_{\th}(x)) = c_{\mu_{\th}}(\th,x)+m_{\th} \leq u(\overline{w}+x) \quad \text{ for all } \ x \in supp( \mu_{\th}). 
\end{align*}
Hence, $\inf\{[u(\overline{w} +x)-c_{\mu_{\th}}(\th,x)] : x \in \supp(\mu_{\th})\}$ is an upper bound for $m_{\th}$.

If the condition in the Corollary does not hold, there exists no $m_{\th}$ and, hence, no $w_{\th}\in \mathcal{W}$ that implements $\mu_{\th}$. Otherwise, if the condition holds, any $w_{\th}$ with associated $m_{\th}$ between the lower and upper bounds implements $\mu_{\th}$ for type $\th$.
\end{proof}

\subsection{Pure Adverse Selection}\label{sec:proofsAS}
\begin{proposition}\label{pureAS}
A mechanism $\{w_\th,\mu_\th\}_{\th\in\Th}$ satisfies truthtelling if and only if there exists an absolutely continuous function $V^{AS}$ such that $V^{AS}(\th) = U(\th, w_\th, \mu_\th)$ and the following conditions are jointly satisfied for all $\th\in\Th$:
\begin{enumerate}
    \item \textbf{Envelope:} 
         \vspace*{-.2em}
    $$V^{AS}(\theta) = V^{AS}(\underline{\theta})-\int_{\underline{\theta}}^{\theta} \frac{\partial}{\partial \th} C(t,\mu_{t})\de t;$$
    
    \item \textbf{Monotonicity:} 
         \vspace*{-.2em}
    $$\int_{\hat{\theta}}^{\theta}\Big[- \frac{\partial C(t,\mu_{t})}{\partial \th }+ \frac{\partial C(t,\mu_{\hat{\theta}})}{\partial \th }\Big]\de t \geq 0 \quad \text{for all } \ \hat{\theta} \in \Th.$$
\end{enumerate}
\end{proposition}
\begin{proof}
Note that $U(\cdot , w_{\hat{\theta}},\mu_{\hat{\th}} )$ is absolutely continuous for all  $ (w_{\hat{\theta}},\mu_{\hat{\th}})$  and $\partial U(\th, w_{\hat{\theta}},\mu_{\hat{\th}})/ \partial \th  = - \partial C(\th, \mu_{\hat{\th}} )/\partial \th$, which by Assumption \ref{Ass:Diff} is bounded by an integrable function $b$.
Hence, by Theorem 2 of \cite{milgrom2002envelope}, any mechanism that induces truthtelling must satisfy
$$V^{AS}(\theta) = V^{AS}(\underline{\theta})-\int_{\underline{\theta}}^{\theta} \frac{\partial}{\partial \th} C(t, \mu_{t})\de t.$$

We have shown that \cref{it:Envelope} is necessary for truthtelling. Now, we show that given that \cref{it:Envelope} holds, item 2 is necessary and sufficient for truthtelling. 
Define $V^{AS}(\theta, \hat \th) :=U(\th, w_{\hat\th}, \mu_{\hat{\th}})$ and $g(\theta,\hat{\theta}) := V^{AS}(\theta)-V^{AS}(\th ,\hat{\theta})$ as $\th$'s loss from misreporting $\hat \th$. 
 $\{w_\th,\mu_\th\}_{\th\in\Th}$ satisfies truthtelling if and only if $g(\theta,\hat \th)\geq 0$ for all $\theta,\hat{\theta}$. Note that  $g(\hat{\theta},\hat{\theta}) = V^{AS}(\hat{\theta})-V^{AS}(\hat{\theta},\hat{\theta})=0$. Hence, the mechanism satisfies truthtelling if and only if
\begin{equation*}
    g(\th, \hat{\theta}) =  g(\th, \hat{\theta})-g(\hat{\theta},\hat{\theta}) 
    = \int_{\hat{\theta}}^{\theta}\frac{\partial g(t,\hat{\theta})}{\partial\theta}\de t 
    % = \int_{\hat{\theta}}^{\theta}\big[V^{\prime}(t)-\frac{\partial V(t,\hat \th)}{\partial \theta}\big]\de t 
    =\int_{\hat{\theta}}^{\theta}\Big[-\frac{\partial C(t, \mu_{t})}{\partial \th}+\frac{\partial C(t, \mu_{\hat \th })}{\partial \th }\Big]\de t \geq 0.
\end{equation*}
\end{proof}

\section{The non-necessity of Corollary \ref{cor:GenMon-MultSeparableCost}'s monotonicity condition}\label{app:nonnencessity}
We illustrate by example that the monotonicity condition in Corollary~\ref{cor:GenMon-MultSeparableCost}, while sufficient for generalized integral monotonicity, is not necessary. In fact, even the implied single-crossing condition at the beginning of the previous proof is not necessary. Specifically, we construct an environment and present an incentive-compatible mechanism in which the single-crossing condition is violated.

\paragraph{Example 7.} Let $u(w)=w$, and $C(\th,\mu) = (1-\th)\int x^{2}\mu(\de x)$. Consider $0<\th_{1}<\th_{2}<1$ and $0= \underline{x}<x_{1}<x_{2}<\overline{x}$. Let
\begin{align*}
    w_{1}(x) = \begin{cases}
0 \ & \text{ if } \ x < x_{1}, \\
(1-\th_{1})x_{1}^{2} \ & \text{ if } \ x \geq x_{1};
    \end{cases}
    \qquad 
    w_{2}(x) = \begin{cases}
0 \ & \text{ if } \ x < x_{2}, \\
(1-\th_{1})x_{1}^{2}+(1-\th_{2})[x_{2}^{2}-x_{1}^{2}] \ &\text{ if } \ x \geq x_{2}.
    \end{cases}
\end{align*}
The contract $w_{2}$ offers a larger bonus but requires a higher output for it. Let
\begin{align*}
    w_{\theta}(x) = \begin{cases}
        w_{1}(x) \ & \text{if } \ \th < \th_{2},\\
        w_{2}(x) \ & \text{if } \ \th \geq \th_{2}.
    \end{cases}
\end{align*}
The contract menu above is constructed such that types below $\th_{2}$ prefer contract $w_{1}$ while types above $\th_{2}$ prefer $w_{2}$. Moreover, on path, types strictly below $\th_{1}$ choose the cheapest distribution, $\delta_{\{\underline{x}\}}$, types above $\th_{1}$ but strictly below $\th_{2}$ choose $\delta_{\{x_{1}\}}$, while types above $\th_{2}$ choose $\delta_{\{x_{2}\}}$. The off-path recommendations are also given according to what would be optimal for each type when facing contract $w_{1}$ or $w_{2}$. The extended effort recommendation schedule is then given by
\begin{align*}
\mu^{*}_{\th,\hat{\th}} = \begin{cases}
    \delta_{\{\underline{x}\}} \ & \text{ if } \ \big[\th<\th_{1}\big] \text{ or } \big[\hat{\theta}\geq \theta_{2} \text{ and } \th < \tilde{\th}\big],\\
    \delta_{\{x_{1}\}} \ & \text{ if } \big[\th\geq \th_{1} \text{ and } \hat{\th}<\th_{2}\big], \\
    \delta_{\{x_{2}\}} \ & \text{ if } \big[\th\geq \tilde{\th} \text{ and } \hat{\th} \geq \th_{2}\big], 
\end{cases}
\end{align*}
where $\tilde{\th} := \th_{1}(x_{1}/x_{2})^{2}+\th_{2}[1-(x_{1}/x_{2})^{2}] < \th_2$.

Note that types strictly below $\tilde{\theta}$ would choose $\delta_{\{\underline{x}\}}$ if they report $\hat{\th}\geq \th_{2}$. That is, it is so costly for them to achieve output $x_{2}$ that they are better off choosing the cheapest distribution and forgoing  the bonus. As a result,
\begin{align*}
    \kappa(\mu^{*}_{\th,\th_{2}}) = 0 < x_{1}^{2} = \kappa(\mu^{*}_{\th,\th_{1}}) \quad \forall \th \in [\th_{1},\tilde{\th}), 
\end{align*}
which is a violation of the single-crossing condition in Corollary \ref{cor:GenMon-MultSeparableCost}. However, the mechanism is by construction incentive-compatible, since each type was assigned their preferred contract and effort recommendations maximized each type's payoff given the contract in place. One can equivalently check directly the generalized integral monotonicity condition using $\mu^{*}_{\th,\hat{\th}}$, where
\begin{align*}
\int_{\hat\th}^{\th} [\kappa(\mu^{*}_{t})-\kappa(\mu^{*}_{t,\hat{\th}})]dt = 
\begin{cases}
    [x_{2}^{2} - x_{1}^{2}][\max\{\th, \th_{2}\} - \th_{2}] 
    & \text{if } \min\{\th, \th_{2}\} > \hat{\th}, \\[8pt]
    [x_{2}^{2} - x_{1}^{2}][\th_{2} - \th] 
    & \text{if } \hat{\th} \geq \th_{2} > \th \geq \tilde{\th}, \\[8pt]
    x_{1}^{2}[\max\{\th, \th_{1}\} - \th_{1}] 
    & \text{if } \hat{\th} \geq \th_{2} > \tilde{\th} > \th, \\[8pt]
    0 
    & \text{otherwise};
\end{cases}
\end{align*}
which is always weakly positive.

\section{Supplementary proofs for Applications}

\subsection{Supplementary proofs for Example~3}
\label{app:log-kl-verification}

The first-order conditions in Appendix~\ref{app:log-kl-relaxed-proof} are necessary. The following lemma gives a sufficient condition under which a solution of these conditions solves the relaxed problem \eqref{eq:log-kl-relaxed-short}. Write \(\Delta_x:=x_n-x_1\), \(\mu_i:=\mu(x_i)\), and \(D(\mu):=D_{\mathrm{KL}}(\mu\Vert\nu)\).

\begin{lemma}\label{lem:log-kl-verification}
Suppose \(f\) is strictly positive. Let \((\mu^*,V^*)\) be feasible in \eqref{eq:log-kl-relaxed-short}, let \(L\) be given by \eqref{eq:log-kl-shadow-cost} along \((\mu^*,V^*)\), and suppose that, for almost every \(\theta\), \(\mu^*_\theta\) has full support and satisfies the first-order condition in Appendix~\ref{app:log-kl-relaxed-proof}. If
\begin{equation}
\Delta_x^2<4h(\theta)\bigl[1+h(\theta)-\Delta_x\bigr]
\qquad\text{for every }\theta\in\Theta,
\label{eq:log-kl-verification-condition}
\end{equation}
then \((\mu^*,V^*)\) solves \eqref{eq:log-kl-relaxed-short}.
\end{lemma}

\begin{proof}
\emph{Step 1: Reduction to a reportwise problem.}
Let \(\bar w(\theta):=e^{V^*(\theta)}\mathcal P_{h(\theta)}(\mu^*_\theta)\) be the expected wage of type \(\theta\) under \((\mu^*,V^*)\), so that \(L(\theta)=\int_\theta^{\overline\theta}f(t)\bar w(t)\,dt\), and define
\[
\mathcal J_\theta(\mu):=\bar w(\theta)\log\mathcal P_{h(\theta)}(\mu)+\rho(\theta)h(\theta)D(\mu)-\sum_ix_i\mu_i .
\]
Let \(\Pi\) denote the objective in \eqref{eq:log-kl-relaxed-short} and let \((\mu,V)\) be feasible. Convexity of the exponential gives
\[
e^{V}\mathcal P_{h}(\mu_\theta)-e^{V^*}\mathcal P_{h}(\mu^*_\theta)
\geq
\bar w\bigl[V-V^*+\log\mathcal P_{h}(\mu_\theta)-\log\mathcal P_{h}(\mu^*_\theta)\bigr],
\]
with all terms evaluated at \(\theta\). Since \(V(\theta)-V^*(\theta)=\int_{\underline\theta}^{\theta}[-h'(t)][D(\mu_t)-D(\mu^*_t)]\,dt\), Fubini's theorem gives \(\int f\bar w(V-V^*)\,d\theta=\int L[-h'][D(\mu_\theta)-D(\mu^*_\theta)]\,d\theta\). Using \(L[-h']/f=\rho h\), we obtain
\[
\Pi(\mu,V)-\Pi(\mu^*,V^*)
\leq
-\int_{\underline\theta}^{\overline\theta}f(\theta)\bigl[\mathcal J_\theta(\mu_\theta)-\mathcal J_\theta(\mu^*_\theta)\bigr]d\theta .
\]
It therefore suffices to show that \(\mu^*_\theta\) minimizes \(\mathcal J_\theta\) on \(\Delta(X)\) for almost every \(\theta\).

\emph{Step 2: Minimizing \(\mathcal J_\theta\).}
Fix \(\theta<\overline\theta\) at which \(\mu^*_\theta\) has full support and satisfies the first-order condition, and drop \(\theta\) from the notation. Note that \(\rho h>0\), because \(L>0\) below \(\overline\theta\), and \(\bar w\geq1\), because \(V^*\geq0\) by the law of motion and \(\mathcal P(\mu)\geq\bigl(\sum_i\nu_i\,\mu_i/\nu_i\bigr)^{1+h}=1\) by Jensen's inequality. Since \(\log\mathcal P=\min_{s>0}\{s\mathcal P-1-\log s\}\),
\[
\min_{\Delta(X)}\mathcal J=\inf_{s>0}\phi(s),
\qquad
\phi(s):=\min_{\Delta(X)}\mathcal G_s-\bar w(1+\log s),
\]
where \(\mathcal G_s(\mu):=s\bar w\,\mathcal P(\mu)+\rho hD(\mu)-\sum_ix_i\mu_i\).
Each \(\mathcal G_s\) is strictly convex, and since \(\rho h>0\) its minimizer \(\mu(s)\) has full support and is characterized by
\begin{equation}
y_i+\rho h\log z_i=x_i+c,
\qquad
z_i:=\mu_i(s)/\nu_i,
\quad
y_i:=(1+h)s\bar w z_i^h,
\label{eq:log-kl-verification-foc}
\end{equation}
for some \(c=c(s)\). By the implicit function theorem \(\mu(s)\) and \(c(s)\) are \(C^1\), and by the envelope theorem \(\phi'(s)=\bar w[\mathcal P(\mu(s))-1/s]\).

Let \(s^*:=1/\mathcal P(\mu^*)\). Because \(s^*\bar w=e^{V^*}\), the first-order condition in Appendix~\ref{app:log-kl-relaxed-proof} is \eqref{eq:log-kl-verification-foc} at \(s^*\), so \(\mu^*=\mu(s^*)\), \(\phi'(s^*)=0\), and \(\phi(s^*)=\mathcal J(\mu^*)\). Conversely, if \(\hat\mu\) minimizes the continuous function \(\mathcal J\) on the compact set \(\Delta(X)\), then \(\hat s:=1/\mathcal P(\hat\mu)\) satisfies \(\phi(\hat s)\leq\mathcal J(\hat\mu)\), so \(\hat s\) minimizes \(\phi\) and \(\phi'(\hat s)=0\). It therefore suffices to show that \(\phi''>0\) at every zero of \(\phi'\): then \(\phi'\) has at most one zero, so \(\hat s=s^*\) and \(\min\mathcal J=\phi(s^*)=\mathcal J(\mu^*)\).

Differentiating \eqref{eq:log-kl-verification-foc} in \(s\) gives \(\mu_i'(s)=\mu_i(m-y_i)/[s(hy_i+\rho h)]\), where \(m:=sc'(s)\) is determined by \(\sum_i\mu_i'(s)=0\). Since \(\tfrac{d}{ds}\mathcal P(\mu(s))=\sum_iy_i\mu_i'(s)/(s\bar w)\),
\[
s^2\phi''(s)
=
\bar w+s\sum_i(y_i-m)\mu_i'(s)
=
\bar w-\sum_i\frac{\mu_i(y_i-m)^2}{hy_i+\rho h}.
\]
The constant \(m\) minimizes \(t\mapsto\sum_i\mu_i(y_i-t)^2/(hy_i+\rho h)\), so the sum is at most \(\sum_i\mu_i(y_i-\bar y)^2/(hy_i+\rho h)\) with \(\bar y:=\sum_i\mu_iy_i\). At a zero of \(\phi'\), \(\mathcal P(\mu(s))=1/s\), so \(\bar y=(1+h)\bar w\). By \eqref{eq:log-kl-verification-foc}, \(y_i\) and \(\log z_i\) both increase with \(z_i\), so \(x_j>x_i\) implies \(0\leq y_j-y_i\leq x_j-x_i\). Hence every \(y_i\geq(1+h)\bar w-\Delta_x\), and Popoviciu's inequality gives \(\sum_i\mu_i(y_i-\bar y)^2\leq\Delta_x^2/4\). Therefore, at a zero of \(\phi'\),
\[
s^2\phi''(s)
\geq
\bar w-\frac{\Delta_x^2/4}{h\bigl((1+h)\bar w-\Delta_x\bigr)+\rho h}>0,
\]
because \(h((1+h)\bar w-\Delta_x)+\rho h-\Delta_x^2/(4\bar w)\) is increasing in \(\bar w\) and \(\rho h\), and hence, since \(\bar w\geq1\) and \(\rho h>0\), exceeds \([4h(1+h-\Delta_x)-\Delta_x^2]/4>0\) by \eqref{eq:log-kl-verification-condition}; in particular, the denominator is positive.
\end{proof}

In the illustration of Example~3, \(\Delta_x=1\), so \eqref{eq:log-kl-verification-condition} reduces to \(h(\theta)>1/2\), which holds because \(h(\theta)=e^{-\theta/2}\geq e^{-1/2}>1/2\). Hence the solution of the boundary-value problem plotted in Figures~\ref{fig:log-kl-short-wages} and~\ref{fig:log-kl-short-distributions} solves the relaxed problem, and Proposition~\ref{prop:log-kl-exact-short} implies that it solves the original problem.

\subsection{Supplementary proofs for Example~5}
\label{app:KL_LL_013_supplement}

\begin{proof}[Proof of Lemma~\ref{lem:KL_LL_013_existence}]
For this proof, write $b_z(t):=w_t(12)$, $d_z(t):=\Delta w_t$, and
$h(t)=4-t$. For a fixed contract $(0,b,b+d)$, define
\[
q(t,b,d):=\frac1{1+e^{b/h(t)}+e^{(b+d)/h(t)}},\qquad
P:=1-q,\qquad s(t,d):=\frac1{1+e^{-d/h(t)}},
\]
and let
\[
g(t,b,d):=\frac{d^2}{h(t)}
\left[2-\frac1d-\frac{q(t,b,d)}{h(t)}
\left(12-\frac bd\right)\right].
\]
The separating equations are $(b_z',d_z')=(-sg,g)$, with
$(b_z(0),d_z(0))=(z,1)$.
For a proposed switching type $T$, let $b_T=b_z(T)$ and $d_T=d_z(T)$,
and evaluate $P_T,s_T$ at $(T,b_T,d_T)$. Define the residuals
\begin{align}
\mathcal F_1(z,T)
&:=\int_T^1\pi_b(t,b_T,d_T)\,\de t
-h(T)P_T(1-1/d_T),\label{eq:KL_LL_013_residual_level}\\
\mathcal F_2(z,T)
&:=\int_T^1[\pi_d(t,b_T,d_T)-s_T\pi_b(t,b_T,d_T)]\,\de t.
\label{eq:KL_LL_013_residual_spread}
\end{align}
We establish a zero of these residuals in the stated rectangle.

\medskip
\emph{Bounds on the separating path.}
Consider the domain
\begin{equation}\label{eq:KL_LL_013_common_domain}
0\leq t\leq\frac7{20},\qquad
\frac1{10}\leq b\leq\frac{61}{500},\qquad
1\leq d\leq\frac{51}{50}.
\end{equation}
On this domain,
\[
\frac{73}{20}\leq h\leq4,\qquad
\frac{29}{100}<q<\frac3{10},\qquad \frac12<s<\frac35.
\]
For example, the upper bound on $q$ follows by applying
$e^u\geq1+u+u^2/2$ to $u=1/40$ and $u=11/40$.
For the lower bound, $e^{61/1825}<25/24$ and
$e^{571/1825}<11/8$ imply $q>12/41>29/100$.
These inequalities, and the bound on $s$, follow from
\begin{equation}\label{eq:KL_LL_013_exp_bound}
e^u\leq1+u+\frac{u^2}{2(1-u/3)},\qquad 0\leq u<3,
\end{equation}
which uses $j!\geq2\cdot3^{j-2}$ for $j\geq2$.
Substituting the preceding bounds into $g$ gives
\begin{align*}
g&>\frac14\left[1-\frac{3/10}{73/20}
\left(12-\frac{1/10}{51/50}\right)\right]>\frac1{200},\\
g&<\frac{(51/50)^2}{73/20}
\left[2-\frac{50}{51}-\frac{29/100}{4}
\left(12-\frac{61}{500}\right)\right]<\frac1{20}.
\end{align*}
For $119/1000\leq z\leq121/1000$, a first-exit argument keeps the
solution in \eqref{eq:KL_LL_013_common_domain} through $7/20$:
$d_z$ increases by less than $7/400$, and $b_z$ decreases by less
than $21/2000$. The other two faces cannot be reached because
$d_z'>0$ and $b_z'<0$. Local existence therefore extends to the
whole interval and yields \eqref{eq:KL_LL_013_separating_bounds}.

\medskip
\emph{Dependence on the initial payment.}
Differentiation gives
\begin{align*}
g_b&=\frac{d^2q}{h^2}
\left[\frac{P(12-b/d)}h+\frac1d\right],\\
g_d&=\frac{2g}{d}+\frac1h+
\frac{d^2Pqs(12-b/d)}{h^3}-\frac{qb}{h^2}.
\end{align*}
The established bounds imply $0<g_b<9/100$ and $|g_d|<42/100$;
also $s_d\leq1/(4h)$. Hence the derivative of $(-sg,g)$ has
maximum row-sum norm less than $3/5$ on
\eqref{eq:KL_LL_013_common_domain}.
Let $u=\partial_z b_z$ and $v=\partial_z d_z$.
Gronwall's inequality gives
$\max\{|u|,|v|\}<e^{21/100}<5/4$.
The second variational equation then implies
\[
|v|\leq\frac9{100}\frac54\frac7{20}e^{147/1000}<\frac1{20}.
\]
The first implies
\[
|u-1|\leq\frac7{20}
\left[\frac{54}{1000}\frac54+
\frac{256}{1000}\frac1{20}\right]<\frac3{100}.
\]
In particular, $u>0$. Since $v'=g_bu+g_dv$ and $v(0)=0$, we
also have $v\geq0$. Thus
\begin{equation}\label{eq:KL_LL_013_initial_sensitivity}
\frac{97}{100}<\partial_z b_z<\frac{103}{100},\qquad
0\leq\partial_z d_z<\frac1{20}.
\end{equation}

\medskip
\emph{Derivatives of the two residuals.}
For this step fix $T\in[1/4,7/20]$ and a contract with
$1/10\leq b\leq61/500$ and $1\leq d\leq51/50$.
For $t\in[T,1]$, put $a:=12-b+s(1-d)$, so that $\pi=Pa$.
Then
\[
3\leq h\leq\frac{15}{4},\quad
\frac7{25}<q<\frac13,\quad \frac23<P<\frac{18}{25},\quad
\frac12<s<\frac35,\quad \frac{59}{5}<a<12.
\]
The bound on $q$ follows from \eqref{eq:KL_LL_013_exp_bound}, using
$e^{61/1500}<21/20$ and $e^{571/1500}<3/2$.
Direct differentiation yields
\begin{align}
\pi_b&=P(qa/h-1),&
\pi_d&=s\pi_b+\frac{Ps(1-s)(1-d)}h,
\label{eq:KL_LL_013_profit_derivatives}\\
\pi_{bb}&=\frac{Pq}{h}\left[(2q-1)\frac ah-2\right],&
\pi_{bd}&=s\pi_{bb}+E,\qquad
E:=\frac{Pq s_d(1-d)}h.
\label{eq:KL_LL_013_profit_second_derivatives}
\end{align}
Consequently,
\[
|\pi_b|<\frac14,\qquad
-\frac13<\pi_{bb}<-\frac3{20},\qquad
|E|<\frac1{7200},\qquad |\pi_{bd}|<\frac{21}{100}.
\]
Holding the contract fixed gives
\[
0\leq s(t)-s(T)
\leq\frac d4\left[\frac1{h(t)}-\frac1{h(T)}\right]
\leq\frac{51}{200}\left(\frac13-\frac4{15}\right)
<\frac9{500}.
\]
Moreover, differentiating $s_d=s(1-s)/h$ gives
\[
|\partial_hs_d|\leq\frac1{4h^2}
+\frac{d|1-2s|}{4h^3},
\qquad |s_d(t)-s_d(T)|<\frac1{40}.
\]
For $D:=Ps(1-s)(1-d)/h$,
\[
D_d=\frac{Ps(1-s)}h
\left[\frac{(qs+1-2s)(1-d)}h-1\right],
\qquad |D_d|<\frac{61}{1000}.
\]

Temporarily regard $\mathcal F_j$ as functions of the endpoint
contract $(b,d)$, with $T$ fixed. The preceding identities imply
\begin{align*}
|(\mathcal F_2)_b|
&\leq\frac34\left(\frac9{500}\frac13+\frac1{7200}\right)
<\frac1{200},\\
|(\mathcal F_2)_d|
&\leq\frac34\left(\frac9{500}\frac{21}{100}
+\frac1{40}\frac14+\frac{61}{1000}\right)<\frac3{50}.
\end{align*}
Combined with \eqref{eq:KL_LL_013_initial_sensitivity}, this gives
$|\partial_z\mathcal F_2|<1/100$.
Write $C=h(T)P_T(1-1/d)$. Then
\[
C_b=P_Tq_T(1-1/d)\geq0,\qquad
C_d=\frac{h(T)P_T}{d^2}+P_Tq_Ts_T(1-1/d)>0.
\]
Since $\pi_{bd}<0$,
\[
(\mathcal F_1)_b<-\frac3{20}(1-T)\leq-\frac{39}{400},
\qquad (\mathcal F_1)_d<0,
\]
and hence $\partial_z\mathcal F_1<-9/100$.
The same bounds give $|(\mathcal F_1)_b|<255/1000$ and
$|(\mathcal F_1)_d|<287/100$. In particular,
\begin{equation}\label{eq:KL_LL_013_contract_sensitivity}
|(\mathcal F_1)_b|+|(\mathcal F_1)_d|<4,\qquad
|(\mathcal F_2)_b|+|(\mathcal F_2)_d|<\frac1{10}.
\end{equation}
Finally, along the separating equations,
$C'=-\pi_b(T,b_T,d_T)$. Differentiating
\eqref{eq:KL_LL_013_residual_level} therefore gives
\[
\partial_T\mathcal F_1
=d_z'(T)\int_T^1
[\pi_{bd}(t,b_T,d_T)-s_T\pi_{bb}(t,b_T,d_T)]\,\de t.
\]
Collecting the bounds, we obtain
\begin{equation}\label{eq:KL_LL_013_residual_sensitivity}
\partial_z\mathcal F_1<-\frac9{100},\qquad
|\partial_T\mathcal F_1|<\frac1{4000},\qquad
|\partial_z\mathcal F_2|<\frac1{100}.
\end{equation}

\medskip
\emph{Existence of a switching type.}
The Taylor calculations below establish
\eqref{eq:KL_LL_013_three_bounds}. Together with
\eqref{eq:KL_LL_013_residual_sensitivity}, the first of those
inequalities gives
\[
|\mathcal F_1(3/25,T)|<\frac1{16000},
\qquad \frac14\leq T\leq\frac7{20}.
\]
Moving $z$ by $1/1000$ changes $\mathcal F_1$ by more than
$9/100000$. Therefore
\[
\mathcal F_1(119/1000,T)>0,
\qquad \mathcal F_1(121/1000,T)<0.
\]
Strict monotonicity and the implicit function theorem yield a unique
continuous $z(T)\in(119/1000,121/1000)$ with
$\mathcal F_1(z(T),T)=0$.
Replacing $3/25$ by $z(T)$ changes $\mathcal F_2$ by less than
$1/100000$. Its two boundary signs in
\eqref{eq:KL_LL_013_three_bounds} consequently imply
\[
\mathcal F_2(z(1/4),1/4)>0,\qquad
\mathcal F_2(z(7/20),7/20)<0.
\]
The intermediate value theorem gives the required interior zero.
\end{proof}

\paragraph{The finite inequalities used in the existence argument.}
The following calculations establish
\begin{equation}\label{eq:KL_LL_013_three_bounds}
 \left|\mathcal F_1\!\left(\frac3{25},\frac3{10}\right)\right|
 <\frac1{20000},\qquad
 \mathcal F_2\!\left(\frac3{25},\frac14\right)>\frac1{40000},\qquad
 \mathcal F_2\!\left(\frac3{25},\frac7{20}\right)<-\frac1{40000}.
\end{equation}
They use a rational comparison path and explicit Taylor remainders.
In particular, none of the estimates relies on numerical solution of an
ordinary differential equation or on numerical quadrature.

Set
\[
 \bar b(t)=\sum_{j=0}^4B_jt^j,
 \qquad \bar d(t)=\sum_{j=0}^4D_jt^j,
\]
where all entries of the following table denote exact rational numbers:
\[
\begin{array}{c|rr}
j&B_j&D_j\\\hline
0&0.12&1\\
1&-0.01607549&0.02859510\\
2&0.00907491&-0.01657844\\
3&0.00509198&-0.00877470\\
4&0.00162947&-0.00265383
\end{array}
\]
Their coefficients imply directly that the comparison path lies in
\[
 0\leq t\leq\frac7{20},\qquad
 \frac1{10}<\bar b(t)<\frac{61}{500},\qquad
 1\leq\bar d(t)<\frac{51}{50}.
\]
For example, discard positive terms to bound $\bar b$ from below and
discard its negative linear term to bound it from above. Factoring $t$
from $\bar d(t)-1$ proves its nonnegativity; discarding its negative
terms gives the upper bound.

Use the functions $g$, $q$, and $s$ defined in the preceding proof.
The elementary Taylor estimates specified below give, for both
$\bar b'+s(t,\bar d)g(t,\bar b,\bar d)$ and
$\bar d'-g(t,\bar b,\bar d)$, the common absolute bound
\[
 \rho(t)=9\cdot10^{-8}
 +5\cdot10^{-9}(1+t+t^2)+11\cdot10^{-9}t^3
 +0.002792t^4+0.000225t^5+0.000076t^6.
\]
Direct integration gives
\[
 \int_0^{7/20}\rho(t)\,\de t<\frac{31}{10^7}.
\]
The vector field $(-sg,g)$ has Lipschitz constant $3/5$ in the maximum
norm on the displayed domain. Since the exact and comparison paths
have the same initial values, Gronwall's inequality and
$e^{21/100}<5/4$ imply
\begin{equation}\label{eq:KL_LL_013_comparison_error}
 \sup_{0\leq t\leq7/20}
 \max\bigl\{|b_{3/25}(t)-\bar b(t)|,
             |d_{3/25}(t)-\bar d(t)|\bigr\}
 <\frac4{10^6}.
\end{equation}

For a fixed proposed switch $T$, let $\overline{\mathcal F}_j(T)$
denote the free-boundary residual evaluated at the contract
$(\bar b(T),\bar d(T))$. Integrating the comparison polynomials as
described below yields the rational bounds
\[
\begin{array}{c|c}
\text{quantity}&\text{strict interval, in units of }10^{-6}\\\hline
\overline{\mathcal F}_1(3/10)&(25,27)\\
\overline{\mathcal F}_2(1/4)&(30,32)\\
\overline{\mathcal F}_2(7/20)&(-28,-26).
\end{array}
\]
The previously established contract-derivative bounds
$|(\mathcal F_1)_b|+|(\mathcal F_1)_d|<4$ and
$|(\mathcal F_2)_b|+|(\mathcal F_2)_d|<1/10$, together with
\eqref{eq:KL_LL_013_comparison_error}, enlarge the first interval by
$16\cdot10^{-6}$ on either side and the other two by
$(2/5)\cdot10^{-6}$. The enlarged intervals establish
\eqref{eq:KL_LL_013_three_bounds}.

\paragraph{Details of the rational Taylor calculations.}
For completeness, the following finite rules specify all the arithmetic
behind the two preceding bounds. They also give a way to check the
calculations without differentiating the ODE repeatedly.

On $|u|\leq a$, represent a function by $(p,e)$ if $p$ is a polynomial
of degree at most six and its uniform distance from the function is at
most $e$. Put $|p|_a=\sum_j|p_j|a^j$. Addition adds polynomials and
errors. For multiplication, form the polynomial product, discard terms
of degree greater than six, and add their coefficient norm to
\[
 |p|_a e_q+|q|_a e_p+e_pe_q.
\]
These rules preserve the asserted error bound.
For a pair $(p,e)$, take $c=p(0)$ and represent $v=g-c$ by
$(p-c,e)$. In the reciprocal rule below use
$\delta=(|p-c|_a+e)/|c|$; in the exponential rule use
$\delta=|p-c|_a+e$. The indicated strict upper bounds on $\delta$
hold in every application below.

If $g=c+v$ and $\sup|v/c|\leq\delta<1$, use
\[
 \frac1g=\frac1c\sum_{j=0}^6(-v/c)^j+R,
 \qquad |R|\leq\frac{\delta^7}{|c|(1-\delta)}.
\]
For exponentials, write $g=c+v$, where $0\leq c\leq1/2$ and
$\sup|v|\leq\delta<1/2$, and use
\[
 \sum_{j=0}^8\frac{c^j}{j!}\leq e^c
 \leq\sum_{j=0}^8\frac{c^j}{j!}+\frac{2c^9}{9!},
 \qquad
 \left|e^v-\sum_{j=0}^6\frac{v^j}{j!}\right|
 \leq\frac{2\delta^7}{7!}.
\]
Both exponential estimates follow from $e^{1/2}<2$ and the Taylor
remainder formula. In each rule errors already attached to an argument
are propagated through the preceding addition and multiplication rules.
All polynomial coefficients can be retained as exact fractions.
Alternatively, after each operation, round each coefficient down to
a multiple of $10^{-24}$ and add its absolute rounding error times
$a^j$ to the remainder; round the remainder upwards to a multiple of
$10^{-24}$. These outward roundings preserve all the displayed bounds.

For the ODE residual, take $a=7/20$, start with the specified
polynomials $h(t)=4-t$, $\bar b(t)$, and $\bar d(t)$, and apply the
rules successively to
\[
 E=e^{\bar b/h},\quad Z=e^{\bar d/h},\quad
 q=[1+E(1+Z)]^{-1},\quad s=Z/(1+Z),\quad
 g=\frac{\bar d^2}{h}
 \left[2-\frac1{\bar d}-\frac qh
 \left(12-\frac{\bar b}{\bar d}\right)\right].
\]
For either component of the resulting residual, the absolute values
of the polynomial coefficients of degrees zero through six are bounded
respectively by
\[
 (5\cdot10^{-9},\,5\cdot10^{-9},\,5\cdot10^{-9},\,
 11\cdot10^{-9},\,0.002792,\,0.000225,\,0.000076),
\]
and the uniform remainder is less than $9\cdot10^{-8}$. These are
precisely the coefficient and remainder bounds defining $\rho$.

For the integrated residuals, fix
$T\in\{1/4,3/10,7/20\}$ and $(b,d)=(\bar b(T),\bar d(T))$.
Set
\[
 t=\frac{1+T}{2}+u,\qquad a=\frac{1-T}{2},\qquad
 h(t)=\frac{7-T}{2}-u.
\]
Apply the same rules in the variable $u$ to
\begin{align*}
 E&=e^{b/h},& Z&=e^{(b+d)/h},&
 q&=(1+E+Z)^{-1},\\
 p_{12}&=Eq,&p_{13}&=Zq,&
 P&=p_{12}+p_{13},\qquad s=p_{13}/P,\\
 \pi&=(12-b)p_{12}+(13-b-d)p_{13},\\
 \pi_b&=P\left(\frac qh[12-b+s(1-d)]-1\right),\\
 \pi_d&=p_{13}\left(\frac{13-b-d-\pi}{h}-1\right).
\end{align*}
For a resulting pair $(p,e)$ representing $\pi_y$, $y=b,d$, use the
exact polynomial integral and its error bound
\[
 \left|\int_T^1\pi_y(t,b,d)\,\de t
 -2\sum_{\substack{0\leq j\leq6\\j\text{ even}}}
 \frac{p_ja^{j+1}}{j+1}\right|\leq2ae.
\]
Evaluate the endpoint terms in $\mathcal F_1$ and $\mathcal F_2$
using the displayed scalar exponential bound and rational interval
arithmetic, and combine them with these integrated intervals. This
gives the three intervals in the table. All inputs and all remainders
in this calculation are rational.

\begin{proof}[Proof of Lemma~\ref{lem:KL_LL_power_log_convexity}]
The case $q=1$ is immediate. Otherwise, work first in the relative
interior of the simplex. Write $z_i=\mu(x_i)/\nu_i$ and
$S_a=\sum_i\nu_i z_i^a$, so $S_1=1$. For a tangent vector $v$,
with $\sum_i v_i=0$, direct differentiation gives
\[
\mathrm D^2H_q(\mu)[v,v]
=\frac{q(q-1)}{S_q}\sum_i\frac{z_i^{q-2}v_i^2}{\nu_i}
-\frac{q^2}{S_q^2}\left(\sum_i z_i^{q-1}v_i\right)^2.
\]
Subtracting $1/S_{2-q}$ from the coefficient of each $v_i$ and
applying the weighted Cauchy--Schwarz inequality yields
\[
\left(\sum_i z_i^{q-1}v_i\right)^2
\leq\left(S_q-\frac1{S_{2-q}}\right)
\sum_i\frac{z_i^{q-2}v_i^2}{\nu_i}.
\]
Consequently,
\begin{equation}\label{eq:KL_LL_013_power_log_hessian}
\mathrm D^2H_q(\mu)[v,v]
\geq\frac{q(q-S_qS_{2-q})}{S_q^2S_{2-q}}
\sum_i\frac{z_i^{q-2}v_i^2}{\nu_i}.
\end{equation}
It remains to show $S_qS_{2-q}\leq q$.
Put $a=q-1$. Since there are at least two outputs,
$\nu_{\min}\leq1/2$, and the assumed bound on $q$ gives
$0<a\leq\nu_{\min}/(1-\nu_{\min})\leq1$.
For $r>0$, convexity of $\sinh$ on $[0,\infty)$ implies
\[
r^a+r^{-a}-2
=4\sinh^2\!\left(\frac{a|\log r|}{2}\right)
\leq a^2(r+r^{-1}-2).
\]
Pairwise expansion therefore gives
\begin{align*}
S_{1+a}S_{1-a}-1
&=\sum_{i<j}\nu_i\nu_jz_iz_j
\left[\left(\frac{z_i}{z_j}\right)^a
+\left(\frac{z_j}{z_i}\right)^a-2\right]\\
&\leq a^2\sum_{i<j}\nu_i\nu_j(z_i-z_j)^2
=a^2(S_2-1)\\
&\leq a^2\frac{1-\nu_{\min}}{\nu_{\min}}\leq a,
\end{align*}
where $S_2=\sum_i\mu(x_i)^2/\nu_i\leq1/\nu_{\min}$.
Equation~\eqref{eq:KL_LL_013_power_log_hessian} is thus
nonnegative. This proves convexity in the relative interior;
continuity of $H_q$ extends it to the closed simplex.
\end{proof}

\begin{proof}[Proof of Lemma~\ref{lem:KL_LL_013_pool_certificate}]
Throughout this proof, abbreviate
\(T:=\theta^P\), \(H_P:=h(T)\),
\(b:=w^P(12)\), and \(d:=\Delta w^P\).
All derivatives in \(H\) hold this pooled contract fixed.
Write \(D_R(H):=R_{13}(H)-R_{12}(H)\).

\emph{Existence and endpoint regularity.}
For \(H>3\), the definition of \(R_x\) gives
\[
R_0(H)<R_{12}(H)<R_{13}(H),\qquad
D_R(H)=\frac d9(H-3)+O((H-3)^2).
\]
Indeed,
\(R_x'(3)=[w^P(x)-\sum_y\mu_y^P(3)w^P(y)]/9\).
Consequently \(B_0(H):=B(H)+(H-3)^{-1}\) extends analytically to
\(H=3\). Define
\[
\omega(H):=(H-3)
\exp\left[-\int_3^H B_0(u)\,\de u\right],
\qquad
N(H):=\frac{1}{\omega(H)}\int_3^H\omega(u)A(u)\,\de u.
\]
Since \(\omega'=-B\omega\), this function solves
\eqref{eq:KL_LL_013_N_ODE}. Every nonzero homogeneous solution is
proportional to \(1/\omega(H)\) and is unbounded at \(3\), so this is
the unique bounded solution. The formula also gives
\[
N(H)=\frac{A(3)}2(H-3)+O((H-3)^2),\qquad
\gamma(1):=\lim_{r\uparrow1}\gamma(r)=\frac{3A(3)}{2d}.
\]
The ratios in \eqref{eq:KL_LL_013_eta_pool} have finite limits at
\(H=3\). Hence all three pooled multipliers extend continuously
to \(r=1\), and \(M:=\int_T^1\gamma(r)\,\de r\) is finite.

\emph{The two boundary conditions.}
Set \(K_r:=\Lambda_r-(1-r)\), and let
\(V_r:=h(r)\log[Z^P(h(r))/3]\).
For \(y\in\{b,d\}\), subscripts \(y\) denote differentiation of
the common contract, whereas primes denote differentiation in \(r\).
Then
\[
(V_r)_b=P_r,\qquad (V_r)_d=P_rs_r,\qquad
V_r'=D_{KL}(\mu_r\Vert\nu).
\]
Direct substitution in \eqref{eq:KL_LL_013_N_ODE} gives
\(\Lambda_r'=\gamma(r)-\eta_0(r)\), and
\eqref{eq:KL_LL_013_L_pool}--\eqref{eq:KL_LL_013_eta_pool}
give the reportwise first-order conditions.
To integrate these conditions, observe that
\[
3\log\left[\sum_x\nu(x)
 \left(\frac{\mu_r(x)}{\nu(x)}\right)^{h(r)/3}\right]
=V_1-V_r.
\]
Thus, along the common-contract family, the reportwise objective is
\[
\mathcal Q_r(\mu_r)
=\pi(r,b,d)+V_r+K_rV_r'
-\eta_0(r)V_r-\gamma(r)(V_1-V_r).
\]
When differentiating this expression in \(y\), hold the multipliers
fixed at their values for the candidate contract. Since
\(K_r'=1+\gamma(r)-\eta_0(r)\), stationarity implies
\[
0=\pi_y(r,b,d)+\bigl[K_r(V_r)_y\bigr]'
-\gamma(r)(V_1)_y.
\]
Integration therefore gives
\[
\int_T^1\pi_y(r,b,d)\,\de r
=K_T(V_T)_y+(M-K_1)(V_1)_y,
\qquad K_1=3(1-1/d).
\]
Let \(\mathcal F_1\) be the left-hand side minus the right-hand side
of \eqref{eq:KL_LL_013_FB_level}, and let
\(\mathcal F_2:=\int_T^1[\pi_d(r,b,d)-s_T\pi_b(r,b,d)]\,\de r\).
The preceding identities yield
\[
\mathcal F_1=\frac{H_PP_T}{d}N(H_P)+(M-K_1)P_1,
\qquad
\mathcal F_2=(M-K_1)P_1(s_1-s_T).
\]
Since \(s_1>s_T\), the two free-boundary equations are equivalent to
\[
N(H_P)=0,\qquad M=3(1-1/d).
\]

\emph{Signs.}
The established candidate bounds give
\(1/10<b<61/500\), \(1<d<51/50\), and \(3<H_P<15/4\).
On \([3,H_P]\), these imply
\[
\frac7{25}<\mu_0^P(H)<\frac13,\qquad
12d-b>\frac{59}{5},\qquad
\frac{\sum_x\mu_x^P(H)w^P(x)}H<\frac25.
\]
For the first lower bound it suffices to use
\(e^{61/1500}<21/20\) and \(e^{571/1500}<3/2\);
both follow from
\(e^u\leq1+u+u^2/[2(1-u/3)]\) for \(0\leq u<3\).
Differentiating \eqref{eq:KL_LL_013_A} now gives
\[
\begin{aligned}
H^2A'(H)
&=\frac{(12d-b)\mu_0^P(H)}H
\left[\frac{\sum_x\mu_x^P(H)w^P(x)}H-2\right]+2d-1\\
&<-\frac{(59/5)(7/25)}4\frac85+\frac{26}{25}<0.
\end{aligned}
\]
Moreover,
\[
A(3)>\frac{(59/5)(7/25)}9-\frac{26/25}{3}
=\frac{23}{1125}>0,
\qquad
A(H_P)=-\frac{(\Delta w)'_{T-}}d<0,
\]
where the last equality follows from
\eqref{eq:KL_LL_013_separating_ODE}.
Thus \(A\) changes sign exactly once. The numerator in the formula
for \(N\) first increases strictly and then decreases strictly to zero
at \(H_P\). It follows that \(N(H)>0\) for \(3<H<H_P\).
In addition, direct addition of
\eqref{eq:KL_LL_013_A} and \eqref{eq:KL_LL_013_B} gives
\[
A(H)+B(H)=\frac d{H^2}
\left\{\mu_0^P(H)
\left[12+\frac{R_0(H)-R_{12}(H)}{D_R(H)}\right]
-2H-\frac1{D_R(H)}\right\}<0.
\]
A first upward crossing of \(N=1\) would consequently have \(N'<0\),
which is impossible. Hence \(0<N<1\) in the interior.
This proves \(\gamma(r)>0\) for \(T<r\leq1\),
\(\gamma(T)=0\), and
\(\widetilde\Phi(r,\Lambda_r)=h(r)[1-N(h(r))]/d>0\).
For the remaining sign, set \(\delta:=1/3-1/H>0\). Then
\[
\frac{R_{12}(H)-R_0(H)}{D_R(H)}
=\frac{1-e^{-b\delta}}{e^{d\delta}-1}\leq\frac bd.
\]
Substitution into \eqref{eq:KL_LL_013_eta_pool} yields
\(\eta_0(r)\geq\mu_0^P(h(r))(12-b/d)/h(r)>0\),
including both endpoints by continuity.

\emph{The stock identity.}
The identity \(N(H_P)=0\) matches the pooled and separating formulas
for \(\Lambda\) at \(T\), while \(\Lambda_1=M\).
Integrating \(\Lambda'=\gamma-\eta_0\) therefore gives
\[
\Lambda_r=\int_r^1\eta_0(t)\,\de t+
\int_T^r\gamma(t)\,\de t,\qquad T\leq r\leq1.
\]
Below \(T\), continuity and \(\Lambda'=-\eta_0\) give
\(\Lambda_r=\int_r^1\eta_0(t)\,\de t\).
These are exactly \eqref{eq:KL_LL_multiplier_stock}, since
\(\Gamma_T=M\delta_{\{1\}}\) and \(\Gamma_R\) has density
\(\gamma(r)\mathbf 1_{[T,1]}(r)\).
Finally \(\Lambda_0=1\), and all densities are continuous and integrable.
\end{proof}

    \end{document}